\documentclass[12pt]{article}
\usepackage[utf8]{inputenc}
\usepackage{mathtools}
\usepackage[margin=1.0in]{geometry}

\usepackage{multirow,multicol}
\usepackage{amsfonts}
\usepackage{amsmath,amssymb,amsthm}
\usepackage[mathscr]{euscript}
\usepackage{bbm}
\usepackage{graphicx,color}
\usepackage{comment}
\usepackage{pifont} 
\usepackage{setspace}
\usepackage{textcomp}
\usepackage{makecell, booktabs}
\usepackage{caption}
\usepackage{tabularx}
\usepackage{xr}
\usepackage{array} 
\usepackage{titlesec}
\usepackage{algorithm}
\usepackage{algorithmic}
\usepackage{float}
\usepackage{subfig}
\usepackage[english]{babel}
\usepackage[toc,page]{appendix}
\usepackage{natbib}
\newtheorem{assumption}{Assumption}

\newtheorem{definition}{Definition}
\newtheorem{lemma}{Lemma}
\newtheorem{proposition}{Proposition}
\newtheorem{theorem}{Theorem}

\newtheorem{remark}{Remark}

\newcommand{\bm}{\boldsymbol}
\newcommand{\cm}[1]{\mbox{\boldmath$\mathscr{#1}$}}

\newcommand{\Fr}{{\mathrm{F}}}
\newcommand{\op}{{\mathrm{op}}}

\definecolor{turquoise}{rgb}{0.03, 0.7, 0.87}

\newcommand{\defeq}{\vcentcolon=}
\newcommand{\norm}[1]{\left\lVert#1\right\rVert}
\DeclareMathOperator*{\vectorize}{vec}
\DeclareMathOperator*{\rank}{rank}
\DeclareMathOperator*{\trace}{tr}
\DeclareMathOperator*{\argmin}{arg\,min}
\DeclareMathAlphabet\mathrsfso{U}{rsfso}{m}{n}

\makeatletter

\numberwithin{equation}{section}
\allowdisplaybreaks
\usepackage[colorlinks,linkcolor=blue,citecolor=blue]{hyperref}
\usepackage{titling}
\usepackage{authblk} 
		
\title{High-dimensional Autoregressive Modeling for Time Series Data with Hierarchical Structures
}

\author{
	\centering
	Lan Li, Shibo Yu, Yingzhou Wang and Guodong Li
	\thanks{Address for correspondence: Guodong Li, Department of Statistics and Actuarial Science, University of Hong Kong, Hong Kong, China. Email: gdli@hku.hk}

	\textit{Department of Statistics and Actuarial Science, University of Hong Kong}
}
\date{\today}

\begin{document}
\maketitle
	
\begin{abstract}
	High-dimensional data, especially time-dependent data with increasingly complex structures, have become ubiquitous in modern applications, and hierarchical relationships among variables are increasingly common.
	However, there is still a lack of supervised learning tools in the literature for them.
	To fill this gap, we introduce a new model-designing framework, and it then combines with unsupervised factor modeling tools to form an efficient and interpretable autoregressive model for high-dimensional time series with hierarchical structures.
	An ordinary least squares estimation is considered, and its non-asymptotic properties are established.
	Moreover, we propose an algorithm to search for estimates, and a boosting method is also suggested for hyperparameter selection.
	Simulation experiments are conducted to evaluate finite-sample performance of the proposed methodology, and its usefulness is demonstrated by an application to the Personality‑120 dataset.
	
\end{abstract}

\newpage
\section{Introduction}
\label{sec:intro}

With the rapid advancement of information technology, multi-dimensional or tensor-valued data are now ubiquitous, and particularly many of them are time-dependent. Examples can be found in medical imaging \citep{zhou2013tensor, li2021tensor}, digital marketing \citep{bi2018multilayer, hao2021sparse}, economics and finance \citep{chen2022factor,wang2024high} and psychology \citep{tang2018bayesian}.
It is arguably the most important task to conduct supervised learning on these data, say tensor regression \citep{lock2018tensor,raskutti2019convex} and tensor autoregression \citep{li2021multi,wang2024high}, while existing methods typically treat all modes in an equal manner, thus mainly for the data with fully juxtaposing structures. 
However, for these observed tensors, some modes may be nested within others, forming hierarchical relationships, and they can be encountered more readily for the data with a larger number of modes \citep{gao2023divide, qiao2025exact}. 
Meanwhile, it is still an open problem to model and then to predict these high-dimensional tensor-valued data, especially time-dependent data, with hierarchical structures.

As an illustrative example, we consider the Personality-120 dataset, which contains 24 questions for each of five broad personality traits, ending up with 120 questions in total. For any personality trait, each of its 24 questions is designed to quantitatively assess a specific psychological aspect, and the answers to those 24 questions jointly can be used to roughly portray an individual's personality under this trait; see more empirical examples in psychometrics \citep{johnson2014measuring,qiao2025exploratory}.
In addition, the dataset is collected over time from participants of different ages and genders, forming 12 demographic groups. As a result, at each time point, there is an observed three-mode tensor with dimensions of $12~(\text{demographic groups})\times 5~(\text{personality traits})\times 24~(\text{questions})$, where the last two modes reflect a hierarchical structure (questions nested within traits), while the 24 questions under each personality trait are arranged in a similar way.
Moreover, the first mode (demographic groups) represents a cross-cutting covariate that does not belong to this hierarchy.
It is challenging to construct an efficient time series model for the above example since it has $12\times 5\times 24=1440$ variables yet with only around 100 time points.
On the other hand, to achieve an appropriate physical interpretation, another layer of difficulties is imposed due to the inherent hierarchical structure between personality traits and questions.

As arguably the most widely used model, autoregression (AR) has been a primary workhorse for many high-dimensional tasks on time series with the observed vector, matrix or tensor at each time point \citep{basu2015regularized,Chen_Xiao_Yang2021,wang2024high}.
The general AR model has $O(P^2)$ parameters, where $P$ is the number of variables, say $P=1440$ in the Personality-120 example, and it hence is necessary to restrict the parameter space such that a reliable estimation can be achieved.
A direct approach in the literature is to assume that the coefficient matrices or tensors are sparse and then to apply sparsity-inducing regularized estimation \citep{basu2015regularized, wilms2023sparse, zheng2025interpretable}. 
The sparse AR model has been widely used, say in genetics and neuroscience \citep{zhou2013tensor, bi2022modeling, wu2023sparse}, where dependence among variables is believed to be sparse.
However, for the Personality-120 example, there exists strong dependence among all $P=1440$ variables, especially each 24 variables from the same personality trait, and it is more reasonable to assume that these  variables are driven by a small number of common latent factors \citep{qiao2025exact}.
Similar scenario can be found for macroeconomic variables, where the factor modeling is usually considered \citep{lam2012,Bai2016}.

Another important method for dimension reduction is to impose low-rank structures to coefficient matrices \citep{negahban2011} or tensors \citep{han2022optimal} and, as one of the most commonly used tools for compressing tensors \citep{kolda2009tensor}, Tucker decomposition \citep{tucker1966some} can be directly applied to the coefficient tensor of AR models \citep{wang2022high,wang2024high,huang2025supervised}.
On the one hand, although the corresponding dimension of parameter spaces can be reduced dramatically, Tucker decomposition is well known not to be able to effectively compress higher-mode tensors, and it loses efficiency very quickly as the mode of tensor increases, say greater than four or five; see \cite{oseledets2011tensor}.
Unfortunately, consider an AR model for tensor-valued time series with mode three, and the mode of its coefficient tensor will be six. 
On the other hand, together with the regression setting, Tucker decomposition will lead to a nice interpretation of supervised factor modeling \citep{wang2024high,huang2025supervised}.
Specifically, it is equivalent to first projecting responses and predictors into a small number of latent factors, termed response and predictor factors, respectively, and then there exists a regression relationship between the two types of factors.
The response factors are used to summarize all predictable components, while the predictor ones contain all useful information of predictors in terms of prediction. However, the above physical interpretation will disappear for the data with hierarchical structures, such as the Personality-120 example; see Section \ref{subsec:sfm} for details.

CP \citep{harshman1970foundations} and tensor train \citep{oseledets2011tensor} decompositions are another two popular methods for compressing tensors, and both of them are more efficient than Tucker decomposition \citep{kolda2009tensor}.
However, it is difficult to interpret fitted models when CP decomposition is applied to coefficient tensors of tensor-on-tensor regression \citep{lock2018tensor} and tensor autoregression.
Tensor train (TT) decomposition may still lead to a supervised factor modeling interpretation when it is applied to tensor regression \citep{liu2020low,qin2025computational} and autoregression \citep{si2024efficient}, while it is limited to data with a very special structure; see Remark \ref{remark:TT} for details.
To the best of our knowledge, for tensor-valued data, especially time-dependent data, with hierarchical structures, it is infeasible to achieve an efficient model with physical interpretations by directly applying an existing tensor decomposition method to coefficient tensors of the corresponding model. 
 
Motivated by the supervised factor modeling interpretation of Tucker tensor regression, this paper introduces a new framework for constructing tensor regression and autoregressive models.
Specifically, two factor modeling approaches are first applied to responses and predictors to extract features, respectively, and a low-rank structure is then imposed to the coefficient tensor so that it admits a regression relationship between response and predictor features. We may even employ different factor modeling methods for responses and predictors; see Section \ref{subsec:sfm} for more details.
Note that, for the data with general hierarchical structures, traditional factor modeling methods can be used to extract features or factors; see, for example, higher-order factor models \citep{yung1999relationship}, group factor models \citep{klami2014group}, and hierarchical factor models \citep{schmid1957development, qiao2025exact}. 
These tools are unsupervised learning methods, and they admit various hierarchical interpretations by either extracting features recursively along the hierarchy, or directly from different groups of variables.
We can apply them to construct tensor regression and autoregressive models within the new framework, while the loading matrices assigned to each lower-level mode are all different, leading to a huge number of parameters and hence excluding high-dimensional settings; see Section \ref{subsec:nested} for details.
On the other hand, for many recent hierarchical data such as the Personality-120 example, they are arranged in tensor forms, i.e., each mode has the same number of variables. Moreover, the variables at each mode are also arranged in similar ways, and the corresponding factors may be formed with the same loading matrices.

\begin{table}[tp]
	\centering
	\caption{Comparison of new-framework models with the higher-order factor model (HOFM), group factor model (GFM) or hierarchical factor model (HFM) being used to extract features, in terms of model complexity and interpretability for hierarchical data. 
		The traditional framework refers to the case with tensor decomposition methods being directly applied.
		The coefficient tensor has $d$ modes, and each has the dimension of $p$ and rank of $r$.}
	\label{tab:intro-comparison}
	\resizebox{0.95\textwidth}{!}{
		\begin{tabular}{@{}c c *{3}{>{\centering\arraybackslash}p{1.5cm}} *{3}{>{\centering\arraybackslash}p{1.5cm}}@{}}
			\toprule
			\multicolumn{1}{c}{\multirow{2}{*}{Models}} & \multicolumn{4}{c}{New framework} & \multicolumn{3}{c}{Traditional framework}  \\
			\cmidrule(lr){2-5} \cmidrule(l){6-8}
			\multicolumn{1}{c}{}  & Ours & HOFM & GFM & HFM & CP & Tucker & TT  \\ 
			\midrule
			Number of parameters: $O(\cdot)$  & $pr^2d$ & $rp^d$ & $rp^d$ & $rp^d$ & $prd$ & $prd+r^d$ & $pr^2d$ \\
			Interpretation: Hierarchical & $\checkmark$ & $\checkmark$ & $\checkmark$ & $\checkmark$ & $\times$ & $\times$ & $\checkmark$ \\
			Interpretation: Crossed  & $\checkmark$ & $\times$ & $\times$ & $\times$ & $\times$ & $\checkmark$ & $\times$\\ 
			\bottomrule
		\end{tabular}
	}
\end{table}

The first contribution of this paper is to overcome the difficulty of traditional factor modeling tools for hierarchical data by considering shared loading matrices within each layer of the hierarchy; see Table \ref{tab:intro-comparison} for a comparison of their model complexity and interpretability for hierarchical data.
Moreover, the real-world hierarchical data most likely have a mixture structure, i.e., some modes are nested, while others are crossed.
We hence apply the new modeling framework to construct an efficient and interpretable regression model for these data with high dimensions in Section \ref{subsec:nested-factorial}, and it is further extended to AR models for time-dependent data in Section \ref{subsec:ar}.
The second contribution is to consider an ordinary least squares estimation in Section \ref{sec:theory}, and its non-asymptotic properties are established. 
This is non-trivial since the corresponding loading matrices are non-orthogonal and, as a cost, an additional term of $\log T$ is added to the corresponding estimation error bound.
Finally, Section \ref{sec:algorithms} proposes an algorithm to search for the estimates, and a boosting-based method is introduced for hyperparameter selection due to the presence of hierarchical structures.

In addition, Section \ref{sec:simulation} conducts simulation experiments to evaluate numerical performance of the proposed methodology, and it is further applied to the Personality-120 dataset in Section \ref{sec:real-data}. A short conclusion and discussion is given in Section \ref{sec:conclusion}. All technical proofs and additional numerical results are relegated to the Supplementary Material.

Throughout the paper, we denote vectors by boldface small letters, e.g., $\bm{x}$; matrices by boldface capital letters, e.g., $\bm{X}$; tensors by boldface Euler capital letters, e.g., $\cm{X}$.
For a $d$-th order tensor $\cm{X} \in \mathbb{R}^{p_1 \times p_2 \times \cdots \times p_d}$, denote by $\vectorize(\cm{X})$, $\cm{X}_{(s)}$, and $[\cm{X}]_s$ its vectorization, mode matricization, and sequential matricization at mode $s$ with $1 \leq s\leq d$, respectively.
The Frobenius norm of $\cm{X}$ is defined as $\norm{\cm{X}}_{\mathrm{F}}=\sqrt{\langle\cm{X},\cm{X}\rangle}$, where $\langle\cdot,\cdot\rangle$ is the inner product.
Moreover, the $(i_1 i_2 \ldots i_d)$-th element of $\cm{X}$ is denoted by $\cm{X}_{i_1i_2\ldots i_d}$,
with $1\leq k\leq d$ and $1\leq i_k \leq p_k$.
We denote $\times_s$ the $s$-th mode matrix product, and $\otimes$ is the Kronecker product.
For a generic matrix $\bm{X}$, we denote by $\bm{X}^\top$, $\norm{\bm{X}}_{\mathrm{F}}$, $\norm{\bm{X}}_2$, and $\vectorize(\bm{X})$ its transpose, Frobenius norm, spectral norm, and a long vector obtained by stacking all its columns, respectively. If $\bm{X}$ is further a square matrix, we then denote its minimum and maximum eigenvalue by $\lambda_{\min}(\bm{X})$ and $\lambda_{\max}(\bm{X})$, respectively. Denote a set of orthonormal matrices by $\mathbb{O}^{p\times r}=\{\bm{X}\in\mathbb{R}^{p\times r}:\bm{X}^\top\bm{X}=\bm{I}_r\}$ for $p\geq r$, where $\bm{I}_r\in\mathbb{R}^{r\times r}$ is the identity matrix. 
For two real-valued sequences $x_k$ and $y_k$, we denote $x_k\gtrsim y_k$ or $x_k\lesssim y_k$ if there exists a $C>0$ such that $x_k\geq Cy_k$ or $x_k\leq Cy_k$ for all $k$, respectively. In addition, $x_k\asymp y_k$ if $x_k\gtrsim y_k$ and $x_k\lesssim y_k$.

\section{Model settings}
\label{sec:method}
\subsection{Tucker tensor regression and supervised factor modeling}
\label{subsec:sfm}
We begin with a general regression model involving tensor-valued responses and predictors:
\begin{equation}
	\label{eq:regmodel}
	\cm{Y}_t = \langle\cm{A},\cm{X}_t\rangle+\cm{E}_t \hspace{2mm}\text{for}\hspace{2mm} 1\leq t\leq T,
\end{equation}
where $\cm{Y}_t$, $\cm{E}_t \in\mathbb{R}^{q_1\times q_2\times \cdots\times q_N}$ are the response and error term, respectively, $\cm{X}_t\in\mathbb{R}^{p_1\times p_2\times\cdots\times p_M}$ is the predictor, $\cm{A}\in\mathbb{R}^{q_1\times\cdots\times q_N\times p_1\times\cdots\times p_M}$ is the coefficient tensor, and $T$ is the sample size.
Suppose that the coefficient tensor $\cm{A}$ has multilinear low ranks $(s_1, \ldots, s_N, r_1, \ldots, r_M)$, i.e., $\text{rank}(\cm{A}_{(n)})=s_{n}$ and $\text{rank}(\cm{A}_{(N+m)})=r_{m}$ for $1\leq n \leq N$ and $1\leq m \leq M$, and then there exists a Tucker decomposition \citep{de2000multilinear,tucker1966some}, 
\begin{equation}\label{tucker}
	\cm{A} = \cm{G} \times_1 \bm{U}_1 \times_2 \cdots \times_N \bm{U}_N \times_{N+1} \bm{V}_1 \times_{N+2} \cdots \times_{N+M} \bm{V}_M,
\end{equation}
where $\cm{G}\in\mathbb{R}^{s_1\times\cdots\times s_N \times r_1\times\cdots\times r_M}$ is the core tensor, and $\bm{U}_n\in\mathbb{R}^{q_n\times s_n}$ with $1\leq n\leq N$ and $\bm{V}_m\in\mathbb{R}^{p_m\times r_m}$ with $1\leq m\leq M$ are factor matrices.
Model \eqref{eq:regmodel}, together with the low-rank coefficient tensor at \eqref{tucker}, is referred to as the Tucker tensor regression model.

For simplicity, we denote $q_{N+m}=p_m$, $s_{N+m}=r_m$ and $\bm{U}_{N+m}=\bm{V}_m$ for $1\leq m\leq M$. 
Note that the Tucker decomposition at \eqref{tucker} is not unique, since 
$\cm{A} = \cm{G} \times_1 \bm{U}_1 \times_2 \cdots \times_{N+M} \bm{U}_{N+M} =\left(\cm{G}\times_1 \bm{O}_1 \times_2 \cdots \times_{N+M} \bm{O}_{N+M}\right) \times_1\left(\bm{U}_1 \bm{O}_1^{-1}\right) \times_2\cdots\times_{N+M}\left(\bm{U}_{N+M} \bm{O}_{N+M}^{-1}\right)$ for any invertible matrices $\boldsymbol{O}_i \in \mathbb{R}^{s_i \times s_i}$ with $1\leq i \leq N+M$.
Without loss of generality, we restrict these factor matrices to be orthonormal, i.e., $\bm{U}_j\in\mathbb{O}^{q_j\times s_j}$ for each $1 \leq j \leq N+M$; see, for example, the higher-order singular value decomposition (HOSVD) in \cite{kolda2009tensor}.
Consequently, the Tucker tensor regression model can be rewritten into
\begin{align}
	\begin{split}\label{eq:reg-tucker}
		\cm{Y}_t& \times_{n=1}^N\bm{U}_n^\top= \langle\cm{G},\cm{X}_t\times_{m=1}^M\bm{V}_m^\top\rangle +\cm{E}\times_{n=1}^N\bm{U}_n^\top \hspace{2mm}\text{or equivalently,}\\
		\vectorize(\cm{Y}_t)& = (\bm{U}_N \otimes \dots \otimes \bm{U}_1) [\cm{G}]_N (\bm{V}_M \otimes \cdots \otimes \bm{V}_1)^\top \vectorize(\cm{X}_t) + \vectorize(\cm{E}_t).	
	\end{split}
\end{align}
Let $\bm{U} = \bm{U}_N \otimes \dots \otimes \bm{U}_1\in\mathbb{O}^{Q\times s}$ and $\bm{V} = \bm{V}_M \otimes \cdots \otimes \bm{V}_1\in\mathbb{O}^{P\times r}$, where $Q\defeq\prod_{n=1}^{N} q_n$, $P\defeq\prod_{m=1}^{M} p_m$, $s\defeq\prod_{n=1}^{N} s_n$, and $r\defeq\prod_{m=1}^{M} r_m$.
Note that $\bm{U}_n$'s and $\bm{V}_m$'s are still not unique, and hence $\bm{U}$ and $\bm{V}$. However, the corresponding column spaces $\text{col}(\bm{U})$ and $\text{col}(\bm{V})$, as well as their projection matrices $\bm{P}_{\bm{U}} = \bm{U}\bm{U}^\top$ and $\bm{P}_{\bm{V}} = \bm{V}\bm{V}^\top$, can be uniquely defined.

Note that model \eqref{eq:reg-tucker} involves dimension reduction on both the responses and predictors. We first consider responses, and project $\vectorize(\cm{Y}_t)$ onto $\text{col}(\bm{U})$ and its orthogonal complement, i.e., $\vectorize(\cm{Y}_t) = \bm{P}_{\bm{U}}\vectorize(\cm{Y}_t) + \bm{P}_{\bm{U}}^\perp\vectorize(\cm{Y}_t)$ with $\bm{P}_{\bm{U}}^\perp = \bm{I}_Q - \bm{U}\bm{U}^{\top}$. It then holds that
\begin{equation*}\label{eq:reg-tucker-sfm}
	\bm{P}_{\bm{U}}\vectorize(\cm{Y}_t) = \bm{U} [\cm{G}]_N \bm{V}^\top \vectorize(\cm{X}_t) + \bm{P}_{\bm{U}}\vectorize(\cm{E}_t) \quad\text{and}\quad \bm{P}_{\bm{U}}^\perp\vectorize(\cm{Y}_t) = \bm{P}_{\bm{U}}^\perp\vectorize(\cm{E}_t),
\end{equation*}
where all predictable information of $\vectorize(\cm{Y}_t)$ is contained in $\text{col}(\bm{U})$.
Moreover,
\begin{equation*}
	\mathrm{cov}\left\{\vectorize(\cm{Y}_t), \vectorize(\cm{X}_t)\; \vert \; \bm{P}_{\bm{V}}\vectorize(\cm{X}_t)\right\}= \bm{0},
\end{equation*}
indicating that $\text{col}(\bm{V})$ contains all useful information of predictors in terms of predicting responses. As a result, define $\bm{f}_t^{\mathrm{response}} =\vectorize(\cm{Y}_t \times_{n=1}^N\bm{U}_n^\top) = \bm{U}^\top \vectorize(\cm{Y}_t)\in\mathbb{R}^s$ and $\bm{f}_t^{\mathrm{predictor}} =\vectorize(\cm{X}_t \times_{m=1}^M\bm{V}_m^\top) = \bm{V}^\top \vectorize(\cm{X}_t)\in\mathbb{R}^r$  to be the \textit{response and predictor factors}, respectively, and then the Tucker tensor regression can be well interpreted from a supervised factor modeling perspective \citep{wang2022high,wang2024high,huang2025supervised}.

The above supervised factor modeling interpretation also motivates us to introduce a new framework to construct a tensor regression model, especially for data with complex structures beyond the capability of Tucker decomposition, or for non-orthogonal loadings in the feature extraction process such as the CP decomposition \citep{harshman1970foundations}.
Specifically, we first extract features or factors from responses and predictors by 
\begin{equation}
	\label{eq:factors}
	\bm{f}_t^{\mathrm{predictor}} = \mathcal{M}(\cm{X}_t)
	\quad\text{and}\quad 
	\bm{f}_t^{\mathrm{response}} = \mathcal{N}(\cm{Y}_t),
\end{equation}
respectively.
Here, $\mathcal{M}(\cdot):\mathbb{R}^{p_1\times\cdots\times p_M} \to \mathbb{R}^r$ and $\mathcal{N}(\cdot):\mathbb{R}^{q_1\times\cdots\times q_N} \to \mathbb{R}^s$ are feature-extracting operators, with $r$ and $s$ being the number of extracted predictor and response features, respectively, and they do not necessarily belong to the same type.
A specific structure is then assumed to the coefficient tensor $\cm{A}$ of model \eqref{eq:regmodel} such that $\bm{f}_t^{\mathrm{response}} = \bm{\Theta} \bm{f}_t^{\mathrm{predictor}}+\widetilde{\bm{e}}_t$, where $\bm{\Theta}\in\mathbb{R}^{s\times r}$ and $\widetilde{\bm{e}}_t\in\mathbb{R}^{s}$ are the corresponding coefficient matrix and compressed error term, respectively.
To this end, if operators $\mathcal{M}$ and $\mathcal{N}$ are linear, i.e., $\mathcal{M}(\cm{X}_t) = \bm{\Lambda}_x^\top \vectorize(\cm{X}_t)$ and $\mathcal{N}(\cm{Y}_t) = \bm{\Lambda}_y^\top \vectorize(\cm{Y}_t)$, with non-degenerating loading matrices $\bm{\Lambda}_x\in\mathbb{R}^{P\times r}$ and $\bm{\Lambda}_y\in\mathbb{R}^{Q\times s}$, we then can set the coefficient tensor to 
\begin{equation}
	\label{eq:A-form}
	[\cm{A}]_N = \bm{\Lambda}_y \left(\bm{\Lambda}_y^\top \bm{\Lambda}_y \right)^{-1} \bm{\Theta} \bm{\Lambda}_x^\top.
\end{equation}
Moreover, if $\bm{\Lambda}_y$ is orthonormal, then $\bm{\Lambda}_y^\top \bm{\Lambda}_y =\bm{I}_s$ and $[\cm{A}]_N = \bm{\Lambda}_y \bm{\Theta} \bm{\Lambda}_x^\top$.

When Tucker decomposition is applied as in \eqref{tucker}, the loading matrices $\bm{\Lambda}_y=\bm{U}$ and $\bm{\Lambda}_x=\bm{V}$ extract features simultaneously from all modes of the response and predictor, respectively. Such simultaneous compression may not be suitable for more complex data, say with hierarchical structures in the next subsection, and a new methodology is required.

\subsection{Purely hierarchical structure}
\label{subsec:nested}
According to technological advances, the available multi-dimensional or tensor-valued data have become more and more complex, and it is very common to observe hierarchical relationships among some modes with their variables being arranged in a similar way.
In order to exploit their structural information, we attempt to design a regression model using the proposed framework at \eqref{eq:factors} and \eqref{eq:A-form}.
This subsection starts from the simple case with purely hierarchical responses and predictors.

\begin{figure}[t]
	\centering
	\includegraphics[width=1\textwidth]{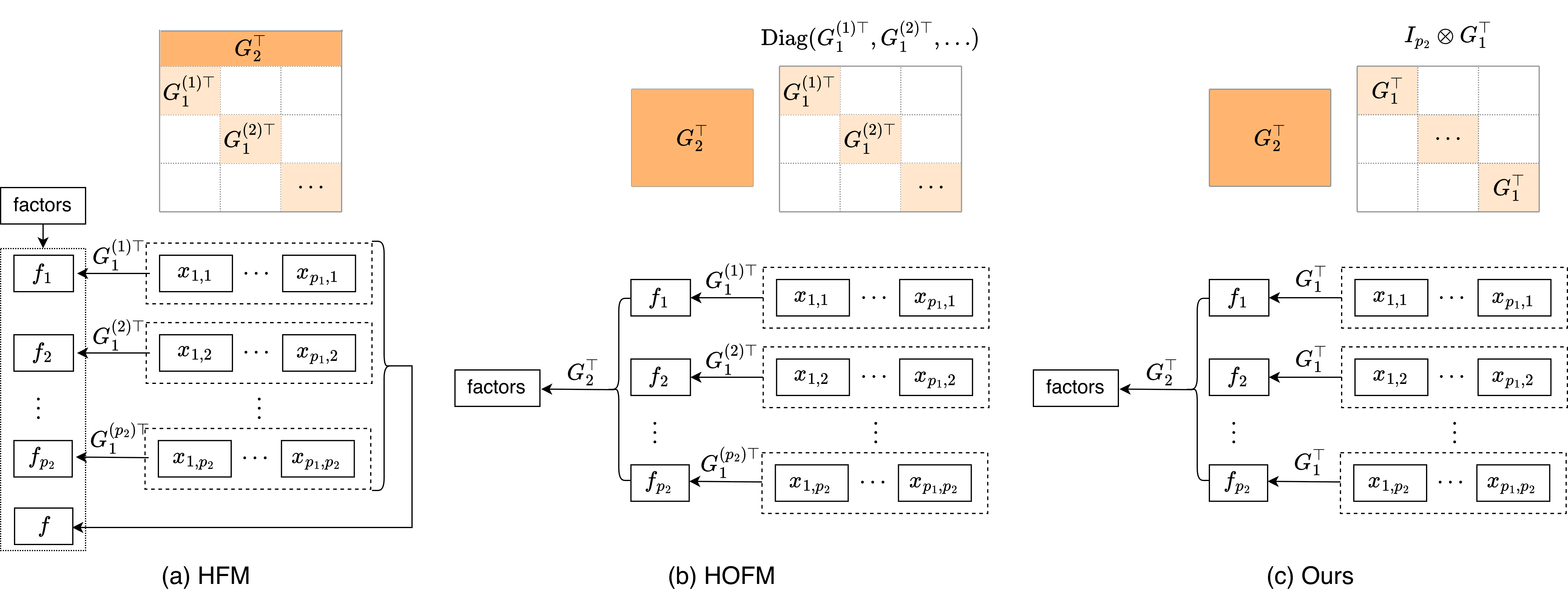} 
	\caption{Three hierarchical feature extraction procedures for a second-order tensor $\cm{X}_t\in \mathbb{R}^{p_1 \times p_2}$: hierarchical factor model (HFM), higher-order factor model (HOFM) and our model.}
	\label{fig:hierarchical_extraction}
\end{figure}

We first consider the feature extraction on predictors, and assume for simplicity that $M=2$, i.e., $\cm{X}_t\in \mathbb{R}^{p_1 \times p_2}$, and the $p_1$ variables of the first mode are nested within each of the $p_2$ variables in its second mode. 
There are two commonly used methods in the literature of factor analysis to extract features from such hierarchical data $\cm{X}_t$: the hierarchical factor model (HFM) in \cite{schmid1957development} and higher-order factor model (HOFM) in \cite{yung1999relationship}.
The HFM directly extracts features from each level of the hierarchy separately and jointly. In contrast, the HOFM performs feature extraction in a sequential manner along the hierarchy, where features extracted from a lower level are used to summarize those from the upper level. The loading matrices for these two methods are illustrated in Figure \ref{fig:hierarchical_extraction}. 
Note that the HFM has more than $p_2$ features, while the HOFM has $O(rP)$ parameters with $P=p_1p_2$, making both of them infeasible in high-dimensional settings.
Moreover, Tucker decomposition method extracts features simultaneously from all modes, i.e. $\mathcal{M}(\cm{X}_t)=\vectorize(\bm{V}_1^\top \cm{X}_t\bm{V}_2)$, and it is not suitable to handle such hierarchical relationships.

Note that the HOFM employs different loadings for each level of the hierarchy.
Such flexibility can maintain its performance, while the corresponding number of parameters is also enlarged dramatically for a larger dimension of $M$.
However, many recent hierarchical data may have a structure similar to the factorial data, and the variables at each level are also arranged in a similar way.
This motivates us to alternatively propose a sequential feature extraction method with shared loadings within each level of the hierarchy.
Specifically, we define the operator $\mathcal{M}(\cm{X}_t;\{\bm{G}_m\}_{m=1}^2) = \bm{G}_2^\top (\bm{I}_{p_2}\otimes\bm{G}_1^\top) \vectorize(\cm{X}_t)$, where $\bm{G}_1\in\mathbb{O}^{p_1\times r_1}$ and $\bm{G}_2\in\mathbb{O}^{p_2 r_1 \times r_2}$ are the component matrices for summarizing predictors, and $r_1$ and $r_2$ are the interim dimensions of features extracted from the corresponding mode.
To interpret this construction, we liken the hierarchy of $\cm{X}_t\in \mathbb{R}^{p_1 \times p_2}$ to a tree structure with $p_2$ branches, each branch containing $p_1$ leaves. For each branch, $\bm{G}_1^\top$ extracts $r_1$ features from its leaves to represent this branch. Overall, we have $r_1 p_2$ features extracted from all leaves, given by $(\bm{I}_{p_2} \otimes \bm{G}_1^\top) \vectorize(\cm{X}_t)$.
Then the subsequent $\bm{G}_2^\top\in\mathbb{R}^{r_2 \times p_2 r_1}$ combines the extracted features from all branches to form $r_2$ features for the entire tree; see Figure \ref{fig:hierarchical_extraction}(c) for the illustration of this hierarchical feature extraction procedure. 
Therefore, the overall loading matrix connecting the $p_1 p_2$ leaves to the final $r_2$ features for the tree is $\bm{G}_{2}^\top (\bm{I}_{p_2} \otimes \bm{G}_1^\top)$, and the column space of this loading can dwell in an arbitrary subspace of $\mathbb{R}^{r_2}$. 

\begin{remark}
	Consider a special case with $p_2=2$, i.e., predictors $\cm{X}_t$ can be divided into two branches, $\bm{x}_{1:p_1,1}$ and $\bm{x}_{1:p_1,2}$. Suppose that the first branch has common features $\bm{G}^{*\top} \bm{x}_{1:p_1,1}\in\mathbb{R}^2$ and branch-specific one $\bm{G}_1^{*\top} \bm{x}_{1:p_1,1}\in\mathbb{R}$, while the second one has the common and branch-specific features of $\bm{G}^{*\top} \bm{x}_{1:p_1,2}\in\mathbb{R}^2$ and  $\bm{G}_2^{*\top} \bm{x}_{1:p_1,2}\in\mathbb{R}^2$, respectively. 
	Note that the HOFM uses loading matrices $\bm{G}_1^{(1)}\in\mathbb{R}^{p_1\times 3}$ and $\bm{G}_1^{(2)}\in\mathbb{R}^{p_1\times 4}$ to extract features from the first and second branches, respectively, while the proposed method employs a common loading matrix $\bm{G}_{1}\in\mathbb{R}^{p_1\times 5}$ to extract features from each branch.
	Shortly speaking, the HOFM concentrates more on branch-specific features, while our method focuses more on common features.
	When the data have an irregular structure or one artificially switches the orders of variables at one level, our method may lead to a misleading interpretation, and we may then have to consider the HOFM or even HFM for the feature extracting. 
	As a result, a huge number of parameters will be involved, making it impossible for high-dimensional data.
\end{remark}

For a general dimension of $M$, we assume that each mode of $\cm{X}_t\in \mathbb{R}^{p_1 \times p_2 \times \cdots \times p_M}$ is nested within its subsequent mode,
and the feature extraction can then be iterated until the last mode of $\cm{X}_t$. Specifically, for each $1\leq m\leq M$, we define the component matrix for summarizing predictors as $\bm{G}_{m}\in\mathbb{O}^{r_{m-1}p_{m}\times r_m}$, where $r_0=1$, $r_m$ is the interim dimension of features extracted from the corresponding mode, and $r_M=r$.
As a result, the feature extraction on predictors at \eqref{eq:factors} can be formulated below,
\begin{equation}
	\label{eq:mappings}
	\bm{f}_t^{\mathrm{predictor}}=\mathcal{M}(\cm{X}_t;\{\bm{G}_m\}_{m=1}^M) = \underbrace{\bm{G}_M^\top (\bm{I}_{p_M}\otimes\bm{G}_{M-1}^\top) \cdots (\bm{I}_{p_2\cdots p_{M}}\otimes\bm{G}_1^\top)}_{\defeq \bm{\Lambda}_x^\top} \vectorize(\cm{X}_t).
\end{equation}

Similarly, if the response tensor $\cm{Y}_t\in \mathbb{R}^{q_1 \times q_2 \times \cdots \times q_N}$ also has such purely hierarchical structure, 
the feature extraction on responses can be formulated into $\bm{f}_t^\mathrm{response}=\mathcal{N}(\cm{Y}_t;\{\bm{H}_n\}_{n=1}^N) = \bm{H}_N^\top (\bm{I}_{q_N}\otimes\bm{H}_{N-1}^\top) \cdots (\bm{I}_{q_2\cdots q_{N}}\otimes\bm{H}_1^\top) \vectorize(\cm{Y}_t)\defeq \bm{\Lambda}_y^\top \vectorize(\cm{Y}_t)$, where $\bm{H}_n\in\mathbb{O}^{s_{n-1}q_n\times s_n}$ with $s_0=1$ and $s_N=s$ are the component matrices  for $1\leq n \leq N$. Note that $\bm{\Lambda}_y$ is orthonormal and, for the tensor regression model at \eqref{eq:regmodel}, its coefficient tensor $\cm{A}$ can then be defined as $[\cm{A}]_N = \bm{\Lambda}_y \bm{\Theta} \bm{\Lambda}_x^\top$ by adopting the method at \eqref{eq:A-form}, where $\bm{\Theta}\in\mathbb{R}^{s\times r}$ is a coefficient matrix.

\begin{remark} \label{remark:TT}
	Consider a special case with $r = s$ and $\bm{\Theta}$ being diagonal, and then the above coefficient tensor $\cm{A}$ will admit a form of tensor train decomposition \citep{oseledets2011tensor,zhou2022optimal,si2024efficient} if predictors are further arranged in a reverse hierarchical order. See Section \ref{appendix:proof-TT} of the Supplementary Material for the detailed illustration.
\end{remark}

\subsection{General hierarchical structure}
\label{subsec:nested-factorial}
This subsection discusses a more practical case with partially hierarchical predictors and responses, and some of their modes may not be strictly nested within the others, possibly with unknown or even multiple hierarchical relationships among the modes. 

We first consider the feature extraction on predictors.
A natural way is first to assume a pseudo-full hierarchical structure to predictors, and the proposed method in the previous subsection can then be applied.
The pseudo-structure should be consistent with the hierarchical structures intrinsic to the data, and the order of its modes is called an \textit{action order}.
For example, we consider the case with $M=3$, and mode 1 is nested within mode 2.
The pseudo-structure can then be represented by \textit{action order} $(1,2,3)$ or $(3,1,2)$.

In general, for the predictor tensor $\cm{X}_t\in \mathbb{R}^{p_1 \times p_2 \times \cdots \times p_M}$, an action order can be denoted by $\alpha = (\alpha_{(1)}, \alpha_{(2)}, \ldots, \alpha_{(M)})$, which is a permutation of set $\{1,2,\ldots,M\}$. 
We then introduce an operator $\mathcal{T}(\cdot;\cdot)$ to permute $\cm{X}_t$ such that its modes follow the action order $\alpha$,
\begin{equation*}
	\mathcal{T}(\cm{X}_t;\alpha)_{i_1\ldots i_M} = (\cm{X}_t)_{i_{\alpha_{(1)}}\dots i_{\alpha_{(M)}}} \hspace{3mm}\text{and}\hspace{3mm}
	\vectorize(\mathcal{T}(\cm{X}_t;\alpha)) = \bm{T}_x(\alpha)\vectorize(\cm{X}_t),
\end{equation*}
where $\bm{T}_x(\alpha)\in\mathbb{R}^{P\times P}$ is the permutation matrix.
Note that $\bm{T}_x(\alpha)=\bm{I}_{P}$ when $\alpha = (1,2,\ldots,M)$.
As a result, by applying \eqref{eq:mappings}, we then can extract features from predictors  
\begin{equation}\label{eq:add1}
\bm{f}_{t}^{\mathrm{predictor}} = \mathcal{M}(\mathcal{T}(\cm{X}_t;\alpha);\{\bm{G}_{m}\in\mathbb{O}^{r_{m-1}p_{\alpha_{(m)}}\times r_{m}}\}_{m=1}^M)\in\mathbb{R}^r \hspace{3mm}\text{with}\hspace{3mm} (r_0,r_M)=(1,r).
\end{equation}

Denote by $\mathcal{P}_x(M)$ the set of all possible action orders for predictors $\cm{X}_t\in \mathbb{R}^{p_1 \times p_2 \times \cdots \times p_M}$.
Assume that $\cm{X}_t$ has $m$ hierarchical structures, which occupy $M^{\prime}$ modes in total, and the set $\mathcal{P}_x(M)$ can then be verified to have $(M-M^{\prime}+m)!$ possible action orders.

\begin{proposition}
	\label{prop:action-equivalence}
	For any $\cm{X}\in \mathbb{R}^{p_1 \times p_2 \times \cdots \times p_M}$, an action order $\alpha=(\alpha_{(1)},\alpha_{(2)},\ldots, \alpha_{(M)})$ is used for the permutation, and factor matrices $\{\bm{G}_{m}\in\mathbb{O}^{r_{m-1}p_{\alpha_{(m)}}\times r_{m}}\}_{m=1}^M$ with $(r_0,r_M)=(1,r)$ are applied to extract features as in \eqref{eq:add1}.
	It then holds that, for a different action order $\alpha^{\prime}=(\alpha_{(1)}^{\prime},\alpha_{(2)}^{\prime},\ldots, \alpha_{(M)}^{\prime})$, there exist $\{\bm{G}_{m}^\prime\in\mathbb{O}^{r'_{m-1}p_{\alpha^{\prime}_{(m)}}\times r'_{m}}\}_{m=1}^M$ with $(r_0^{\prime},r_M^{\prime})=(1,r)$ such that $\mathcal{M}(\mathcal{T}(\cm{X};\alpha);\{\bm{G}_{m}\}_{m=1}^M) = \mathcal{M}(\mathcal{T}(\cm{X};\alpha^{\prime});\{\bm{G}_{m}^\prime\}_{m=1}^M)$.
\end{proposition}

The above proposition proves a nice equivalence of all elements inside $\mathcal{P}_x(M)$; however, it is a theoretical result only. In real applications, each action order may have a totally different number of parameters, and a wrongly selected order may end up with a huge number of parameters.
As a result, we may choose the one with the smallest size,
\begin{equation}\label{eq:add2}
\alpha^*=\argmin_{\alpha\in \mathcal{P}_x(M)} d_{x}(\alpha) \hspace{3mm}\text{with}\hspace{3mm} d_{x}(\alpha)=\sum_{m=1}^{M}r_{m-1}r_mp_m.
\end{equation}
On the other hand, for a certain predictor tensor $\cm{X}_t$, more than one action orders may be more practical, while the summarized features $\bm{f}_{t}^{\mathrm{predictor}}$ have a fixed dimension of $r$ regardless which action orders are used.
This motivates us to choose the best set of action orders,
\begin{equation}\label{eq:add3}
\mathcal{P}_x(M, K_x)\defeq \{\alpha_1^*,\ldots,\alpha_{K_x}^*\} =\argmin_{1\leq K\leq |\mathcal{P}_x(M)|,\alpha_k\in \mathcal{P}_x(M)} \sum_{k=1}^K d_{x}(\alpha_k) \hspace{3mm}\text{subject to}\hspace{3mm} \sum_{k=1}^Kr_M(\alpha_k)=r,
\end{equation}
where $|\mathcal{P}_x(M)|$ is the cardinality of $\mathcal{P}_x(M)$, and $r_M(\alpha_k)$ is the rank of $r_M$ with respect to $\alpha_k$.
Note that the number of parameters for $\mathcal{P}_x(M, K_x)$ is smaller than that for $\alpha^*$ from \eqref{eq:add2}.

For the selected $\alpha^*$ from \eqref{eq:add2}, we can extract features from predictors as in \eqref{eq:add1}.
This paper focuses on $\mathcal{P}_x(M, K_x)$, the selected set of action orders.
For each $1\leq k\leq K_x$, denote $\alpha_k^*=(\alpha_{(1)}^k,\alpha_{(2)}^k,\ldots, \alpha_{(M)}^k)$ with ranks $(r_{0,k},r_{1,k},\ldots,r_{M,k})$, and 
let $\bm{f}_{k,t}^{\mathrm{predictor}} = \mathcal{M}(\mathcal{T}(\cm{X}_t;\alpha_k^*);\{\bm{G}_{m,k}\in\mathbb{O}^{r_{m-1,k}p_{\alpha^k_{(m)}}\times r_{m,k}}\}_{m=1}^M)\in\mathbb{R}^{r_{M,k}}$.
As a result, the extracted features from predictors can be denoted by
$\bm{f}_{t}^{\mathrm{predictor}} = \left( \bm{f}_{1,t}^{\mathrm{predictor}\top}, \ldots, \bm{f}_{K_x,t}^{\mathrm{predictor}\top} \right)^\top=\bm{\Lambda}_x^\top \vectorize(\cm{X}_t)\in\mathbb{R}^r$, where $r=\sum_{k=1}^{K_x}r_{M,k}$, and the loading matrix $\bm{\Lambda}_x$ has the form of
\begin{equation}
	\label{eq:lambda-x}
	\bm{\Lambda}_x = 
	\begin{bmatrix}
		\bm{G}_{M,1}^\top (\bm{I}_{p_{\alpha^1_{(M)}}}\otimes\bm{G}_{M-1,1}^\top) \cdots (\bm{I}_{p_{\alpha^1_{(2)}}\cdots p_{\alpha^1_{(M)}}}\otimes\bm{G}_{1,1}^\top) \bm{T}_x(\alpha_1)\\
		\bm{G}_{M,2}^\top (\bm{I}_{p_{\alpha^2_{(M)}}}\otimes\bm{G}_{M-1,2}^\top) \cdots (\bm{I}_{p_{\alpha^2_{(2)}}\cdots p_{\alpha^2_{(M)}}}\otimes\bm{G}_{1,2}^\top) \bm{T}_x(\alpha_2)\\
		\cdots\\
		\bm{G}_{M,K_x}^\top (\bm{I}_{p_{\alpha^{K_x}_{(M)}}}\otimes\bm{G}_{M-1,K_x}^\top) \cdots (\bm{I}_{p_{\alpha^{K_x}_{(2)}}\cdots p_{\alpha^{K_x}_{(M)}}}\otimes\bm{G}_{1,K_x}^\top) \bm{T}_x(\alpha_{K_x})
	\end{bmatrix}^\top.
\end{equation} 
The above matrix is no longer orthogonal in general.
The philosophy of the above feature extraction is analogous to that of CP decomposition, which concatenates multiple separately extracted features. 
Note that, comparing with Tucker decomposition, CP decomposition requires much less parameters by sacrificing orthogonality of loading matrices.

The feature extraction from responses $\cm{Y}_t\in\mathbb{R}^{q_1\times q_2\times\cdots\times q_N}$ can be conducted in a similar way. 
Denote the corresponding action order by $\beta = (\beta_{(1)}, \beta_{(2)}, \ldots, \beta_{(N)})$, which is a permutation of set $\{1,2,\ldots,N\}$, and let $\mathcal{P}_y(N)$ be the set of all possible action orders.
By a method similar to \eqref{eq:add2} and \eqref{eq:add3}, we can define the best action order $\beta^*$ and the best set of action orders $\mathcal{P}_y(N, K_y)\defeq \{\beta_1^*,\ldots,\beta_{K_y}^*\} =\argmin_{1\leq K\leq |\mathcal{P}_y(N)|,\beta_{k}\in \mathcal{P}_y(N)} \sum_{k=1}^K d_{y}(\beta_{k})$ with $d_y(\beta)=\sum_{n=1}^{N}s_{n-1}s_nq_n$, respectively.
Moreover, for each $1\leq k\leq K_y$, let $\beta_{k}^*=(\beta_{(1)}^{k},\beta_{(2)}^{k},\ldots, \beta_{(N)}^{k})$ with ranks $(s_{0,k},s_{1,k},\ldots,s_{N,k})$, component matrices be $\bm{H}_{n,{k}}\in\mathbb{O}^{s_{n-1,{k}}q_{\beta^{k}_{(n)}}\times s_{n,{k}}}$ with $1\leq n\leq N$, and the corresponding extracted features be $\bm{f}_{{k},t}^{\mathrm{response}} = \mathcal{N}(\mathcal{T}(\cm{Y}_t;\beta^*_{k});\{\bm{H}_{n,{k}}\}_{n=1}^N)$. As a result, the total extracted features from responses can be represented as $\bm{f}_{t}^{\mathrm{response}} = \left( \bm{f}_{1,t}^{\mathrm{response}\top}, \ldots, \bm{f}_{K_y,t}^{\mathrm{response}\top} \right)^\top\defeq \bm{\Lambda}_y^\top\vectorize(\cm{Y}_t)\in\mathbb{R}^s$, where $s=\sum_{k=1}^{K_y}s_{N,k}$, and the loading matrix $\bm{\Lambda}_y$ can be defined similar to \eqref{eq:lambda-x} with the permutation matrix $\bm{T}_y(\beta)\in\mathbb{R}^{Q\times Q}$.


From \eqref{eq:A-form}, the coefficient tensor of model \eqref{eq:regmodel} can be designed to have the form of 
\begin{align}
	\label{eq:A-form-nested}
	[\cm{A}_{\mathrm{LR}}]_N = \bm{\Lambda}_y \left(\bm{\Lambda}_y^\top \bm{\Lambda}_y \right)^{-1} \bm{\Theta}_{\mathrm{LR}} \bm{\Lambda}_x^\top
\end{align}
with $\bm{\Theta}_{\mathrm{LR}}\in \mathbb{R}^{s \times r}$ being the coefficient matrix, leading to an efficient and interpretable tensor regression model for data with hierarchical structures.

\subsection{Autoregressive model for hierarchical time series}
\label{subsec:ar}
Consider an autoregressive model for hierarchical time series $\{\cm{Y}_t\}$ with $\cm{Y}_t\in \mathbb{R}^{q_1 \times q_2 \times \cdots \times q_N}$,
\begin{align}
	\label{model:AR(P)}
	\cm{Y}_t = \left\langle \cm{A}_1, \cm{Y}_{t-1} \right\rangle + \left\langle \cm{A}_2, \cm{Y}_{t-2} \right\rangle + \cdots +
	\left\langle \cm{A}_L, \cm{Y}_{t-L} \right\rangle + \cm{E}_t, 
\end{align}
where $L$ is the order, $\cm{A}_l \in \mathbb{R}^{q_1 \times \cdots \times q_N \times q_1 \times \cdots \times q_N}$ with $1\leq l\leq L$ are coefficient tensors, and $\cm{E}_t \in \mathbb{R}^{q_1 \times q_2 \times \cdots \times q_N}$ is the error term. This subsection attempts to introduce an efficient and interpretable version of model \eqref{model:AR(P)} by adopting the idea proposed in Section \ref{subsec:nested-factorial}.

Note that model \eqref{model:AR(P)} can be vectorized into the form of $\vectorize(\cm{Y}_t) = \sum_{l=1}^{L} \bm{A}_l \vectorize(\cm{Y}_{t-l}) + \vectorize(\cm{E}_t)$, where $\bm{A}_l = [\cm{A}_l]_N$ is the mode-$N$ sequential matricization of $\cm{A}_l$. We then can state the necessary and sufficient condition for the strict stationarity of model \eqref{model:AR(P)} below. 
\begin{assumption}
	\label{assump:stationarity}
	The determinant of the matrix polynomial $\cm{A}(z) = \bm{I}_Q-\bm{A}_1 z-\cdots-\bm{A}_L z^L$ is not equal to zero for all $z \in \mathbb{C}$ and $|z| \le 1$.
\end{assumption}

We first consider the case with $L=1$, and the coefficient tensor $\cm{A}_1$ can be directly assumed to have the decomposition in \eqref{eq:A-form-nested} with the same definitions of $\bm{\Lambda}_x$ and $\bm{\Lambda}_y$, where $M = N$, $p_n = q_n$ for $1\leq n\leq N$, and $\cm{X}_t = \cm{Y}_{t-1}$.
As a result, model \eqref{model:AR(P)} can then be rewritten into an interpretable form:
\begin{equation*}
	\label{model:AR(1)}
	\underbrace{\bm{\Lambda}_y^\top \vectorize(\cm{Y}_t)}_{\bm{f}_t^{\mathrm{response}}}  
	= \bm{\Theta}_1 \underbrace{\bm{\Lambda}_x^\top \vectorize(\cm{Y}_{t-1})}_{\bm{f}^{\mathrm{predictor}}_{t}} 
	+ \bm{\Lambda}_y^\top \vectorize(\cm{E}_t). 
\end{equation*}
Note that the high-order tensor $\cm{Y}_t$ is summarized into features $\bm{f}_t^{\mathrm{response}}\in\mathbb{R}^s$, while the information from $\cm{Y}_{t-1}$ is compressed into $\bm{f}^{\mathrm{predictor}}_{t}\in\mathbb{R}^r$. The coefficient matrix $\bm{\Theta}_1\in\mathbb{R}^{s\times r}$ characterizes the linear relationship between features $\bm{f}^{\mathrm{response}}_t$ and $\bm{f}^{\mathrm{predictor}}_{t}$. 

For a general order $L$, we apply \eqref{eq:A-form} to each coefficient tensor $\cm{A}_l$, resulting in:
\begin{equation}
	\label{model:AR(p)-feature}
	[\cm{A}_l]_N = \bm{\Lambda}_y \left(\bm{\Lambda}_y^\top \bm{\Lambda}_y \right)^{-1} \bm{\Theta}_l \bm{\Lambda}_x^\top\quad \text{for} \quad 1\leq l\leq L.
\end{equation}
It then holds that $\bm{f}_t^{\mathrm{response}} = \sum_{l=1}^L \bm{\Theta}_l \bm{f}^{\mathrm{predictor}}_{t+1-l} + \bm{\Lambda}_y^\top \vectorize(\cm{E}_t)$. From \eqref{model:AR(p)-feature}, the mode-$N$ sequential matricization of each $\cm{A}_l$ shares the same loading matrices to extract features from responses and predictors, respectively.
The different coefficient matrices $\bm{\Theta}_l \in\mathbb{R}^{s\times r}$ allow flexible relationships between response and predictor features at different lags. Note that the number of parameters for model \eqref{model:AR(p)-feature} is $d_\mathrm{AR} = \sum_{k=1}^{K_x} d_{x}(\alpha_k) + \sum_{k=1}^{K_y} d_{y}(\beta_{k}) + Lsr$.

\section{High-dimensional estimation}
\label{sec:theory}
\subsection{Ordinary least squares approach}
\label{subsec:l2loss}
This paper considers the ordinary least squares (OLS) method to estimate the proposed tensor autoregressive model at \eqref{model:AR(P)} and \eqref{model:AR(p)-feature} with hierarchical time series.

Suppose that both sets of action orders $\mathcal{P}_x(M, K_x)=\{\alpha_1, \ldots, \alpha_{K_x}\}$ and $\mathcal{P}_y(N, K_y)=\{\beta_{1}, \ldots, \beta_{K_y}\}$ are known and, for all $1\leq k\leq K_x$ and $1\leq k^\prime\leq K_y$, the corresponding ranks $\{r_{m,k}\}_{m=1}^M$ and $\{s_{n,k^\prime}\}_{n=1}^N$ are also given.
Let $\bm{p}=(p_1, \ldots, p_M)$, $\bm{q}=(q_1, \ldots, q_N)$, $\bm{r}_{k}=(r_{1,k}, \ldots, r_{M,k})$ and $\bm{s}_{k^\prime}=(s_{1,k^\prime}, \ldots, s_{N,k^\prime})$, where $1\leq k\leq K_x$ and $1\leq k^\prime\leq K_y$. 
Note that $\sum_{k=1}^{K_x} r_{M,k}=r$, $\sum_{k=1}^{K_y}s_{N,k}=s$, $M=N$ and $\bm{p}=\bm{q}$.
We then denote the parameter spaces of loading matrices from predictor and response features respectively by
\begin{equation*}
	\begin{split}
		&\mathcal{R}_x(\bm{p},\{\bm{r}_{k}\}_{k=1}^{K_x}, \mathcal{P}_x(M, K_x)) = \{ \bm{\Lambda}_x\in\mathbb{R}^{\prod_{m=1}^{M}p_m\times r}: \bm{\Lambda}_x ~\text{has the form of \eqref{eq:lambda-x}} \}, \hspace{3mm}\text{and}\\
		&\mathcal{R}_y(\bm{q},\{\bm{s}_{k}\}_{k=1}^{K_y}, \mathcal{P}_y(N, K_y)) = \{ \bm{\Lambda}_y\in\mathbb{R}^{\prod_{n=1}^{N}q_n\times s}: \bm{\Lambda}_y~\text{is defined as in Section \ref{subsec:nested-factorial}} \}.
	\end{split}
\end{equation*}

For autoregressive models at \eqref{model:AR(P)} and \eqref{model:AR(p)-feature}, the coefficient tensor can be denoted by $\cm{A}_{\mathrm{AR}}\in\mathbb{R}^{q_1\times\cdots \times q_N\times q_1\times\cdots\times q_N\times L}$ with $[\cm{A}_{\mathrm{AR}}]_N = ([\cm{A}_1]_N, [\cm{A}_2]_N, \ldots, [\cm{A}_L]_N)$ being the mode-$N$ sequential matricization. 
Let $\bm{\Theta}_{\mathrm{AR}} = (\bm{\Theta}_1, \ldots, \bm{\Theta}_L)\in\mathbb{R}^{s\times rL}$, and it holds that $[\cm{A}_{\mathrm{AR}}]_N = \bm{\Lambda}_y (\bm{\Lambda}_y^\top \bm{\Lambda}_y )^{-1} \bm{\Theta}_{\mathrm{AR}} (\bm{I}_L\otimes\bm{\Lambda}_x^\top)$.
Moreover, by reparameterizing $(\bm{\Lambda}_y^\top \bm{\Lambda}_y )^{-1} \bm{\Theta}_{\mathrm{AR}}$ into $\bm{\Theta}_{\mathrm{AR}}$, we have $[\cm{A}_{\mathrm{AR}}]_N = \bm{\Lambda}_y \bm{\Theta}_{\mathrm{AR}} (\bm{I}_L\otimes\bm{\Lambda}_x^\top)$, and then the parameter space can be defined as
\begin{equation*}
	\label{eq:parameter-space-consolidated}
	\begin{split}
		\mathcal{S}_{\mathrm{AR}} = \{
		\cm{A} \in &\mathbb{R}^{q_1 \times \cdots \times q_N \times p_1 \times \cdots \times p_M \times L}:
		~[\cm{A}]_N = \bm{\Lambda}_y \bm{\Theta} (\bm{I}_L\otimes\bm{\Lambda}_x^\top),
		~\bm{\Theta}\in\mathbb{R}^{s\times rL},\\ 
		&~~~~~\bm{\Lambda}_y \in \mathcal{R}_y(\bm{q},\{\bm{s}_{k}\}_{k=1}^{K_y}, \mathcal{P}_y(N, K_y)), 
		~\bm{\Lambda}_x \in \mathcal{R}_x(\bm{p},\{\bm{r}_{k}\}_{k=1}^{K_x}, \mathcal{P}_x(M, K_x)) \}.
	\end{split}
\end{equation*}


For an observed time series $\{\cm{Y}_{1-L},\ldots,\cm{Y}_0,\cm{Y}_1,\ldots,\cm{Y}_T\}$ generated by the model at \eqref{model:AR(P)} and \eqref{model:AR(p)-feature}, the predictor tensor $\cm{X}_t\in \mathbb{R}^{q_1 \times \cdots \times q_N \times L}$ can be formed by stacking $\cm{Y}_{t-1}, \dots, \cm{Y}_{t-L}$ together, and the OLS estimator is hence defined as
\begin{equation}
	\label{eq:least-square-est}
	\widehat{\cm{A}}_{\mathrm{AR}} = \argmin_{\cm{A}\in\mathcal{S}_{\mathrm{AR}}} \mathcal{L}_T(\cm{A}) \quad\text{with}\quad
	\mathcal{L}_T(\cm{A}) = \frac{1}{T} \sum_{t=1}^{T} \norm{\vectorize(\cm{Y}_t) - [\cm{A}]_N \vectorize(\cm{X}_t)}_2^2.
\end{equation}
Moreover, for the tensor regression model at \eqref{eq:regmodel} and \eqref{eq:A-form-nested}, by letting $L=1$, we can similarly define its parameter space and OLS estimator, denoted by $\mathcal{S}_{\mathrm{LR}}$ and $\widehat{\cm{A}}_{\mathrm{LR}}$, respectively.

Denote by $\cm{A}^*$ the true coefficient tensor of the tensor regression or autoregressive model in the previous section, and it depends on the best set of action orders, which is unknown in real applications.
In fact, $\mathcal{P}_x(M, K_x)$ and $K_x$ in this section are only the running set and numbers of action orders for predictors, respectively, and the situation is the same for $\mathcal{P}_y(N, K_y)$ and $K_y$.
Fortunately, from Proposition \ref{prop:action-equivalence}, we can project the true coefficient tensor $\cm{A}^*$ into the running sets $\mathcal{P}_x(M, K_x)$ and $\mathcal{P}_y(N, K_y)$, with running numbers $K_x$ and $K_y$.
Such projection has no error, while the number of parameters will be enlarged. 
Without loss of generality, we will assume $\mathcal{P}_x(M, K_x)$ and $\mathcal{P}_y(N, K_y)$ to be known in establishing theoretical properties, and Section \ref{sec:algorithms} will discuss the selection of these hyperparameters.

\subsection{Non-asymptotic properties for tensor regression settings}
\label{subsec:theorems}
As a preliminary discussion for tensor autoregressive models, this subsection first establishes non-asymptotic properties of the OLS estimator $\widehat{\cm{A}}_\mathrm{LR}$ with tensor regression settings, i.e., $\{(\cm{Y}_t, \cm{X}_t)\}_{t=1}^T$ are independent and identically distributed ($i.i.d.$) observations.
Moreover, to simplify notations, we denote $\vectorize(\cm{X}_t)$ and $\vectorize(\cm{E}_t)$ by $\bm{x}_t$ and $\bm{e}_t$, respectively.

\begin{assumption}
	\label{assump:reg-input}
	Predictor $\bm{x}_t$ follows a $\sigma^2$-sub-Gaussian distribution with $\mathbb{E}(\bm{x}_t)=\bm{0}$ and $\mathrm{var}(\bm{x}_t)=\bm{\Sigma}_x$. 
	There exist constants $0<c_{x}<C_{x}<\infty$ such that 
	$c_x \leq \lambda_{\min}(\bm{\Sigma}_x) \leq  \lambda_{\max}(\bm{\Sigma}_x)\leq C_x$,
	where the two quantities $c_x$ and $C_x$ depend on the dimensions of $p_j$'s, and they may shrink to zero or diverge to infinity as the dimensions increase.
\end{assumption}

\begin{assumption}
	\label{assump:reg-error}
	Error term $\bm{e}_t$, conditional on predictor $\bm{x}_t$, follows  $\kappa^2$-sub-Gaussian distribution with mean zero.
\end{assumption}

\begin{assumption}
	\label{assump:reg-core}
	It holds that $\|\bm{\Theta}_{\mathrm{LR}}\|_{\mathrm{op}}<g_1$, where $\|\cdot\|_{\mathrm{op}}$ is the operator norm.
\end{assumption}

Assumption \ref{assump:reg-input} guarantees the well-conditioning of predictors, while Assumption \ref{assump:reg-error} controls the tail behavior of noises by the sub-Gaussianity. The two assumptions are standard high-dimensional settings, and they have been widely used in the literature \citep{wainwright2019high, si2024efficient, cai2025efficient}.
Assumption \ref{assump:reg-core} imposes an upper bound on the operator norm of the coefficient matrix $\bm{\Theta}_{\mathrm{LR}}\in\mathbb{R}^{s\times r}$. It is used to handle the non-orthogonality of loading matrices $\bm{\Lambda}_x$ and $\bm{\Lambda}_y$, and we can remove it with the cost of a large model complexity. This assumption is mild since both dimensions $s$ and $r$ are typically small by carefully choosing the action orders.
Denote the model complexity by $d_\mathrm{LR} = \sum_{k=1}^{K_x}d_x(\alpha_k) + \sum_{k=1}^{K_y}d_y(\beta_k)+sr$, and the estimation error bound is stated below.

\begin{theorem}\label{thm:reg}
	Suppose that Assumptions \ref{assump:reg-input} - \ref{assump:reg-core}  hold, and the true coefficient tensor $\cm{A}_\mathrm{LR}^* \in \mathcal{S}_{\mathrm{LR}}$.
	If the sample size $T \gtrsim (\sigma^4/c_x^2) d_\mathrm{LR}\log \{(\sigma^4/c_x^2) d_\mathrm{LR}\}$, then,
	\begin{align*}
		\|\cm{\widehat{A}}_{\mathrm{LR}}-\cm{A}^*_\mathrm{LR}\|_{\mathrm{F}}\lesssim \frac{\kappa \sigma}{c_x}\sqrt{\frac{d_\mathrm{LR} \log T}{T}}
	\end{align*}
	with probability at least $1 - 2\exp\left\{ -C_1d_\mathrm{LR}\log T \right\}$, where $C_1$ is a positive constant.
\end{theorem}

Let $r_{\max} \defeq \max_{m,k} r_{m,k}$ and $s_{\max} \defeq \max_{n,k^\prime} s_{n,k^\prime}$, and the model complexity can be simplified as $d_\mathrm{LR} \leq K_x r_{\max}^2  \sum_{m=1}^{M}p_m + K_y s_{\max}^2 \sum_{n=1}^{N}q_n + K_xK_yr_{\max}s_{\max}$.
From the above theorem, the OLS estimator $\widehat{\cm{A}}_{\mathrm{LR}}$ achieves an estimation error bound of order $O_p(\sqrt{d_\mathrm{LR} / T})$, up to a logarithmic factor, when the sample size $T\gtrsim d_\mathrm{LR}/(c_x/\sigma^2)^2$. This rate aligns with the general optimal rate for low-rank regression established in \cite{han2022optimal}. The model complexity $d_\mathrm{LR}$ grows linearly with the dimensions of each mode $\{p_m\}$ and $\{q_n\}$,
and quadratically with the interim ranks $\{r_{m,k}\}$ and $\{s_{n,k^\prime}\}$. 




\subsection{Non-asymptotic properties for tensor autoregressive models}
\label{subsec:theorems-auto}
This subsection states non-asymptotic properties of the OLS estimator $\cm{\widehat{A}}_{\mathrm{AR}}$ for tensor autoregressive models at \eqref{model:AR(P)} and \eqref{model:AR(p)-feature}, and we first introduce some conditions below.

\begin{assumption}
	\label{assump:errorauto}
	Let $\bm{e}_t =\bm{\Sigma}_e^{1/2}\bm{\xi}_t$ with $\bm{\Sigma}_e=\mathrm{var}(\bm{e}_t)$. Random vectors $\{\bm{\xi}_t\}$ are $i.i.d.$ with mean zero and covariance matrix $\bm{I}_Q$, and the entries of $\bm{\xi}_t\in\mathbb{R}^Q$ are independent and $\kappa^2$-sub-Gaussian distributed. Moreover, there exist constants $0<c_{e}<C_{e}<\infty$ such that $c_{e} \leq \lambda_{\min}(\bm{\Sigma}_e) \leq \lambda_{\max}(\bm{\Sigma}_e)\leq C_e$, where the two quantities depend on the dimensions of $q_j$'s, and they may shrink to zero or diverge to infinity as the dimensions increase.
\end{assumption}


\begin{assumption}
	\label{assump:auto-core}
	It holds that $\|\bm{\Theta}_{\mathrm{AR}}\|_{\mathrm{op}}<g_2$, where $\|\cdot\|_{\mathrm{op}}$ is the operator norm.
\end{assumption}

The sub-Gaussian assumption has also been commonly used in the literature of high-dimensional time series \citep{zheng2020finite, wang2024high, huang2025supervised}.

We next establish non-asymptotic properties of the OLS estimator $\cm{\widehat{A}}_{\mathrm{AR}}$, and they rely on the temporal and cross-sectional dependence of $\{\cm{Y}_t\}$ \citep{basu2015regularized}. To this end, for the matrix polynomial of model \eqref{model:AR(P)}, we define the quantities below,
\begin{align*}
	\mu_{\mathrm{min}}(\cm{A}) = \min_{|z|=1} \lambda_{\mathrm{min}}(\bar{\cm{A}}(z)\cm{A}(z))
	\quad \text{and} \quad
	\mu_{\mathrm{max}}(\cm{A}) = \max_{|z|=1} \lambda_{\mathrm{max}}(\bar{\cm{A}}(z)\cm{A}(z)),
\end{align*}
where $\bar{\cm{A}}(z)$ denotes the conjugate transpose of $\cm{A}(z)$ with $z\in\mathbb{C}$.
Moreover, the AR$(L)$ model at \eqref{model:AR(P)} can be rewritten into an equivalent AR$(1)$ form with a companion matrix $\bm{B}\in\mathbb{R}^{QL\times QL}$; see the supplementary file for details.
As a result, the corresponding matrix polynomial is given by $\cm{B}(z) = \bm{I}_{QL} - \bm{B}z$, and hence two more quantities of $\mu_{\mathrm{min}}(\cm{B}) = \min_{|z|=1} \lambda_{\mathrm{min}}(\bar{\cm{B}}(z)\cm{B}(z))$ and $\mu_{\mathrm{max}}(\cm{B}) = \max_{|z|=1} \lambda_{\mathrm{max}}(\bar{\cm{B}}(z)\cm{B}(z))$.
Note that $\mu_{\mathrm{min}}(\cm{B})$, $\mu_{\mathrm{max}}(\cm{B})$ are not necessarily the same as $\mu_{\mathrm{min}}(\cm{A})$, $\mu_{\mathrm{max}}(\cm{A})$, respectively, and it can be verified that $\mu_{\mathrm{min}}(\cm{B}) \leq \mu_{\mathrm{max}}(\cm{A})$; see \cite{basu2015regularized} and Lemma \ref{lemma:AR-spectral} in the supplementary file for details.
In addition, let $\kappa_{U, B}=C_{e}/\mu_{\mathrm{min}}(\cm{B})$, 
$\kappa_{L, A}=c_{e}/\mu_{\mathrm{max}}(\cm{A})$ and 
$\kappa^\prime_{U, B}=C_{e}/\mu^{1/2}_{\mathrm{min}}(\cm{B})$, and the tensor autoregressive model at \eqref{model:AR(P)} and \eqref{model:AR(p)-feature} has the complexity of
$d_\mathrm{AR} = \sum_{k=1}^{K_x} d_{x}(\alpha_k) + \sum_{k=1}^{K_y} d_{y}(\beta_{k}) + Lsr$.

\begin{theorem}\label{thm:autoreg}
	Suppose that Assumptions \ref{assump:stationarity}, \ref{assump:errorauto} and \ref{assump:auto-core} hold, and the true coefficient tensor $\cm{A}_{\mathrm{AR}}^\ast \in\mathcal{S}_{\mathrm{AR}}$. If the sample size $T \gtrsim (\kappa^4\kappa_{U,B}^2/\kappa_{L,A}^2) d_\mathrm{AR} \log\{\kappa^4\kappa_{U,B}^2/\kappa_{L,A}^2) d_\mathrm{AR}\}$, then
	\begin{align*}
		\|{\cm{\widehat{A}}_{\mathrm{AR}}-\cm{A}_{\mathrm{AR}}^\ast}\|_{\mathrm{F}} \lesssim \frac{\kappa^2\kappa_{U,B}'}{\kappa_{L,A}} \sqrt{\frac{d_\mathrm{AR}\log T}{T}}
	\end{align*}
	with probability at least $1-2\exp\left\{-C_1 d_\mathrm{AR}\log T \right\}$, where $C_1>0$ is a positive constant.
\end{theorem}

The model complexity $d_\mathrm{AR}$ can be simplified as $d_\mathrm{AR} \leq (K_x r_{\max}^2 + K_y s_{\max}^2) \sum_{n=1}^{N}q_n + LK_xK_yr_{\max}s_{\max}$.
Moreover, when $N$, $K_x$, $K_y$, $s_{\max}$ and $r_{\max}$ are fixed and the quantities of $c_e$, $C_e$, $\kappa^2$, $\mu_{\mathrm{min}}(\cm{B})$, and $\mu_{\mathrm{max}}(\cm{A})$ are all bounded away from zero and infinity, the estimation error bound has a form of $\|{\cm{\widehat{A}}_{\mathrm{AR}}-\cm{A}_{\mathrm{AR}}^\ast}\|_{\mathrm{F}} = O_p(\sqrt{d_{\mathrm{AR}} / T})$, up to a logarithmic term.

\section{Algorithms and hyperparameter selection}
\label{sec:algorithms}
\subsection{Algorithms}
\label{subsec:algorithms}
This subsection introduces an alternating least squares (ALS) method to search for the estimate $\widehat{\cm{A}}_{\mathrm{AR}}$ in Section \ref{subsec:l2loss}.
Specifically, for each $\cm{A}\in\mathcal{S}_{\mathrm{AR}}$ at \eqref{eq:least-square-est}, its mode-$N$ sequential matricization has the form of $[\cm{A}]_N = \bm{\Lambda}_y \bm{\Theta} (\bm{I}_L\otimes\bm{\Lambda}_x^\top)$, where $\bm{\Theta} = (\bm{\Theta}_1, \ldots, \bm{\Theta}_L)$.
Let
\begin{align*}
	\bm{\Lambda}_{x,k} &= \bm{T}_x^{\top}(\alpha_k) (\bm{I}_{p_{\alpha^k_{(2)}}\cdots p_{\alpha^k_{(M)}}}\otimes\bm{G}_{1,k}) \cdots (\bm{I}_{p_{\alpha^k_{(M)}}}\otimes\bm{G}_{M-1,k}) \bm{G}_{M,k} \in\mathbb{R}^{P \times r_{M,k}} \hspace{3mm}\text{and}\\
	\bm{\Lambda}_{y,k^\prime}&= \bm{T}_y^{\top}(\beta_{k^\prime}) (\bm{I}_{q_{\beta^{k^\prime}_{(2)}}\cdots q_{\beta^{k^\prime}_{(N)}}}\otimes\bm{H}_{1,k^\prime}) \cdots (\bm{I}_{q_{\beta^{k^\prime}_{(N)}}}\otimes\bm{H}_{N-1,k^\prime}) \bm{H}_{N,k^\prime} \in\mathbb{R}^{Q \times s_{N,k^\prime}}
\end{align*}
for $1\leq k\leq K_x$ and $1\leq k^\prime \leq K_y$, and it then holds that
$\bm{\Lambda}_x = (\bm{\Lambda}_{x,1}, \ldots, \bm{\Lambda}_{x,K_x})\in\mathbb{R}^{P\times r}$ and $\bm{\Lambda}_y = (\bm{\Lambda}_{y,1}, \ldots, \bm{\Lambda}_{y,K_y})\in\mathbb{R}^{Q\times s}$; see \eqref{eq:lambda-x} for details. Moreover, for each $1\leq l\leq L$, 
$[\cm{A}_l]_N=\bm{\Lambda}_y \bm{\Theta}_l \bm{\Lambda}_x^\top =\sum_{k^\prime=1}^{K_y}\sum_{k=1}^{K_x}\bm{\Lambda}_{y,k^\prime} \bm{\Theta}_{k^\prime,k,l} \bm{\Lambda}_{x,k}^\top$, where $\bm{\Theta}_{k^\prime,k,l}\in \mathbb{R}^{s_{N,k^\prime}\times r_{M,k}}$ with $1\leq k\leq K_x$ and $1\leq k^\prime \leq K_y$ are the corresponding submatrices of $\bm{\Theta}_l$.
As a result, it is equivalent to minimizing the loss function of $\mathcal{L}_T(\cm{A})$ at \eqref{eq:least-square-est} with respect to two blocks of parameters: 
$\{\bm{\Lambda}_{x,k}(\bm{G}_{m,k}, 1\leq m\leq M), 1\leq k\leq K_x\}$, $\{\bm{\Lambda}_{y,k}(\bm{H}_{n,k}, 1\leq n\leq N), 1\leq k^\prime \leq K_y\}$, and $\bm{\Theta} $, and the first two blocks of parameters correspond to $\bm{\Lambda}_x$ and $\bm{\Lambda}_y$, respectively.

\begin{algorithm}[t]
	\caption{ALS for tensor autoregression}
	\label{alg:als-auto-seq}
	\begin{algorithmic}
		\STATE \textbf{Input:} Time series $\{\bm{y}_t,-L+1\leq t\leq T\}$, initial coefficients $[\cm{A}^{(0)}]_N = \bm{\Lambda}_y^{(0)} \bm{\Theta}^{(0)} (\bm{I}_L\otimes\bm{\Lambda}_x^{(0)\top})$ with component matrices $\{ \{ \bm{H}^{(0)}_{n,k} \}_{n=1}^N \}_{k=1}^{K_y}$ and $\{ \{ \bm{G}^{(0)}_{m,k} \}_{m=1}^M \}_{k=1}^{K_x}$.
		\REPEAT
		\FOR{$k=1,2,\ldots, K_x$}
		\STATE Sequentially update $\bm{G}_{m,k}$ for $1\leq m \leq M$ while keeping all other components being fixed at the values from the last iteration.
		\STATE Standardize each $\bm{G}_{m,k}$ for $1\leq m \leq M$ such that each column has unit norm.
		\ENDFOR
		\FOR{$k=1,2,\ldots, K_y$}
		\STATE Sequentially update $\bm{H}_{n,k}$ for $1\leq n \leq N$ while keeping all other components being fixed at the values from the last iteration.
		\STATE Standardize each $\bm{H}_{n,k}$ for $1\leq n \leq N$ such that each column has unit norm.
		\ENDFOR
		\STATE Update $\bm{\Theta} = (\bm{\Theta}_1, \ldots, \bm{\Theta}_L)$.
		\UNTIL{convergence}
		\STATE Call the SSVD algorithm to orthogonalize all $\bm{G}_{m,k}$'s and $\bm{H}_{n,k}$'s, and then update $\bm{\Theta}$.
		\STATE \textbf{Return } $\widehat{\cm{A}}_{\mathrm{AR}}$ such that $[\widehat{\cm{A}}_{\mathrm{AR}}]_N = \bm{\Lambda}_y \bm{\Theta} (\bm{I}_L\otimes\bm{\Lambda}_x^{\top})$.
	\end{algorithmic}
\end{algorithm}

Let $\bm{y}_t = \vectorize(\cm{Y}_t)$, and the tensor autoregressive model at \eqref{model:AR(P)} and \eqref{model:AR(p)-feature} has the form of $\bm{y}_t =  \bm{\Lambda}_{y} \bm{\Theta}_{1}
\bm{\Lambda}_{x}^\top \bm{y}_{t-1} +\cdots+ \bm{\Lambda}_{y} \bm{\Theta}_{L}
\bm{\Lambda}_{x}^\top \bm{y}_{t-L} +\bm{e}_t$ or
\begin{equation}
	\label{eq:loss-function-AR-L-block}
	\begin{split}
		\bm{y}_t = \sum_{k=1}^{K_x}\sum_{k^\prime=1}^{K_y} \bm{\Lambda}_{y,k^\prime} \bm{\Theta}_{k,k^\prime,1}
			\bm{\Lambda}_{x,k}^\top \bm{y}_{t-1} +\cdots+ \sum_{k=1}^{K_x}\sum_{k^\prime=1}^{K_y} \bm{\Lambda}_{y,k^\prime} \bm{\Theta}_{k,k^\prime,L}
			\bm{\Lambda}_{x,k}^\top \bm{y}_{t-L} +\bm{e}_t,
	\end{split}
\end{equation}
where $\bm{e}_t=\vectorize(\cm{E}_t)$.
Note that, when updating one block of parameters while the other two blocks are fixed, the model at \eqref{eq:loss-function-AR-L-block} can be rewritten into a standard linear regression.
This motivates us to update the three blocks of parameters in an alternating way: we first update each block column of $\bm{\Lambda}_x$, i.e., $\bm{\Lambda}_{x,1}$ up to $\bm{\Lambda}_{x,K_x}$, and
then update $\bm{\Lambda}_y$ in a similar way, and update the coefficient matrix $\bm{\Theta}$, which forms an iteration; see Algorithm \ref{alg:als-auto-seq} for details.

%
%
%

In addition, when updating $\bm{\Lambda}_{x,k}(\bm{G}_{m,k}, 1\leq m\leq M)$ for each $1\leq k\leq K_x$, we sequentially update each of its component matrices $\bm{G}_{m,k}$ with $1\leq m \leq M$. Interestingly, the update of each $\bm{G}_{m,k}$ can be formulated as a least squares problem by fixing all other components, and it can be verified to have a solution with closed form; see Section \ref{subsec:als-rule} of the supplementary file for details.
The situation is the same when sequentially updating the component matrices of $\bm{\Lambda}_{y,k}(\bm{H}_{n,k}, 1\leq n\leq N)$ for each $1\leq k\leq K_y$, and hence all updating steps at Algorithm \ref{alg:als-auto-seq} have solutions with closed forms.
Moreover, these component matrices are all orthogonal with unit norm, i.e., they are orthonormal. To guarantee the convergence, we give up the orthogonality during iterations, and these component matrices will be orthogonalized at the end by a SSVD algorithm detailed at Section \ref{appendix:alg-SSVD} of the supplementary file. 

For the initialization, we may simply consider random numbers around zero.
In addition, the algorithm to search for the estimate $\widehat{\cm{A}}_{\mathrm{LR}}$ in Section \ref{subsec:l2loss} can be designed similarly.

\subsection{Hyperparameter selection}
\label{subsec:rank-selection}

We first consider the case that the sets of action orders $\mathcal{P}_x(M, K_x)$ and $\mathcal{P}_y(N, K_y)$ are known, and their cardinalities, $K_x$ and $K_y$, are small; see, for example, the number of all possible action orders is not too large. 
A validation method can then be used to select the ranks $\{r_{m,k}, 1\leq m\leq M, 1\leq k\leq K_x\}$ and $\{s_{n,k^\prime}, 1\leq n\leq N, 1\leq k^{\prime}\leq K_y\}$.

Specifically, for an observed time series $\{\cm{Y}_t, 1-L\leq t\leq T\}$, we set the training and validation sets to be the first $T_0+L$ and last $T-T_0$ observations, respectively.
A rolling forecasting method with the fixed starting point is then employed, and we will choose the ranks with minimum validation errors. 
Note that, for an action order from $\mathcal{P}_x(M, K_x)$ or $\mathcal{P}_y(N, K_y)$, its associated ranks will be set to zero if it is not involved in extracting features.
Moreover, it is impossible to search among all possible ranks since there are many interim ranks. We suggest selecting these ranks sequentially, i.e., we choose a rank among $1\leq r_{m,k} \leq R_{\max}$ or $1\leq s_{n,k^\prime} \leq S_{\max}$, while all the other ranks are kept unchanged.
This will depend on the order of ranks, and the selected ranks are also not optimal globally. Practically, we may iterate the procedure several times to alleviate this effect to some extent.

We next introduce a boosting method to select the sets of action orders, $\mathcal{P}_x(M, K_x)$ and $\mathcal{P}_y(N, K_y)$, as well as the maximum possible ranks, $R_{\max}$ and $S_{\max}$, in a data-driven manner.
To this end, the Bayesian information criterion (BIC) is given below,
\begin{equation*}
	\label{eq:BIC}
	\mathrm{BIC}=T\log\left\{ \mathcal{L}_T(\cm{\widehat{A}}) \right\}+ d_\mathrm{AR}\log(T) ,
\end{equation*}
where $d_\mathrm{AR} = \sum_{k=1}^{K_x} d_{x}(\alpha_k) + \sum_{k=1}^{K_y} d_{y}(\beta_{k}) + Lsr$ is the model complexity.
Moreover, for an action order $\alpha = (\alpha_{(1)}, \alpha_{(2)}, \ldots, \alpha_{(M)})\in\mathcal{P}_x(M)$ with ranks $\bm{r} = (r_{1}, \ldots, r_{M})$, we denote $\bm{\Lambda}_{x,\alpha}(\bm{r})= \bm{T}_x^{\top}(\alpha)(\bm{I}_{p_{\alpha_{(2)}}\cdots p_{\alpha_{(M)}}}\otimes\bm{G}_{1}) \cdots (\bm{I}_{p_{\alpha_{(M)}}}\otimes\bm{G}_{M-1}) \bm{G}_{M} \in\mathbb{R}^{P \times r_{M}}$; see \eqref{eq:lambda-x} for more details.
Consider the action order $\alpha$ with two groups of ranks, $\bm{r}_1$ and $\bm{r}_2$, and it is equivalent to the case with ranks $\bm{r}_3=\bm{r}_1+\bm{r}_2$, up to a linear transformation; see the proposition below.

\begin{proposition}
	\label{prop:additive}
	Consider two vectors of ranks $\bm{r}_{1}=(r_{1,1},\ldots,r_{M,1})$ and $\bm{r}_{2}=(r_{1,2},\ldots,r_{M,2})$, and there exist a matrix $\bm{O}\in\mathbb{R}^{r_{M,3}\times (r_{M,1}+r_{M,2})}$ and the corresponding component matrices such that ${\bm{\Lambda}}_{x,\alpha}(\bm{r}_3) = \bm{O} ({\bm{\Lambda}}_{x,\alpha}(\bm{r}_1), {\bm{\Lambda}}_{x,\alpha}(\bm{r}_2))$, where $\bm{r}_3=\bm{r}_1+\bm{r}_2=(r_{1,3},\ldots,r_{M,3})$.
\end{proposition}

The results are similar to $\bm{\Lambda}_{y}$.
Let $\mathcal{P}_x(M, K_x)=\{\alpha_1(i_1),\ldots,\alpha_{K_x}(i_{K_x})\}$ and $\mathcal{P}_y(N, K_y)=\{\beta_1(j_1),\ldots,\beta_{K_y}(j_{K_y})\}$ be the running sets of action orders, where $\alpha_k(i_k)$ refers to the action order $\alpha_k$ with ranks $(r_{1,k},\ldots,r_{M,k})=(i_k,\ldots,i_k)$, and $\beta_k(j_k)$ to the action order $\beta_k$ with ranks $(s_{1,k},\ldots,s_{N,k})=(j_k,\ldots,j_k)$.
We next introduce a procedure to update $\mathcal{P}_x(M, K_x)$ and $\mathcal{P}_y(N, K_y)$ by considering action orders with all ranks being one: 
\begin{itemize}
	\item [(i.)] Fit a tensor autoregressive model at \eqref{model:AR(P)} and \eqref{model:AR(p)-feature} with the running sets of $\mathcal{P}_x(M, K_x)$ and $\mathcal{P}_y(N, K_y)$ to the observed time series, and denote the fitted values by $\{\widehat{\cm{Y}}_t\}$.
	\item [(ii.)] Consider a regression model with the response and predictors being $\cm{Y}_t-\widehat{\cm{Y}}_t$ and $\cm{Y}_{t-1},\cdots,\cm{Y}_{t-L}$, respectively, and we try each possible action order $\alpha\in\mathcal{P}_x(M)$ with ranks $(r_1,\ldots,r_M)=(1,\ldots,1)$ and $\beta\in\mathcal{P}_y(N)$ with ranks $(s_1,\ldots,s_N)=(1,\ldots,1)$.
	Denote by $\alpha^*$ and $\beta^*$ the action orders with minimum mean square errors.
	\item [(iii.)] Update the sets of $\mathcal{P}_x(M, K_x)$ and $\mathcal{P}_y(N, K_y)$. Specifically, we add $\alpha^*(1)$ to $\mathcal{P}_x(M, K_x)$ and replace $K_x$ by $K_x+1$ if $\alpha^*$ is a new action order. Otherwise, we simply replace $\alpha_k(i_k)$ by $\alpha_k(i_k+1)$ if $\alpha^*=\alpha_k$. $\mathcal{P}_y(N, K_y)$ can be updated in the same way.
\end{itemize}

The above procedure stops when the BIC no longer decreases, and it ends up with the sets of action orders, as well as the associated maximum ranks.

\section{Simulation studies}
\label{sec:simulation}
This section conducts two simulation experiments to evaluate action order misspecification, and to check finite-sample performance of the proposed methodology, respectively.

\begin{figure}[t!]
	\centering
	\includegraphics[width=1\textwidth]{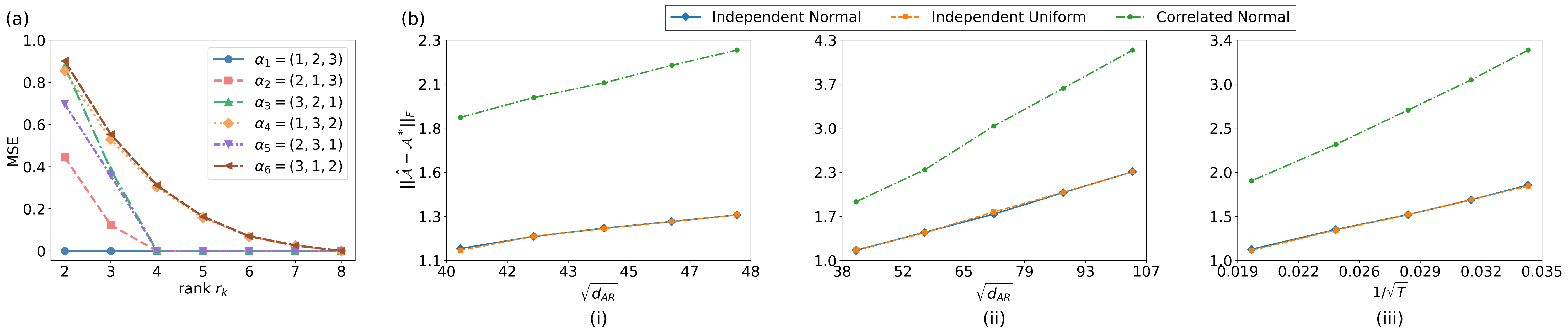}
	\caption{(a) Averaged MSEs under different action orders with varying rank $r_k$. (b) Averaged estimation errors $\|\cm{\widehat{A}}_{\mathrm{AR}}-\cm{A}^*_\mathrm{AR}\|_{\mathrm{F}}$ under three hyperparameter settings: (i) varying dimension of $q$, (ii) varying rank of $r$, and (iii) varying sample size of $T$. Three distributions of $\cm{E}_t$ are considered, as specified in the legend.}
	\label{fig:sim2}	
\end{figure}

The first experiment is to evaluate the impact of action order misspecification. Consider a true action order $\alpha^*$ with loading matrices $\{\bm{G}_{m}^*\}_{m=1}^M$. For another action order $\alpha_k$, by Proposition \ref{prop:action-equivalence}, there exists a set of loading matrices $\{\bm{G}_{m,k}\}_{m=1}^M$ such that the feature extraction mappings under both action orders are equivalent, i.e., $\mathcal{M}(\mathcal{T}(\cm{X}_t;\alpha^*);\{\bm{G}_{m}^*\}_{m=1}^M) = \mathcal{M}(\mathcal{T}(\cm{X}_t;\alpha_k);\{\bm{G}_{m,k}\}_{m=1}^M)$, but it may lead to larger ranks and hence large model complexity when $\alpha_k\neq\alpha^*$. 
To numerically verify this point, we fix $M = 3$ and specify the tensor dimensions as $(p_1, p_2, p_3) = (8,9,10)$. 
Note that there are six possible action orders in total, and the true action order is given by $\alpha^* = (1,2,3)$, with true ranks $r_1^* = r_2^* = r_3^* = 2$. The data are generated by $\bm{f}_t = \mathcal{M}(\mathcal{T}(\cm{X}_t;\alpha^*);\{\bm{G}_{m}^*\}_{m=1}^3)\in\mathbb{R}^{r_3^*}$, where entries of $\cm{X}_t\in\mathbb{R}^{p_1\times p_2\times p_3}$ are $i.i.d.$ standard normal, and
$\bm{G}_{m}^*$'s come from QR decomposition on random matrices with $i.i.d$ standard normal entries. 
We then directly regress $\bm{f}_t$ on $\cm{X}_t$, leading to a model without noise and $\bm{\Lambda}_y$ and $\bm{\Theta}$ being fixed as identity matrices.
The sample size is set to $T=1000$, and we compare the accuracy of feature extraction with each action order $\alpha_k$, where $\alpha_1 = \alpha^*$. 
When fitting the models with different action orders, we fix $r_{1,k} = r_{2,k} = r_k$ and $r_{3,k} = 2$ for all $k$ for simplicity. Algorithm \ref{alg:als-auto-seq} is applied to estimate $\bm{G}_{m,k}$'s, and the rank $r_k$ varies from 2 to 8. 
Figure \ref{fig:sim2}(a) gives mean squared errors (MSE), averaged over 50 replications.

It can be seen that, for the true action order $\alpha_1$, the MSE is zero when $r_k$ is as small as two, and this is as expected since the model is correctly specified without noise. 
Moreover, for the misspecified action orders, the MSEs are substantially higher when $r_k$ is small, but decrease gradually to zero as $r_k$ increases.
This is consistent with theoretical results at Proposition \ref{prop:action-equivalence}. On one hand, it confirms that loading matrices with different action orders can represent the same feature extraction mapping, given sufficiently large ranks. On the other hand, since the model complexity $d= \sum_{i=1}^{3}p_i r_i^{2}$ grows quadratically with respect to $r_k$,  the misspecification may lead to larger model complexity and computational burden.

The second experiment is to check whether the estimation error, $\|\cm{\widehat{A}}_{\mathrm{AR}}-\cm{A}^*_\mathrm{AR}\|_{\mathrm{F}}$, has a convergence rate of $\sqrt{d_{\mathrm{AR}} / T}$, where $d_{\mathrm{AR}} = (K_x r^2 + K_y s^2)Nq + LK_x K_y rs$, since Theorem \ref{thm:autoreg} only provides an upper bound of $O_p(\sqrt{d_{\mathrm{AR}}  / T})$, up to a logarithmic factor.
To this end, we generate the sequences $\{\cm{Y}_t\}$ with $\cm{Y}_t\in\mathbb{R}^{q\times q\times q}$, i.e. $N=3$, by using the autoregressive model at \eqref{model:AR(P)} with order $L=2$, and error terms $\{\cm{E}_t\}$ are $i.i.d.$ across all time points. The entries of  $\cm{E}_t$ follow (i) $i.i.d.$ uniform distribution on $[-0.5, 0.5]$, (ii) $i.i.d.$ standard normal distribution, or (iii) correlated Gaussian distribution with $\vectorize(\cm{E}_t) \sim \mathcal{N}(\bm{0}, \bm{\Sigma}_e)$ where $\bm{\Sigma}_e = (0.5^{|i-j|})_{1\leq i, j \leq q^3}$.
The coefficient tensors, $\cm{A}_1$, $\cm{A}_2\in\mathbb{R}^{q\times q\times q \times q\times q\times q}$, have a low-rank structure at \eqref{model:AR(p)-feature} with $\bm{\Lambda}_x$ and $\bm{\Lambda}_y$ being defined as in \eqref{eq:lambda-x} and $K_x=K_y=2$.
The two action orders are $\{ (1,2,3), (2,1,3) \}$ for both responses and predictors, and their ranks are $r_{m,1} = r_{m,2} = r$ and $s_{n,1} = s_{n,2} = s$ with $s=r-1$.
To generate the component matrices $\{ \{ \bm{G}_{n,k} \}_{n=1}^3 \}_{k=1}^2$ and $\{ \{ \bm{H}_{n,k^\prime} \}_{n=1}^3 \}_{k^\prime=1}^2$, we apply QR decomposition to random matrices with $i.i.d.$ standard normal entries to ensure orthogonality. The coefficient matrices $\bm{\Theta}_1$ and $\bm{\Theta}_2$ are generated by $i.i.d.$ uniform entries from $0$ to $4$, and $\cm{A}_1$ and $\cm{A}_2$ are then computed accordingly. 
To ensure stationarity, we rescale the coefficient tensors by a common factor such that $\|([\cm{A}_1]_{(1)}, [\cm{A}_2]_{(1)})\|_{\mathrm{F}} = 0.8$.

Finally, for the model complexity of $d_{\mathrm{AR}}$, the three hyperparameters, $r$, $q$ and $T$, may vary, while all the others have already been fixed in the above.
As a result, we (i) vary $q \in \{ 10, 11, 12, 13, 14 \}$ with $(r, T) = (3, 2500)$ being fixed, (ii) vary $r \in \{ 3, 4, 5, 6, 7 \}$ with $(q, T) = (10, 2500)$ being fixed, or (iii) vary $T \in \{ 833, 1000, 1250, 1670, 2500 \}$ with $(q, r) = (10, 3)$ being fixed. 
Algorithm \ref{alg:als-auto-seq} is applied to search for $\cm{\widehat{A}}_{\mathrm{AR}}$, and Figure \ref{fig:sim2}(b) presents the estimation errors $\|\cm{\widehat{A}}_{\mathrm{AR}}-\cm{A}^*_\mathrm{AR}\|_{\mathrm{F}}$, averaged over 200 replications, for each combination of error distributions and hyperparameter settings.
The linearity observed in Figure \ref{fig:sim2}(b) confirms the convergence rate, i.e., the squared estimation error depends linearly on $q$ and quadratically on $r$.
It thereby demonstrates the efficiency of our proposed methodology.

This paper also conducts a simulation experiment to compare our feature extracting methods with two traditional ones, hierarchical factor models (HFM) and higher-order factor models (HOFM), under the new model-designing framework. The results are provided in the supplementary file due to the limited space, and the consistently better performance can be observed. It perhaps is due to the fact that both HFM and HOFM involve a large number of parameters, demonstrating the necessity of shared components within each mode.

\section{Real data analysis}
\label{sec:real-data}
This section applies the proposed methodology to the Personality-120 example, i.e. the International Personality Item Pool (IPIP) NEO-120 personality inventory \citep{johnson2014measuring}.
The data span from 2015 to 2025, and they are available at the Hugging Face Datasets repository (\url{https://huggingface.co/datasets/ecorbari/IPIP120-SCORES}). There are five major personality traits (Neuroticism, Extraversion, Conscientiousness, Agreeableness, and Openness to Experience), and each trait is assessed by 24 questions arranged according to psychometrical conventions \citep{johnson2014measuring,kajonius2019assessing}. 
The answers to all questions are recorded on a 1-5 Likert scale, and the reverse-worded questions are reverse-coded to ensure that higher scores indicate higher levels of the corresponding personality traits.
Participants are categorized by gender (two groups) and age (six groups: 10--19, 20--29, 30--39, 40--49, 50--59, and 60+), resulting in 12 demographic groups. 
The monthly averaged scores are calculated, forming a tensor-valued time series $\{\cm{Y}_t\in\mathbb{R}^{12\times 5\times 24}, 1\leq t\leq T=115\}$. 
There is only one missing entry in the data, and it is imputed by using spline interpolation.
For each sequence, it is first detrended by the first-order differencing to ensure stationarity, and we then standardize it to have mean zero and variance one.

\begin{figure}[t!]
	\centering
	\includegraphics[width=1\textwidth]{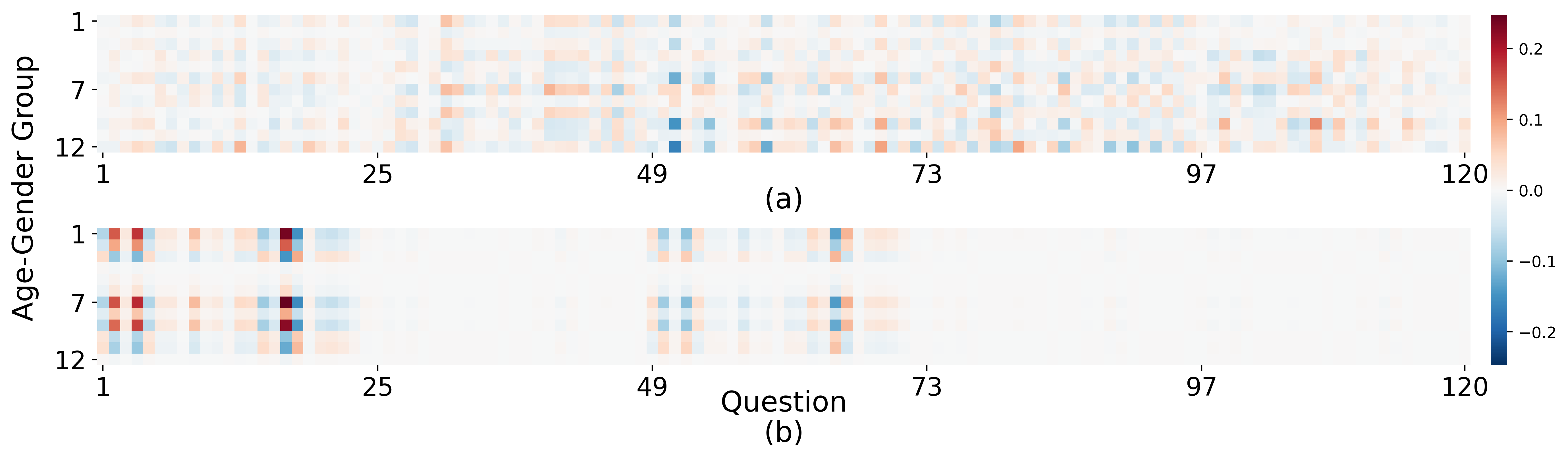}
	\caption{Heatmaps of loading matrices for two predictor factors. The upper and lower panels correspond to action orders $\alpha_1=(3,2,1)$ and $\alpha_2=(1,3,2)$, respectively.}
	\label{fig:lambda-x}
\end{figure}

Tensor autoregressive model at \eqref{model:AR(P)} is fitted to the sequence, and we simply set the order $L=1$, ending up with the coefficient tensor $\cm{A}\in\mathbb{R}^{12\times 5\times 24\times 12\times 5\times 24}$.
Note that there are only two possible action orders $\{(3,2,1),(1,3,2)\}$. The estimating procedure in Section \ref{sec:algorithms} is employed, and the selected interim ranks are $\bm{r}_1 = (6,5,1)$, $\bm{r}_2 = (1,1,1)$, $\bm{s}_1 = (6,1,1)$, and $\bm{s}_2 = (1,1,1)$, i.e., there is one predictor and one response factors for each of two action orders.  
We focus on the two predictor factors, and Figure \ref{fig:lambda-x} gives their loading matrices.

\begin{figure}[t!]
	\centering
	\includegraphics[width=1\textwidth]{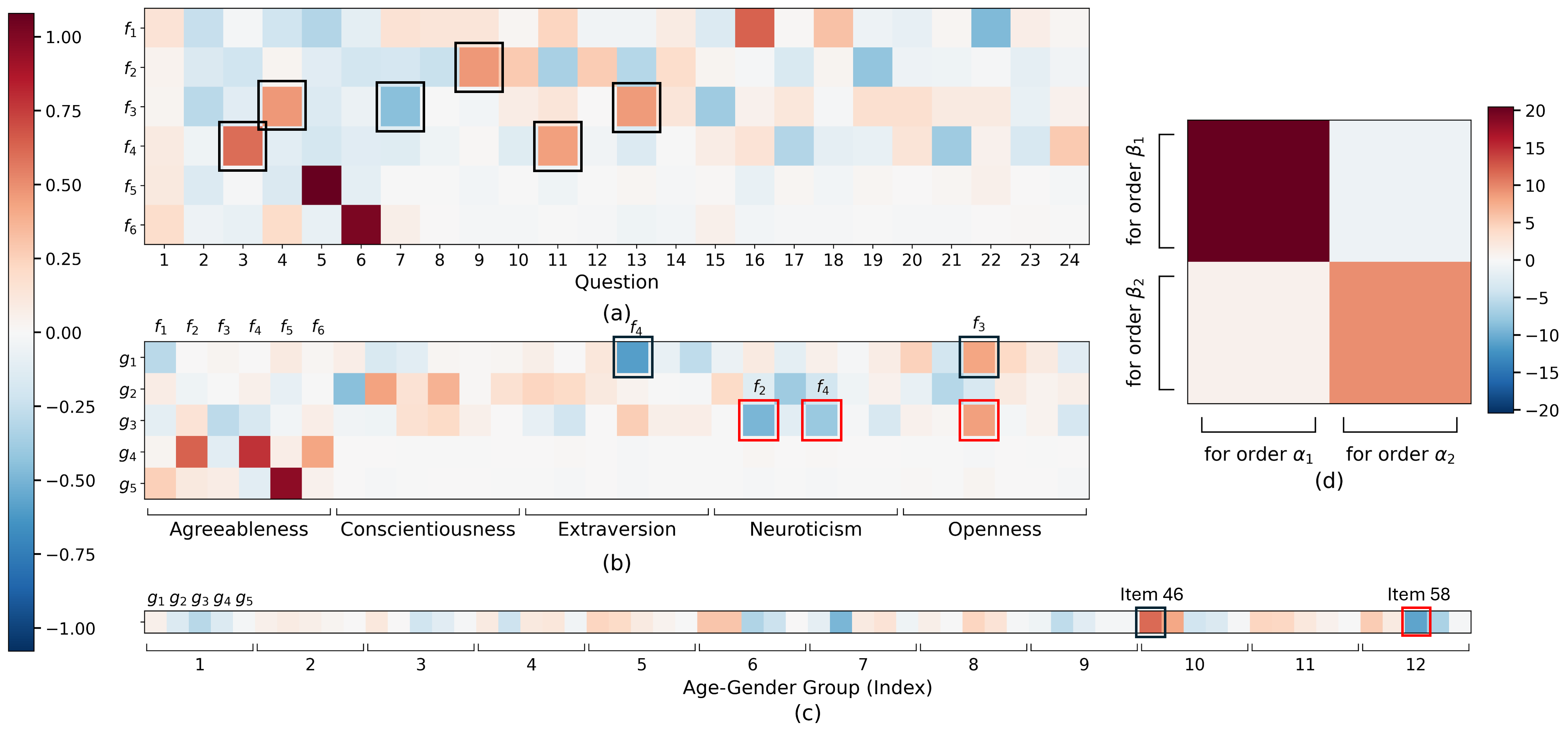}
	\caption{Heatmaps of reshaped estimated component matrices of the predictor factor for action order $\alpha_1=(3,2,1)$ at (a)--(c) and that of the estimated coefficient matrix of $\bm{\Theta}$ at  (d). } 
\label{fig:real}
\end{figure}

There are different interpretations for the two action orders. For $\alpha_1=(3,2,1)$, it first summarizes the 24 questions into six factors within each trait, and then extracts five features from $5\times 6=30$ factors for each demographic group, i.e., the five features can represent specific characteristics of each demographic group.
Finally, the predictor factor is formed from $12\times 5=60$ features; see also Figure \ref{fig:real} for the illustration. 
On the other hand, for action order $\alpha_2=(1,3,2)$, it first summarizes the demographic groups, and then the two modes in the hierarchical structure.
Since the effect of demographic groups on the questions is integrated in the first step, this order thus captures the common trend across demographic groups, and this trend concentrates on two personality traits, Agreeableness and Extraversion; see Figure \ref{fig:lambda-x}(b) for details.
Note that interpretations on two action orders can be well illustrated by the difference between the two loading matrices in Figure \ref{fig:lambda-x}.

We next give more details on the feature extraction process, and the promax rotation, a type of oblique rotation, is applied to the estimated component matrices.
This method allows for correlated factors, enhancing interpretability by simplifying factor structures, and it is commonly used in psychometrics \citep{hendrickson1964promax}.
Figure \ref{fig:real}(d) gives the estimated coefficient matrix of $\bm{\Theta}$, and we may argue that the action order $(3,2,1)$ dominates the feature extracting process.
As a result, we focus on the predictor factor corresponding to $\alpha_1=(3,2,1)$ again, and its component matrices are presented in Figure \ref{fig:real}(a)--(c).

It can be seen from Figure \ref{fig:real}(c) that two factors, namely Item 46 (Factor $g_1$ for females aged 40-49) and Item 58 (Factor $g_3$ for females aged 60+), have relatively large absolute loadings, indicating their importance in the final feature extraction.
For Item 46, Factor $f_4$ of Extraversion trait and $f_3$ of Openness trait exhibit large absolute loadings; see Figure \ref{fig:real}(b).
By identifying the significant cells corresponding to $f_3$ and $f_4$ in Figure \ref{fig:real}(a), we can pinpoint important questions, i.e. Question 3 and 11 of Extraversion trait and Question 4, 7 and 13 of Openness trait.
For the five questions, positive loadings are on ``Avoid contacts with others,'' ``Like to get lost in thought,'' ``Prefer variety to routine,'' and ``Do not like poetry'', while a negative one is on ``Take control of things''.
These questions highlight a blend of social introversion (low extraversion in friendliness and assertiveness) with cognitive openness (high imagination and adventurousness) for middle-aged females.
We can similarly interpret Item 58. Refer to the supplementary file for more details on model interpretations.


Finally, we compare the proposed methodology with existing methods in the literature in terms of forecasting performance; see Table \ref{tab:intro-comparison}. 
Note that models with Tucker decomposition, GFM and HFM will involve a large number of parameters, which makes them infeasible for this dataset.
Moreover, Tucker decomposition cannot capture the nested structures presented in the data. As a result, the three models are removed from our comparison.
For the two tensor decomposition based models, we adopt the ALS method in \cite{zhou2013tensor} for CP-based tensor autoregression and the Riemannian gradient descent method in \cite{si2024efficient} and \cite{qin2025computational} for TT-based model. For the HOFM under supervised learning frameworks, we use an ALS method similar to Algorithm \ref{alg:als-auto-seq} for estimation and a coordinate descent algorithm for rank selection.
Ranks are selected by their respective methods, and it leads to $r_{\mathrm{CP}} = 1$ for the CP, $\bm{r}_{\mathrm{TT}} = (1, 2, 2, 3, 2, 1, 1)$ for the TT, and $\bm{r}_{\mathrm{HOFM}} = (3, 1, 6)$ for predictors and $\bm{s}_{\mathrm{HOFM}} = (4, 1, 1)$ for responses in the HOFM.

\begin{table}[t!]
	\centering
	\caption{Mean squared forecast errors (MSFE) and mean absolute forecast errors (MAFE) of our model and three competing methods. The smallest numbers in each row are in bold.}
	\label{tab:real-data}
	\begin{tabular}{>{\centering\arraybackslash}p{2cm} >{\centering\arraybackslash}p{2cm} >{\centering\arraybackslash}p{2cm} >{\centering\arraybackslash}p{2cm} >{\centering\arraybackslash}p{2cm}}
		\toprule
		\multicolumn{1}{c}{\multirow{2}{*}{Models}} & \multicolumn{2}{c}{New framework} & \multicolumn{2}{c}{Traditional framework} \\
		\cmidrule(lr){2-3} \cmidrule(l){4-5}
		& \multicolumn{1}{c}{Ours} & \multicolumn{1}{c}{HOFM} & \multicolumn{1}{c}{CP} & \multicolumn{1}{c}{TT} \\
		\hline
		MSFE  & \textbf{1530.4}  & 1644.6 & 1577.1 & 1571.1 \\
		MAFE  & \textbf{1153.0}  & 1185.8 & 1156.0 & 1164.9 \\
		\bottomrule
	\end{tabular}
\end{table}

We divide the 115 observations into a training set with the first 103 months and a testing set with the last 12 months. 
The rolling forecast method is employed, i.e., we fit models to historical data up to month $t$ and then forecast the data at month $t+1$ for $t=103,\ldots,114$. The one-step-ahead forecast accuracy is assessed by the mean squared forecast errors (MSFE) and mean absolute forecast errors (MAFE); see Table \ref{tab:real-data} for the results. It can be seen that our model outperforms all competing methods in both MSFE and MAFE metrics. This may be due to the existence of hierarchical structures in the data, which can be effectively captured by our model. Moreover, the superior performance of our method against TT-based model highlights the advantage of our model by considering multiple action orders for partially hierarchical data. The better performance compared to HOFM-based method demonstrates the benefit of improved efficiency in feature extraction by sharing components within each mode in loading matrices. In sum, the proposed model provides an efficient and interpretable approach for modeling tensor-valued time series with hierarchical structures.

\section{Conclusion and discussion}
\label{sec:conclusion}
With technological advancements, high-dimensional data with hierarchical structures have been observed more frequently, while it is usually infeasible to construct inference tools for them due to a huge number of parameters. 
Fortunately, most of these data come with regular structures, say in tensor forms, and the variables within each mode are arranged in a similar way, leading to a seemingly factorial but hierarchical structure.
This makes it possible to design efficient and interpretable inference tools for them, and this paper focuses on regression and autoregressive models for supervised learning problems.
Theoretical properties have been established for the corresponding ordinary least squares estimation, and an alternating algorithm is introduced to search for estimates.
Applications to the Personality-120 example have demonstrated usefulness of the proposed methodology, and a nice interpretation is given to the use of multiple action orders for partially hierarchical structures. 

This paper can be extended in three possible directions.
First, missing data will be presented when we consider more detailed demographic groups, and this actually is common in psychometrics \citep{johnson2014measuring}.
It is an important task to adjust the proposed methodology to handle the case with missing data.
Secondly, due to the availability of more and more large-size datasets, it becomes urgent to consider more sophisticated models for better performance, and neural networks have played a key role in the literature \citep{chen2025recurrent}. This motivates us to design a suitable neural network with theoretical justifications for these hierarchical data.
Finally, it is well known that, comparing with AR models, the autoregressive moving average (ARMA) model can achieve more accurate prediction, and it is interesting to extend the AR model here to an ARMA-type one; see, e.g., \cite{zheng2025interpretable}.

\bibliography{TR}

@article{wu2023sparse,
  title={Sparse Kronecker product decomposition: a general framework of signal region detection in image regression},
  author={Wu, Sanyou and Feng, Long},
  journal={Journal of the Royal Statistical Society Series B: Statistical Methodology},
  volume={85},
  number={3},
  pages={783--809},
  year={2023},
  publisher={Oxford University Press US}
}

@article{chen2025recurrent,
  title={Recurrent Neural Networks for Nonlinear Time Series},
  author={Chen, Xiao and Chen, Yu and Shen, Zhouyu and Xiu, Dacheng},
  journal={Chicago Booth Research Paper},
  number={26-01},
  year={2025}
}

@article{hendrickson1964promax,
  title={Promax: A quick method for rotation to oblique simple structure},
  author={Hendrickson, Alan E and White, Paul Owen},
  journal={British Journal of Statistical Psychology},
  volume={17},
  number={1},
  pages={65--70},
  year={1964},
  publisher={Wiley Online Library}
}

@article{schmid1957development,
  title={The development of hierarchical factor solutions},
  author={Schmid, John and Leiman, John M},
  journal={Psychometrika},
  volume={22},
  number={1},
  pages={53--61},
  year={1957},
  publisher={Springer}
}

@article{liu2020low,
  title={Low-rank tensor train coefficient array estimation for tensor-on-tensor regression},
  author={Liu, Yipeng and Liu, Jiani and Zhu, Ce},
  journal={IEEE Transactions on Neural Networks and Learning Systems},
  volume={31},
  number={12},
  pages={5402--5411},
  year={2020},
  publisher={IEEE}
}

@article{gao2023divide,
	title={Divide-and-conquer: a distributed hierarchical factor approach to modeling large-scale time series data},
	author={Gao, Zhaoxing and Tsay, Ruey S},
	journal={Journal of the American Statistical Association},
	volume={118},
	number={544},
	pages={2698--2711},
	year={2023},
	publisher={Taylor \& Francis}
}

@article{chen2022factor,
  title={Factor models for high-dimensional tensor time series},
  author={Chen, Rong and Yang, Dan and Zhang, Cun-Hui},
  journal={Journal of the American Statistical Association},
  volume={117},
  number={537},
  pages={94--116},
  year={2022},
  publisher={Taylor \& Francis}
}

@article{zhou2013tensor,
	title={Tensor regression with applications in neuroimaging data analysis},
	author={Zhou, Hua and Li, Lexin and Zhu, Hongtu},
	journal={Journal of the American Statistical Association},
	volume={108},
	number={502},
	pages={540--552},
	year={2013},
	publisher={Taylor \& Francis}
}

@article{qiao2025exact,
	title={Exact Exploratory Bi-factor Analysis: A Constraint-Based Optimization Approach},
	author={Qiao, Jiawei and Chen, Yunxiao and Ying, Zhiliang},
	journal={Psychometrika},
	pages={1--16},
	year={2025},
	publisher={Cambridge University Press}
}

@article{klami2014group,
	title={Group factor analysis},
	author={Klami, Arto and Virtanen, Seppo and Lepp{\"a}aho, Eemeli and Kaski, Samuel},
	journal={IEEE Transactions on Neural Networks and Learning Systems},
	volume={26},
	number={9},
	pages={2136--2147},
	year={2014},
	publisher={IEEE}
}

@article{yung1999relationship,
	title={On the relationship between the higher-order factor model and the hierarchical factor model},
	author={Yung, Yiu-Fai and Thissen, David and McLeod, Lori D},
	journal={Psychometrika},
	volume={64},
	number={2},
	pages={113--128},
	year={1999},
	publisher={Springer-Verlag}
}

@article{bi2022modeling,
	title={Modeling Pregnancy Outcomes Through Sequentially Nested Regression Models},
	author={Bi, Xuan and Feng, Long and Li, Cai and Zhang, Heping},
	journal={Journal of the American Statistical Association},
	volume={117},
	number={538},
	pages={602--616},
	year={2022},
	publisher={Taylor \& Francis}
}

@article{de2000multilinear,
	title={A multilinear singular value decomposition},
	author={De Lathauwer, Lieven and De Moor, Bart and Vandewalle, Joos},
	journal={SIAM journal on Matrix Analysis and Applications},
	volume={21},
	number={4},
	pages={1253--1278},
	year={2000},
	publisher={SIAM}
}

@article{tang2018bayesian,
  title={Bayesian tensor factorization for multi-way analysis of multi-dimensional EEG},
  author={Tang, Yunbo and Chen, Dan and Wang, Lizhe and Zomaya, Albert Y and Chen, Jingying and Liu, Honghai},
  journal={Neurocomputing},
  volume={318},
  pages={162--174},
  year={2018},
  publisher={Elsevier}
}

@article{lock2018tensor,
	title={Tensor-on-tensor regression},
	author={Lock, Eric F},
	journal={Journal of Computational and Graphical Statistics},
	volume={27},
	number={3},
	pages={638--647},
	year={2018},
	publisher={Taylor \& Francis}
}

@article{harshman1970foundations,
	title={Foundations of the PARAFAC procedure: Models and conditions for an “explanatory” multi-modal factor analysis},
	author={Harshman, Richard A and others},
	journal={UCLA working papers in phonetics},
	volume={16},
	number={1},
	pages={84},
	year={1970},
	publisher={Los Angeles, CA}
}

@article{raskutti2019convex,
	title={Convex regularization for high-dimensional multiresponse tensor regression},
	author={Raskutti, Garvesh and Yuan, Ming and Chen, Han},
	year={2019},
  volume = {47},
  journal = {The Annals of Statistics},
  number = {3},
  publisher = {Institute of Mathematical Statistics},
  pages = {1554 -- 1584},
  doi = {10.1214/18-AOS1725},
  URL = {https://doi.org/10.1214/18-AOS1725}
}

@article{wilms2023sparse,
	title={Sparse identification and estimation of large-scale vector autoregressive moving averages},
	author={Wilms, Ines and Basu, Sumanta and Bien, Jacob and Matteson, David S},
	journal={Journal of the American Statistical Association},
	volume={118},
	number={541},
	pages={571--582},
	year={2023},
	publisher={Taylor \& Francis}
}

@article{qiao2025exploratory,
	title={Exploratory Hierarchical Factor Analysis with an Application to Psychological Measurement},
	author={Qiao, Jiawei and Chen, Yunxiao and Ying, Zhiliang},
	journal={arXiv preprint arXiv:2505.09043},
	year={2025}
}

@article{johnson2014measuring,
	title={Measuring thirty facets of the Five Factor Model with a 120-item public domain inventory: Development of the IPIP-NEO-120},
	author={Johnson, John A},
	journal={Journal of Research in Personality},
	volume={51},
	pages={78--89},
	year={2014},
	publisher={Elsevier}
}

@article{qin2025computational,
  title     = {Computational and statistical guarantees for tensor-on-tensor regression with tensor train decomposition},
  author    = {Qin, Zhen and Zhu, Zhihui},
  journal   = {IEEE Transactions on Pattern Analysis and Machine Intelligence},
  year      = {2025},
  publisher = {IEEE}
}

@article{wang2022high,
  title     = {High-dimensional vector autoregressive time series modeling via tensor decomposition},
  author    = {Wang, Di and Zheng, Yao and Lian, Heng and Li, Guodong},
  journal   = {Journal of the American Statistical Association},
  volume    = {117},
  number    = {539},
  pages     = {1338--1356},
  year      = {2022},
  publisher = {Taylor \& Francis}
}

@article{han2022optimal,
  title     = {An optimal statistical and computational framework for generalized tensor estimation},
  author    = {Han, Rungang and Willett, Rebecca and Zhang, Anru R},
  journal   = {The Annals of Statistics},
  volume    = {50},
  number    = {1},
  pages     = {1--29},
  year      = {2022},
  publisher = {Institute of Mathematical Statistics}
}

@article{basu2015regularized,
  title   = {Regularized estimation in sparse high-dimensional time series models},
  author  = {Basu, Sumanta and Michailidis, George},
  journal = {The Annals of Statistics},
  volume  = {43},
  pages   = {1535--1567},
  year    = {2015}
}

@article{zheng2020finite,
  author  = {Zheng, Yao and Cheng, Guang},
  title   = {Finite-time analysis of vector autoregressive models under linear restrictions},
  journal = {Biometrika},
  volume  = {108},
  number  = {2},
  pages   = {469-489},
  year    = {2020},
  month   = {08},
  issn    = {0006-3444},
  doi     = {10.1093/biomet/asaa065},
  url     = {https://doi.org/10.1093/biomet/asaa065},
  eprint  = {https://academic.oup.com/biomet/article-pdf/108/2/469/37938985/asaa065.pdf}
}

@inproceedings{melnyk2016estimating,
  author    = {Melnyk, Igor and Banerjee, Arindam},
  title     = {Estimating structured vector autoregressive models},
  year      = {2016},
  publisher = {JMLR},
  booktitle = {Proceedings of the 33rd International Conference on International Conference on Machine Learning - Volume 48},
  pages     = {830–839},
  numpages  = {10},
  location  = {New York, NY, USA},
  series    = {ICML'16}
}

@article{zhang2018tensor,
  title     = {Tensor SVD: Statistical and computational limits},
  author    = {Zhang, Anru and Xia, Dong},
  journal   = {IEEE Transactions on Information Theory},
  volume    = {64},
  number    = {11},
  pages     = {7311--7338},
  year      = {2018},
  publisher = {IEEE}
}

@book{vershynin2019high,
  place      = {Cambridge},
  series     = {Cambridge Series in Statistical and Probabilistic Mathematics},
  title      = {High-Dimensional Probability: An Introduction with Applications in Data Science},
  publisher  = {Cambridge University Press},
  author     = {Vershynin, Roman},
  year       = {2018},
  collection = {Cambridge Series in Statistical and Probabilistic Mathematics}
}

@book{wainwright2019high,
  place      = {Cambridge},
  series     = {Cambridge Series in Statistical and Probabilistic Mathematics},
  title      = {High-Dimensional Statistics: A Non-Asymptotic Viewpoint},
  publisher  = {Cambridge University Press},
  author     = {Wainwright, Martin J.},
  year       = {2019},
  collection = {Cambridge Series in Statistical and Probabilistic Mathematics}
}

@article{oseledets2011tensor,
  author  = {Oseledets, I. V.},
  journal = {SIAM Journal on Scientific Computing},
  title   = {Tensor-Train Decomposition},
  year    = {2011},
  number  = {5},
  pages   = {2295-2317},
  volume  = {33},
  doi     = {10.1137/090752286},
  eprint  = {https://doi.org/10.1137/090752286},
  url     = {https://doi.org/10.1137/090752286}
}

@article{si2024efficient,
  title         = {An efficient tensor regression for high-dimensional data},
  author        = {Yuefeng Si and Yingying Zhang and Yuxi Cai and Chunling Liu and Guodong Li},
  year          = {2024},
  journal       = {arXiv preprint arXiv:2205.13734},
}

@article{zhou2022optimal,
  title     = {Optimal high-order tensor svd via tensor-train orthogonal iteration},
  author    = {Zhou, Yuchen and Zhang, Anru R and Zheng, Lili and Wang, Yazhen},
  journal   = {IEEE Transactions on Information Theory},
  volume    = {68},
  number    = {6},
  pages     = {3991--4019},
  year      = {2022},
  publisher = {IEEE}
}

@article{wang2024high,
  title    = {High-dimensional low-rank tensor autoregressive time series modeling},
  journal  = {Journal of Econometrics},
  volume   = {238},
  number   = {1},
  pages    = {105544},
  year     = {2024},
  issn     = {0304-4076},
  doi      = {https://doi.org/10.1016/j.jeconom.2023.105544},
  url      = {https://www.sciencedirect.com/science/article/pii/S0304407623002609},
  author   = {Di Wang and Yao Zheng and Guodong Li}
}

@article{huang2025supervised,
	title={Supervised factor modeling for high-dimensional linear time series},
	author={Huang, Feiqing and Lu, Kexin and Zheng, Yao and Li, Guodong},
	journal={Journal of Econometrics},
	volume={249},
	pages={105995},
	year={2025},
	publisher={Elsevier}
}

@article{cai2025efficient,
  title   = {An Efficient and Interpretable Autoregressive Model for High-Dimensional Tensor-Valued Time Series},
  author  = {Cai, Yuxi and Li, Lan and Wang, Yize and Li, Guodong},
  journal = {arXiv preprint arXiv:2506.01658},
  year    = {2025}
}

@article{li2021tensor,
  title={Tensor quantile regression with application to association between neuroimages and human intelligence},
  author={Li, Cai and Zhang, Heping},
  journal={The Annals of Applied Statistics},
  volume={15},
  number={3},
  pages={1455},
  year={2021}
}

@article{hao2021sparse,
  title={Sparse tensor additive regression},
  author={Hao, Botao and Wang, Boxiang and Wang, Pengyuan and Zhang, Jingfei and Yang, Jian and Sun, Will Wei},
  journal={Journal of Machine Learning Research},
  volume={22},
  number={64},
  pages={1--43},
  year={2021}
}

@article{li2021multi,
  title={Multi-linear tensor autoregressive models},
  author={Li, Zebang and Xiao, Han},
  journal={arXiv preprint arXiv:2110.00928},
  year={2021}
}

@article{bi2018multilayer,
	ISSN = {00905364, 21688966},
	URL = {https://www.jstor.org/stable/26542903},
	author = {Xuan Bi and Annie Qu and Xiaotong Shen},
	journal = {The Annals of Statistics},
	number = {6B},
	pages = {3308--3333},
	publisher = {Institute of Mathematical Statistics},
	title = {MULTILAYER TENSOR FACTORIZATION WITH APPLICATIONS TO RECOMMENDER SYSTEMS},
	urldate = {2024-07-01},
	volume = {46},
	year = {2018}
}

@book{talagrand2005generic,
  title     = {The generic chaining: upper and lower bounds of stochastic processes},
  author    = {Talagrand, Michel},
  year      = {2005},
  publisher = {Springer Science \& Business Media}
}

@article{tucker1966some,
  title   = {Some mathematical notes on three-mode factor analysis},
  author  = {Tucker, Ledyard},
  year    = {1966},
  journal = {Psychometrika},
  volume  = {31},
  number  = {3},
  pages   = {279-311},
  url     = {https://EconPapers.repec.org/RePEc:spr:psycho:v:31:y:1966:i:3:p:279-311}
}

@article{kolda2009tensor,
  author  = {Kolda, Tamara G. and Bader, Brett W.},
  journal = {SIAM Review},
  title   = {Tensor Decompositions and Applications},
  year    = {2009},
  number  = {3},
  pages   = {455-500},
  volume  = {51},
  doi     = {10.1137/07070111X},
  eprint  = {https://doi.org/10.1137/07070111X},
  url     = {https://doi.org/10.1137/07070111X}
}

@article{zheng2025interpretable,
  title={An interpretable and efficient infinite-order vector autoregressive model for high-dimensional time series},
  author={Zheng, Yao},
  journal={Journal of the American Statistical Association},
  volume={120},
  number={549},
  pages={212--225},
  year={2025},
  publisher={Taylor \& Francis}
}

@article{dirksenTailBoundsGeneric2015a,
  title     = {Tail Bounds via Generic Chaining},
  author    = {Dirksen, Sjoerd},
  year      = {2015},
  month     = jan,
  journal   = {Electronic Journal of Probability},
  volume    = {20},
  number    = {none},
  pages     = {1--29},
  publisher = {{Institute of Mathematical Statistics and Bernoulli Society}},
  issn      = {1083-6489, 1083-6489},
  doi       = {10.1214/EJP.v20-3760},
  urldate   = {2025-08-29}
}

@article{talagrand2001majorizing,
  author  = {Talagrand, Michel},
  journal = {The Annals of Probability},
  number  = {1},
  title   = {Majorizing Measures without Measures},
  volume  = {29},
  year    = {2001},
  pages   = {411--417},
  url     = {http://www.jstor.org/stable/2652928}
}

@article{banerjee2015estimation,
  title={Estimation with norm regularization},
  author={Banerjee, Arindam and Chen, Sheng and Fazayeli, Farideh and Sivakumar, Vidyashankar},
  journal={arXiv preprint arXiv:1505.02294},
  year={2015}
}

@book{luetkepohl2005intro,
  author  = {Luetkepohl, Helmut},
  year    = {2005},
  month   = {01},
  pages   = {},
  title   = {The New Introduction to Multiple Time Series Analysis},
  isbn    = {978-3-540-40172-8},
  journal = {New Introduction to Multiple Time Series Analysis},
  doi     = {10.1007/978-3-540-27752-1}
}

@Article{Lam2012,
  author  = {Lam, Clifford and Yao, Qiwei},
  journal = {Annals of Statistics},
  title   = {FACTOR MODELING FOR HIGH-DIMENSIONAL TIME SERIES: INFERENCE FOR THE NUMBER OF FACTORS},
  year    = {2012},
  pages   = {694--726},
  volume  = {40},
}

@Article{Bai2016,
  author    = {Bai, Jushan and Wang, Peng},
  journal   = {Annual Review of Economics},
  title     = {Econometric analysis of large factor models},
  year      = {2016},
  pages     = {53--80},
  volume    = {8},
  publisher = {Annual Reviews},
}

@Article{Chen_Xiao_Yang2021,
  author  = {Rong Chen and Han Xiao and Dan Yang},
  journal = {Journal of Econometrics},
  title   = {Autoregressive models for matrix-valued time series},
  year    = {2021},
  pages   = {539 - 560},
  volume  = {222},
}

@Article{Negahban2011,
  author    = {Negahban, Sahand and Wainwright, Martin J},
  journal   = {Annals of Statistics},
  title     = {Estimation of (near) low-rank matrices with noise and high-dimensional scaling},
  year      = {2011},
  pages     = {1069--1097},
  volume    = {39},
  publisher = {JSTOR},
}

@article{kajonius2019assessing,
  author  = {Kajonius, Petri J. and Johnson, John A.},
  title   = {{Assessing the Structure of the Five Factor Model of Personality (IPIP-NEO-120) in the Public Domain}},
  journal = {Europe's Journal of Psychology},
  year    = {2019},
  volume  = {15},
  number  = {2},
  pages   = {260--275},
  month   = {jun},
  doi     = {10.5964/ejop.v15i2.1671},
  url     = {https://doi.org/10.5964/ejop.v15i2.1671}
}

\appendix
\setcounter{algorithm}{0}
\renewcommand{\thealgorithm}{S.\arabic{algorithm}}
\renewcommand{\theHalgorithm}{S.\arabic{algorithm}}
\renewcommand{\theremark}{S.\arabic{remark}}
\renewcommand{\theHremark}{S.\arabic{remark}}
\setcounter{remark}{0}
\numberwithin{definition}{section}
\numberwithin{lemma}{section}
\numberwithin{proposition}{section}
\numberwithin{theorem}{section}
\numberwithin{corollary}{section}
\numberwithin{assumption}{section}
\numberwithin{condition}{section}
\numberwithin{example}{section}
\numberwithin{innercustom}{section}
\numberwithin{figure}{section}
\setcounter{definition}{0}
\setcounter{lemma}{0}
\setcounter{proposition}{0}
\setcounter{theorem}{0}
\setcounter{corollary}{0}
\setcounter{assumption}{0}
\setcounter{condition}{0}
\setcounter{example}{0}
\setcounter{innercustom}{0}
\setcounter{figure}{0}

\title{Supplementary Material for ``High-dimensional Autoregressive Modeling for Time Series Data with Hierarchical Structures"}

\author{\centering
	Lan Li, Shibo Yu, Yingzhou Wang and Guodong Li
	
	\textit{Department of Statistics and Actuarial Science, University of Hong Kong}}
\date{}
\makeatletter
\renewcommand{\@thanks}{}
\makeatother
\maketitle

\begin{abstract}
	\vspace{-2mm}
	This supplementary material provides technical details and additional empirical evidence for the main paper. We first introduce the tensor notation and algebra used throughout the paper in Section \ref{appendix:notation}, and then present additional results on model properties in Section \ref{appendix:sec-alg}, including proofs of propositions, derivations of the closed-form ALS update rules, and the SSVD procedure used for orthogonalization. We also provide complete proofs of the main theoretical guarantees in Section \ref{appendix:sec-theory}, together with auxiliary lemmas in Section \ref{appendix:sec-lemma}. Finally, we include additional details for the simulation studies in Section \ref{appendix:simu} and further empirical results for the Personality-120 data analyses in Section \ref{appendix:real-data}, which complement the findings reported in the main text.

\end{abstract}

\section{Tensor Notation and Algebra} \label{appendix:notation}
Tensors, a.k.a. multidimensional arrays, are natural high order extensions of matrices, and the order of a tensor is known as the dimension, way or mode; see \cite{kolda2009tensor} for a review of basic tensor algebra.
A $d$-th order tensor $\cm{X}\in\mathbb{R}^{p_1\times p_2\times\cdots \times p_d}$ is defined as a $d$-dimensional array, where each order is called a mode, and its element is denoted by $\cm{X}_{i_1i_2\ldots i_d}$ or $\cm{X}_{i_1,i_2,\ldots, i_d}$ for $1\le i_{l}\le p_{l}$ with $1\leq l\leq d$.

For a $d$-th order tensor $\cm{X}\in\mathbb{R}^{p_1\times p_2\times\cdots \times p_d}$ and a matrix $\bm{Y}\in\mathbb{R}^{q_k\times p_k}$, their mode-$k$ multiplication is defined as $\cm{X}\times_k\bm{Y}\in \mathbb{R}^{p_1\times\cdots \times p_{k-1}\times q_k\times p_{k+1}\times \cdots \times p_d}$ with elements of
\[
(\cm{X}\times_k\bm{Y})_{i_1,\ldots, i_{k-1},j,i_{k+1},\ldots, i_d}=\sum_{i_k=1}^{p_k}\cm{X}_{i_1,\ldots, i_d}\bm{Y}_{ji_k},
\]
where $1\leq j \leq q_k$, and $1\leq i_l\leq p_l$ with $1\leq l\leq d$ and $l\neq k$.
Given another $(d-m)$-th order tensor $\cm{Z}\in\mathbb{R}^{p_{m+1}\times p_{m+2}\times\cdots \times p_d}$, the generalized inner product of $\cm{X}$ and $\cm{Z}$ is defined as $\langle\cm{X},\cm{Z}\rangle\in \mathbb{R}^{p_{1}\times\cdots \times p_{m}}$ with elements of
\[
\langle\cm{X},\cm{Z}\rangle_{i_{1},\ldots, i_m}=\sum_{i_{m+1}=1}^{p_{m+1}}\sum_{i_{m+2}=1}^{p_{m+2}}\cdots\sum_{i_d=1}^{p_d}\cm{X}_{i_1,i_2,\ldots, i_d}\cm{Z}_{i_{m+1},i_{m+2},\ldots, i_d} \text{ for $1\leq i_l\leq p_l$ with $1\leq l\leq m$}.
\]
When $m=0$, it will reduce to a scalar, and then the generalized inner product becomes an inner product.
Moreover, the Frobenius norm of $\cm{X}$ is defined as $\norm{\cm{X}}_{\mathrm{F}}=\sqrt{\langle\cm{X},\cm{X}\rangle}$.

Matricization or unfolding is an operation to reshape a tensor into matrices of different sizes, and this paper will involve two types of matricization: \textit{mode matricization} and \textit{sequential matricization}.
Mode-$s$ matricization sets the $s$-th mode as rows, and columns enumerate the rest modes.
Specifically, let $p_{-s}=\prod_{i=1,i\neq s}^dp_i$, and the tensor $\cm{X}$ is reshaped into a $p_s$-by-$ p_{-s}$ matrix, denoted by $\cm{X}_{(s)}$, where the element $\cm{X}_{i_1,i_2,\ldots,i_d}$ is mapped to $(i_s,j)$-th element of $\cm{X}_{(s)}$ with
\[
j=1+\sum_{k=1,k\neq s}^d(i_k-1)J_k \hspace{2mm}\text{and}\hspace{2mm} J_k=\prod_{l=1,l\neq s}^{k-1}p_l.
\]
Sequential matricization reshapes tensor $\cm{X}$ into a $\prod_{i=1}^sp_i$-by-$\prod_{i=s+1}^dp_i$ matrix, denoted by $[\cm{X}]_{s}$, where rows enumerate all indices from modes 1 to $s$, and columns enumerate modes $s+1$ to $d$. Specifically, $\cm{X}_{i_1,i_2,\ldots,i_d}$ is mapped to $(i,j)$-th element of $[\cm{X}]_{s}$ with
\[
i=(i_s-1)p_1\cdots p_{s-1}+(i_{s-1}-1)p_1\cdots p_{s-2}+\cdots+i_1 \hspace{2mm}\text{and}\hspace{2mm} j=(i_d-1)p_{s+1}\cdots p_{d-1}+\cdots+i_{s+1}.
\]
We also define the Reshape$(\cdot; \cdot)$ operation such that $\text{Reshape}(\bm{a}; r, p)\in \mathbb{R}^{r\times p}$ for $\bm{a}\in\mathbb{R}^{rp}$ and $\vectorize(\text{Reshape}(\bm{a}; r, p))=\bm{a}$.

\section{Additional Details on Model Properties and Algorithm}
\label{appendix:sec-alg}
This section provides additional details supporting the results presented in Section~\ref{sec:theory}. Specifically, we elaborate on the derivation of certain conclusions and describe the \textbf{S}equential \textbf{S}ingular \textbf{V}alue \textbf{D}ecomposition (SSVD) procedure used to orthogonalize the components in our proposed model, as implemented in Algorithm~\ref{alg:als-auto-seq}.

\subsection{Proof of Proposition \ref{prop:action-equivalence}}
We restate Proposition \ref{prop:action-equivalence} in a more convenient way:
\begin{proposition}
	Specify any action order $\alpha=(\alpha_{(1)},\alpha_{(2)},\dots, \alpha_{(M)})\in \mathcal{P}(M)$ and another target action order $\alpha'=(\alpha'_{(1)},\alpha'_{(2)},\dots, \alpha'_{(M)})\in \mathcal{P}(M)$. Then, to any set of core matrices $\{\bm{G}_{m}\in \mathbb{O}^{r_{m-1}p_m \times r_m}\}_{m=1}^M$ there corresponds a set of core matrices $\{\bm{G}'_{m}\in \mathbb{O}^{r_{m-1}'p'_{m}\times r_m'}\}_{m=1}^M$ such that $\mathcal{M}(\mathcal{T}(\cdot;\alpha);\{\bm{G}_{m}\}_{m=1}^M) = \mathcal{M}(\mathcal{T}(\cdot;\alpha');\{\bm{G}'_{m}\}_{m=1}^M)$.
\end{proposition}
\begin{proof}
We begin by noting that the group action of switching the order of two adjacent elements in the sequence of actions $\alpha$ makes the permutation set $\mathcal{P}(M)$ a permutation group. This means that any permutation of the action order can be achieved by a series of adjacent swaps and thus we may assume $\alpha=(1,2,3,\dots, M)$ and $\alpha'=(1,2,\dots,k+1,k,k+2,\dots, M)$ without loss of generality. We abbreviate the operator $\mathcal{T}(\cdot; \alpha')$ as $\mathcal{T}$.

Let $\bm{G}_M^\top (\bm{I}_{p_M}\otimes \bm{G}_{M-1}^\top)(\bm{I}_{p_M p_{M-1}}\otimes \bm{G}_{M-2}^\top)\cdots(\bm{I}_{p_M p_{M-1}\cdots p_2}\otimes \bm{G}_{1}^\top)$ be the matrix corresponding to the operator $\mathcal{M}(\mathcal{T}(\cdot;\alpha);\{\bm{G}_{m}\}_{m=1}^M)$, where each $\bm{G}_m \in \mathbb{O}^{p_m r_{m-1}\times r_m}$ is the orthogonal matrix corresponding to the $m$-th action in the order $\alpha$, and $r_m$ is the extraction size for the $m$-th interim feature. By assumption, $p_1\times p_2 \times \cdots \times p_M$ is the size of the input tensor to be processed.
We are left to show that there exist orthogonal matrices $\bm{G}_m' \in \mathbb{O}^{p'_{m}r_{m-1}'\times r_m'}, m=1,2,\dots, M$ with $r_m' \leq p'_m r_{m-1}'$ such that 
\begin{align}
\label{eq:action-equivalence-main}
	\begin{split}
	\underbrace{\bm{G}_M'^\top (\bm{I}_{p_M}\otimes \bm{G}_{M-1}'^\top)(\bm{I}_{p_M p_{M-1}}\otimes \bm{G}_{M-2}'^\top)\dots}_{=:\bm{A}'} (\bm{I}_{p_M p_{M-1}\dots p_{k+2}}\otimes \bm{H}_{k}^\top)
	(\bm{I}_{p_M p_{M-1}\dots p_{k+2} p_{k}}\otimes \bm{H}_{k+1}^\top)\\
	\underbrace{(\bm{I}_{p_M p_{M-1}\dots p_{k+1} p_{k}}\otimes \bm{G}_{k-1}'^\top)\cdots
	(\bm{I}_{p_M p_{M-1}\cdots p_2}\otimes \bm{G}_{1}'^\top)}_{=:\bm{C}'}\bm{T}\\
	= \underbrace{\bm{G}_M^\top (\bm{I}_{p_M}\otimes \bm{G}_{M-1}^\top)(\bm{I}_{p_M p_{M-1}}\otimes \bm{G}_{M-2}^\top)\dots}_{=:\bm{A}}(\bm{I}_{p_M p_{M-1}\dots p_{k+2}}\otimes \bm{G}_{k+1}^\top)
	(\bm{I}_{p_M p_{M-1}\dots p_{k+1}}\otimes \bm{G}_{k}^\top)\\
	\underbrace{(\bm{I}_{p_M p_{M-1}\dots p_{k+1} p_{k}}\otimes \bm{G}_{k-1}^\top)\cdots
	(\bm{I}_{p_M p_{M-1}\cdots p_2}\otimes \bm{G}_{1}^\top)}_{=:\bm{C}}
	\end{split}
\end{align}
where $\bm{T}$ is the matrix corresponding to the operator $\mathcal{T}$, and $\bm{H}_k=\bm{G}'_{k+1}, \bm{H}_{k+1}=\bm{G}'_k$ for notational convenience.

Denote $\mathcal{C}$ and $\mathcal{C}'$ as the operators corresponding to the matrices $\bm{C}$ and $\bm{C}'$, respectively, i.e., $\text{vec}(\mathcal{C}\cm{X})=\bm{C}\text{vec}(\cm{X})$ and $\text{vec}(\mathcal{C}'\cm{X}') = \bm{C}'\text{vec}(\cm{X}')$. Note that we omit the parentheses between the operator and the tensor for simplicity. We first want to show that for any $1\leq i_{k-1} \leq r_{k-1}$, and $1\leq i_m\leq p_m$ for $m=k, k+1,\dots,M$, and for all $\cm{X}\in \mathbb{R}^{p_1\times p_2\times\cdots\times p_M}$, the following holds:
\begin{equation}
\label{eq:action-equivalence-1}
(\mathcal{CT}\cm{X})_{i_{k-1}, i_{k+1}, i_k, i_{k+2}, \dots, i_M}=(\mathcal{C}\cm{X})_{i_{k-1}, i_k, i_{k+1}, i_{k+2}, \dots, i_M}
\end{equation}
and thereby $\mathcal{C'}$ can be constructed as $\mathcal{C}$, i.e., $\bm{G}_m'=\bm{G}_m$ for $m=1,2,\dots,k-1$.

We begin by first observing that $(\bm{I}_{p_M p_{M-1}\cdots p_2}\otimes \bm{G}_{1}^\top) \vectorize(\mathcal{T}\cm{X}) = \vectorize(\bm{G}_1^\top [\mathcal{T}\cm{X}]_1)$. Let $\cm{X}^{(1)} \in \mathbb{R}^{r_1\times p_2\times\cdots\times p_M}, \cm{X}'^{(1)}\in \mathbb{R}^{r_1\times\cdots p_{k-1}\times p_{k+1} \times p_k\times p_{k+2}\cdots\times p_M}$ be the tensors such that $\vectorize(\cm{X}^{(1)}) = (\bm{I}_{p_M p_{M-1}\cdots p_2}\otimes \bm{G}_{1}^\top) \vectorize(\cm{X})$, and $\vectorize(\cm{X}'^{(1)}) = (\bm{I}_{p_M p_{M-1}\cdots p_2}\otimes \bm{G}_{1}^\top) \vectorize(\mathcal{T}\cm{X})$. Then, we have
\begin{align*}
(\cm{X}'^{(1)})_{i_1, \dots, i_{k-1}, i_{k+1}, i_{k}, i_{k+2}, \dots, i_M} &= \sum_{j_1=1}^{p_1} (\bm{G}_1^\top)_{i_1, j_1} (\mathcal{T}\cm{X})_{j_1, i_2,\dots, i_{k-1}, i_{k+1}, i_{k}, i_{k+2}, \dots, i_M}\\
(\cm{X}^{(1)})_{i_1, i_2, \dots, i_M} &= \sum_{j_1=1}^{p_1} (\bm{G}_1^\top)_{i_1, j_1} \cm{X}_{j_1, i_2, \dots, i_M}.
\end{align*}
But by the definition of $\mathcal{T}$, $(\mathcal{T}\cm{X})_{i_1, i_2,\dots, i_{k-1}, i_{k+1}, i_{k}, i_{k+2}, \dots, i_M} = \cm{X}_{i_1, i_2, \dots, i_M}$ for all applicable $i_m$, and thus we have
\begin{align*}
(\cm{X}'^{(1)})_{i_1, \dots, i_{k-1}, i_{k+1}, i_{k}, i_{k+2}, \dots, i_M} &= (\cm{X}^{(1)})_{i_1, i_2, \dots, i_M}.
\end{align*}

Now if we combine the first two modes and let $\cm{X}^{(2)}\in \mathbb{R}^{r_1p_2\times p_3\cdots\times p_M}$ to be the tensor such that $\vectorize(\cm{X}^{(2)}) = \vectorize(\cm{X}^{(1)})$ and apply the same reasoning, we can show \eqref{eq:action-equivalence-1}.

The intuition of the first step is that, despite the change of order of certain modes, the operator $\mathcal{C}$ changes the values of the modes that are not involved in the swap in the same way. Next, we want to show that there exist orthogonal matrices $\bm{H}_{k+1}\in\mathbb{R}^{p_{k+1}r_{k-1}\times s_k}$ and $\bm{H}_{k}\in\mathbb{R}^{s_k p_{k}\times r_{k+1}}$ such that the following holds:
\begin{equation}
	\label{eq:action-equivalence-2}
	(\bm{I}_{p_M p_{M-1}\dots p_{k+2}}\otimes \bm{H}_{k}^\top)
	(\bm{I}_{p_M p_{M-1}\dots p_{k+2} p_{k}}\otimes \bm{H}_{k+1}^\top)\bm{T}'
	= (\bm{I}_{p_M p_{M-1}\dots p_{k+2}}\otimes \bm{G}_{k+1}^\top)
	(\bm{I}_{p_M p_{M-1}\dots p_{k+1}}\otimes \bm{G}_{k}^\top),
\end{equation}
where $\bm{T}'$ is a square matrix corresponding to the operator $\mathcal{T}':=\mathcal{T}(\cdot; (k-1,k+1,k,k+2,\dots, M))$, and $s_k :=r'_{k-1}< p_{k+1}r_{k-1}$ is the surrogate of $r_k$ to be determined later.

Redefine $\cm{X}\in\mathbb{R}^{r_{k-1}\times p_k\times p_{k+1}\times p_{res}}$ as the tensor sharing the same vectorization as the output of $\mathcal{C}$, where $p_{res}=p_{k+2}\cdots p_M$, and $\cm{X}':=\mathcal{T}'\cm{X}\in\mathbb{R}^{r_{k-1}\times p_{k+1}\times p_k\times p_{res}}$. 
Denote 
\[\bm{m}:=\vectorize(\bm{G}_{k+1}^\top \text{Reshape}(\bm{G}_k^\top [\cm{X}]_2; r_kp_{k+1}, p_{res}))\in \mathbb{R}^{r_{k+1}p_{res}}\] and 
\[\bm{m}':=\vectorize(\bm{H}_{k}^\top \text{Reshape}(\bm{H}_{k+1}^\top [\cm{X}']_2; s_kp_{k}, p_{res}))\in \mathbb{R}^{r_{k+1}p_{res}}.\]
In this step, we take the $l$-th element of the vector $\bm{m}$ and $\bm{m}'$, denoted by $\bm{m}_l$ and $\bm{m}'_l$, and suppose $l=a\cdot r_{k+1}+b$, where $a=\left\lfloor \frac{l-1}{r_{k+1}}\right\rfloor $ and $b\in \{1,\dots, r_{k+1}\}$ are the quotient and positive remainder of the division of $l$ by $r_{k+1}$, respectively. It can be shown that
\begin{align}
	\bm{m}_l &= \bm{G}_{k+1, b}^\top \vectorize\left(\sum\nolimits_{i=1}^{r_{k-1}}\bm{G}_{k,i}^\top \cm{X}_{i,:,:,a}\right) = \trace\left(\widetilde{\bm{G}}_{k+1,b}\sum\nolimits_{i=1}^{r_{k-1}}\bm{G}_{k,i}^\top \cm{X}_{i,:,:,a}\right) \label{eq:action-equivalence-ml}\\ 
	\bm{m}'_l &= \bm{H}_{k, b}^\top \vectorize\left(\sum\nolimits_{i=1}^{r_{k-1}}\bm{H}_{k+1,i}^\top \cm{X}'_{i,:,:,a}\right) 
	= \trace\left(\widetilde{\bm{H}}_{k,b}\sum\nolimits_{i=1}^{r_{k-1}}\bm{H}_{k+1,i}^\top \cm{X}_{i,:,:,a}^\top\right) \notag\\
	&\hspace*{61.8mm}=\trace\left(\sum\nolimits_{i=1}^{r_{k-1}}\bm{H}_{k+1,i}\widetilde{\bm{H}}_{k,b}^\top \cm{X}_{i,:,:,a}\right) \label{eq:action-equivalence-ml2}
\end{align}
where $\widetilde{\bm{G}}_{k+1,b}\in\mathbb{R}^{p_{k+1}\times r_k}$ and $\widetilde{\bm{H}}_{k,b}\in\mathbb{R}^{p_k\times s_k}$ can be vectorized into $\bm{G}_{k+1, b}$ and $\bm{H}_{k,b}$, which stand for the $b$-th column of $\bm{G}_{k+1}$ and $\bm{H}_{k}$, respectively; and $\bm{G}_{k,i}\in\mathbb{R}^{p_k\times r_k}$ and $\bm{H}_{k+1,i}\in\mathbb{R}^{p_{k+1}\times s_k}$ for $i\in\{1,2,\dots, r_{k-1}\}$ are formed by the rows of $\bm{G}_k$ and $\bm{G}_{k+1}$ such that \eqref{eq:action-equivalence-ml} and \eqref{eq:action-equivalence-ml2} hold. We claim that 
\begin{equation}
	\label{eq:column-orthogonality}
\langle\widetilde{\bm{G}}_{k+1,b}, \widetilde{\bm{G}}_{k+1,c}\rangle = \langle \widetilde{\bm{H}}_{k,b}, \widetilde{\bm{H}}_{k,c}\rangle = \delta_{bc} = \begin{cases}
1, & b=c\\
0, & b\neq c
\end{cases}
\end{equation}
holds for all $b,c\in \{1,\dots, r_{k+1}\}$, which is equivalent to the orthogonality of $\bm{G}_{k+1}$ and $\bm{H}_{k}$. The former part is assumed, and we later prove the latter. Thus, we need to show that to any orthogonal matrices $\bm{G}_{k+1}$ and $\bm{G}_k$ there correspond an orthogonal matrix $\bm{H}_{k+1}$ and a set of matrices $\widetilde{\bm{H}}_{k,b}$ satisfying \eqref{eq:column-orthogonality} such that $\bm{m}_l=\bm{m}'_l$, i.e.,
\begin{equation}
	\label{eq:action-equivalence-3}
	\widetilde{\bm{G}}_{k+1,b}\bm{G}_{k,i}^\top = \bm{H}_{k+1,i}\widetilde{\bm{H}}_{k,b}^\top \in \mathbb{R}^{p_{k+1}\times p_k}
\end{equation}
holds for all $(b,i)\in \{1,\dots, r_{k+1}\}\times \{1,\dots, r_{k-1}\}$, where $\bm{G}_{k,i}\in\mathbb{R}^{p_k\times r_k}$ and $\bm{H}_{k+1,i}\in\mathbb{R}^{p_{k+1}\times s_k}$ are defined as above.

Let $\bm{B}_b=\big[\bm{G}_{k,1}\widetilde{\bm{G}}_{k+1,b}^\top; \dots; \bm{G}_{k,r_{k-1}}\widetilde{\bm{G}}_{k+1,b}^\top\big]^\top \in \mathbb{R}^{p_{k+1}r_{k-1}\times p_k}$ be the LHS of \eqref{eq:action-equivalence-3} stacked up, and set $V=\text{span}\left\{\text{col}(\bm{B}_b): b=1,\dots, r_{k+1} \right\}\subset \mathbb{R}^{p_{k+1}r_{k-1}}$. Set
\begin{equation}
\label{eq:action-equivalence-interim-rank}
s_k = \dim V.
\end{equation}

By the construction of $\bm{B}_b$ we know that 
\begin{equation}
\bm{B}_b=\left[\begin{array}{c}
\widetilde{\bm{G}}_{k+1, b} \bm{G}_{k,1}^\top \\
\widetilde{\bm{G}}_{k+1, b} \bm{G}_{k,2}^\top \\
\vdots \\
\widetilde{\bm{G}}_{k+1, b} \bm{G}_{k,r_{k-1}}^\top
\end{array}\right]=\left(\bm{I}_{r_{k-1}} \otimes \widetilde{\bm{G}}_{k+1, b}\right)\left[\begin{array}{c}
\bm{G}_{k,1}^\top \\
\bm{G}_{k,2}^\top \\
\vdots \\
\bm{G}_{k, r_{k-1}}^\top
\end{array}\right] \defeq (\bm{I}\otimes \widetilde{\bm{G}}_{k+1, b}) \bm{B}_\natural,
\end{equation}
where $\bm{B}_\natural \in \mathbb{R}^{r_kr_{k-1}\times p_k}$ is not guaranteed to be orthogonal.
Therefore, $\rank(\bm{B}_b)\leq r_kr_{k-1}$ and $s_k\leq r_{k-1}\min \{r_kr_{k+1},p_{k+1}\}$.

Construct the matrix $\bm{H}_{k+1}=\left[\bm{H}_{k+1,1}^\top;\dots;\bm{H}_{k+1,r_{k-1}}^\top\right]^\top \in \mathbb{O}^{p_{k+1} r_{k-1}\times s_k}$ so that $\bm{H}_{k+1}\bm{H}_{k+1}^\top=\mathcal{P}_V$, i.e., columns of $\bm{H}_{k+1}$ constitute an orthonormal basis of $V$, by, for example, Gram-Schmidt orthogonalization, and $\bm{H}_{k+1,i}\in\mathbb{R}^{p_{k+1}\times s_k}$ by chopping $\bm{H}_{k+1}$ accordingly.

Consequently, set
\begin{equation}\label{eq:action-equivalence-Gkb}
\widetilde{\bm{H}}_{k,b} = \bm{B}_b^\top \bm{H}_{k+1} \in \mathbb{R}^{p_k\times s_k}.
\end{equation}
It remains to verify the orthogonality condition \eqref{eq:column-orthogonality} and the central equality \eqref{eq:action-equivalence-3} before we can conclude the proof by setting $\bm{G}_m'=\bm{G}_m$ for $m=k+2,\dots,M$ so that $\bm{A}'=\bm{A}$ in \eqref{eq:action-equivalence-main}.

Note that $\bm{H}_{k+1} \widetilde{\bm{H}}_{k,b}^\top = \bm{H}_{k+1}\bm{H}_{k+1}^\top \bm{B}_b = \bm{B}_b$, and taking the $i$-th block (component) out from both sides, we have
\begin{equation}
	\bm{H}_{k+1,i} \widetilde{\bm{H}}_{k,b}^\top = \left[\bm{B}_b\right]_{\text{block i}} = \widetilde{\bm{G}}_{k+1,b} \bm{G}_{k,i}^\top,
\end{equation}
whence \eqref{eq:action-equivalence-3} holds. Similarly, $\trace(\widetilde{\bm{H}}_{k,b} \widetilde{\bm{H}}_{k,c}^\top) = \trace(\bm{B}_b^\top \bm{H}_{k+1} \bm{H}_{k+1}^\top \bm{B}_c) = \trace(\bm{B}_b^\top \bm{B}_c)$, but 
\begin{align*}
	\trace(\bm{B}_b^\top \bm{B}_c) &= \trace\left(\left(\bm{I}_{r_{k-1}}\otimes \widetilde{\bm{G}}_{k+1,b}\right)\bm{B}_\natural \bm{B}_\natural^\top \left(\bm{I}_{r_{k-1}}\otimes \widetilde{\bm{G}}_{k+1,c}\right)^\top \right) \\
	&=\trace\left((\bm{I}_{r_{k-1}}\otimes \widetilde{\bm{G}}_{k+1,b}^\top \widetilde{\bm{G}}_{k+1,c})\bm{B}_\natural \bm{B}_\natural^\top\right) \\
	&=\sum_{i=1}^{r_{k-1}}\trace\left(\widetilde{\bm{G}}_{k+1,b}^\top \widetilde{\bm{G}}_{k+1,c} \bm{G}_{k,i}^\top \bm{G}_{k,i}\right) \\
	&=\trace\left(\widetilde{\bm{G}}_{k+1,b}^\top \widetilde{\bm{G}}_{k+1,c}\sum_i \bm{G}_{k,i}^\top \bm{G}_{k,i}\right) \\
	&=\trace\left(\widetilde{\bm{G}}_{k+1,b}^\top \widetilde{\bm{G}}_{k+1,c}\bm{G}_k^\top \bm{G}_k\right) \\
	&=\delta_{bc},
\end{align*}
which confirms the orthogonality condition \eqref{eq:column-orthogonality}.
\end{proof}

\subsection{ALS update rules} \label{subsec:als-rule}

Recall that the model is given by
\begin{equation*}
	\bm{y}_t = \bm{\Lambda}_y\bm{\Theta}(\bm{I}_L\otimes \bm{\Lambda}_x^\top) \bm{x}_t + \bm{e}_t,
\end{equation*}
and the update rule comprises two parts: updating the component matrices sequentially by converting to a regression problem (minimization), and imposing the orthogonality constraints on the updated matrices (orthogonalization).

When updating $\bm{G}_{m,k}$ in $\bm{\Lambda}_x$, denote $\bm{y}_{t,k}=\bm{y}_t-\sum_{l=1}^{L}\bm{\Lambda}_y \sum_{\gamma\neq k}^{K_x}\bm{\Theta}_{l,\gamma} \bm{\Lambda}_{x,\gamma}^\top$ as the response variable with the contributions of other action orders deducted, where $\bm{\Lambda}_{x,\gamma}$ is the $\gamma$'s block column of $\bm{\Lambda}_x$ and $\bm{\Theta}_{l,\gamma}$ is the $\gamma$'s block column of $\bm{\Theta}_l$ corresponding to the action order $k$. The update for $\bm{G}_{m,k}$ is then converted to the following OLS problem of linear regression:
\begin{equation*}
	\bm{y}_{t,k}=\bigg[\sum_{l=1}^{L} \bm{B}_{l,m,k}(\bm{Y}_{t-l,m,k}^\top\otimes \bm{I}_{r_m})\bigg]\vectorize(\bm{G}_{m,k}^\top)+ \bm{e}_t,
\end{equation*}
where $\bm{B}_{l,m,k} = \bm{\Lambda}_y \bm{\Theta}_{l,k}\bm{G}_{M,k}^\top (\bm{I}_{p_{\alpha^k_{(M)}}}\otimes\bm{G}_{M-1,k}^\top) \cdots (\bm{I}_{p_{\alpha^k_{(m+2)}}\cdots p_{\alpha^k_{(M)}}}\otimes\bm{G}_{m+1,k}^\top)$ accounts for the matrices before $\bm{G}_{m,k}$, and $\bm{Y}_{t-l,m,k}=\text{Reshape}((\bm{I}_{p_{\alpha^k_{(m)}}\cdots p_{\alpha^k_{(M)}}}\otimes \bm{G}_{m-1,k}^\top)\cdots (\bm{I}_{p_{\alpha^k_{(2)}}\cdots p_{\alpha^k_{(M)}}}\otimes \bm{G}_{1,k}^\top)\bm{T}_x(\alpha_k)\bm{y}_{t-l}; r_{m-1}p_{\alpha^k_{(m)}}, p^{(m)})$, and $p^{(m)}:= p_{\alpha^k_{(m+1)}}\cdots p_{\alpha^k_{(M)}}$.

To update $\bm{H}_{n,k^\prime}$ in $\bm{\Lambda}_y$, we define $\bm{g}_{t}=\bm{\Theta}(\bm{I}_L\otimes \bm{\Lambda}_x^\top) \bm{x}_t$, and thus the model becomes
\begin{equation*}
 \bm{y}_t = \bm{\Lambda}_y \bm{g}_t + \bm{e}_t =: (\cm{H}_1,\ldots,\cm{H}_N) (\vectorize(\bm{g}_{t,1})^\top, \ldots, \vectorize(\bm{g}_{t,N})^\top)^\top + \bm{e}_t,
 \end{equation*}
where $\cm{H}_{k^\prime} = \bm{T}^\top_y(\beta_{k^\prime})(\bm{I}_{q_{\beta^{k'}_{(2)}}\cdots q_{\beta^{k'}_{(N)}}}\otimes\bm{H}_{1,k'})\cdots \bm{H}_{N,k'}=:\bm{H}_{n,k'}^\mathrm{fore}(\bm{I}_{q_{\beta^{k'}_{(n+1)}}\cdots q_{\beta^{k'}_{(N)}}}\otimes\bm{H}_{n,k'}) \bm{H}_{n,k'}^\mathrm{aft}$ for $1\leq n\leq N$. Similarly, we can define $\bm{y}_{t,k^\prime}=\bm{y}_t-\sum_{l=1}^{L}\sum_{\gamma\neq k^\prime}^{K_y} \bm{\Lambda}_{y,\gamma} \bm{\Theta}_{l,\gamma} \bm{\Lambda}_x^\top \bm{y}_{t-l}$ as the response variable faced by the action order $k^\prime$ via the coordinate descent, where $\bm{\Theta}_{l,\gamma}$ is the $\gamma$'s block row of $\bm{\Theta}_l$, and the update for $\bm{H}_{n,k^\prime}$ is converted to the following OLS problem of linear regression:
\begin{equation*}
\bm{y}_{t,k^\prime} = \bm{H}_{n,k^\prime}^\mathrm{fore}(\bm{Y}_{t,n,k^\prime}^\top\otimes \bm{I}_{q_{\beta^{k^\prime}_{(n)}}s_{n-1,k'}})\vectorize(\bm{H}_{n,k^\prime})+ \bm{e}_t,
\end{equation*}
where $\bm{Y}_{t,n,k^\prime}=\text{Reshape}(\bm{H}_{n,k'}^\mathrm{aft}\bm{g}_{t,k^\prime}; s_{n,k'}, q_{\beta^{k^\prime}_{(n+1)}}\cdots q_{\beta^{k^\prime}_{(N)}})$.

For the coefficient matrix $\bm{\Theta}$, it is a bilinear regression problem equivalent to solving
\begin{equation*}
	\bm{y}_t = \big[(\bm{x}_t^\top (\bm{I}_L\otimes \bm{\Lambda}_x))\otimes \bm{\Lambda}_y\big]\vectorize(\bm{\Theta}) + \bm{e}_t.
\end{equation*}
Each of these OLS problems can then be solved efficiently with closed forms.

To satisfy the orthogonality assumptions required by the model, an additional orthogonalization step is therefore necessary.
Specifically, for any set of invertible matrices $\bm{O}_{m,k}\in\mathbb{R}^{r_{m,k}\times r_{m,k}}$, where $1 \leq m \leq M$, the following identity holds with proof provided in Section \ref{appendix:proof-orth-eq}:
\begin{equation}
	\label{alg:orth}
	\begin{split}
		\bm{G}_{M,k}^\top (\bm{I}_{p_{\alpha^k_{(M)}}}\otimes\bm{G}_{M-1,k}^\top)& \cdots (\bm{I}_{p_{\alpha^k_{(2)}}\cdots p_{\alpha^k_{(M)}}}\otimes\bm{G}_{1,k}^\top) =\\
		&(\bm{O}_{M,k}^\top)^{-1} \widetilde{\bm{G}}_{M,k}^\top (\bm{I}_{p_{\alpha^k_{(M)}}}\otimes\widetilde{\bm{G}}_{M-1,k}^\top) \cdots (\bm{I}_{p_{\alpha^k_{(2)}}\cdots p_{\alpha^k_{(M)}}}\otimes\widetilde{\bm{G}}_{1,k}^\top)
	\end{split}	
\end{equation}
where the transformed components $\widetilde{\bm{G}}_{m,k}$ are defined recursively as $\widetilde{\bm{G}}_{1,k} = \bm{G}_{1,k}\bm{O}_{1,k}$, $\widetilde{\bm{G}}_{2,k} = (\bm{I}_{p_{\alpha^k_{(2)}}}\otimes\bm{O}_{1,k}^{-1})\bm{G}_{2,k}\bm{O}_{2,k}$, $\ldots$, $\widetilde{\bm{G}}_{M,k} = (\bm{I}_{p_{\alpha^k_{(M)}}}\otimes\bm{O}_{M-1,k}^{-1})\bm{G}_{M,k}\bm{O}_{M,k}$.
By appropriately selecting the matrices $\bm{O}_{m,k}$ for $1 \leq m \leq M$, we can enforce orthogonality across all components $\{ \bm{G}_{m,k} \}_{m=1}^M$. We apply this orthogonalization procedure to all block columns of $\bm{\Lambda}_x$ and $\bm{\Lambda}_y$, and denote the whole process by the operator $\mathrm{SSVD}(\cdot)$, detailed in Section \ref{appendix:alg-SSVD}, which is applied after the updates of all $\{ \{ \bm{H}_{n,k^\prime} \}_{n=1}^N \}_{k^\prime=1}^{K_y}$ and $\{ \{ \bm{G}_{m,k} \}_{m=1}^M \}_{k=1}^{K_x}$ have been completed.

\subsection{Proof of \eqref{alg:orth}}
\label{appendix:proof-orth-eq}
We begin by verifying the identity for the last two terms on the left-hand side of \eqref{alg:orth}. Let $\bm{O}_{1,k}\in\mathbb{R}^{r_{1,k}\times r_{1,k}}$ be an arbitrary invertible matrix. Consider the following product:
\begin{align*}
	&(\bm{I}_{p_{\alpha^k_{(3)}}\cdots p_{\alpha^k_{(M)}}}\otimes\bm{G}_{2,k}^\top)(\bm{I}_{p_{\alpha^k_{(2)}}p_{\alpha^k_{(3)}}\cdots p_{\alpha^k_{(M)}}}\otimes\bm{G}_{1,k}^\top)\\
	=&~\bm{I}_{p_{\alpha^k_{(3)}}\cdots p_{\alpha^k_{(M)}}}\otimes[\bm{G}_{2,k}^\top (\bm{I}_{p_{\alpha^k_{(2)}}}\otimes\bm{G}_{1,k}^\top)]\\
	=&~\bm{I}_{p_{\alpha^k_{(3)}}\cdots p_{\alpha^k_{(M)}}}\otimes[\bm{G}_{2,k}^\top (\bm{I}_{p_{\alpha^k_{(2)}}}\otimes((\bm{O}_{1,k}^{-1})^\top\bm{O}_{1,k}^\top\bm{G}_{1,k}^\top))]\\
	=&~\bm{I}_{p_{\alpha^k_{(3)}}\cdots p_{\alpha^k_{(M)}}}\otimes[\bm{G}_{2,k}^\top(\bm{I}_{p_{\alpha^k_{(2)}}}\otimes\bm{O}_{1,k}^{-1})^\top (\bm{I}_{p_{\alpha^k_{(2)}}}\otimes(\bm{O}_{1,k}^\top\bm{G}_{1,k}^\top))]\\
	=&~(\bm{I}_{p_{\alpha^k_{(3)}}\cdots p_{\alpha^k_{(M)}}}\otimes[(\bm{I}_{p_{\alpha^k_{(2)}}}\otimes\bm{O}_{1,k}^{-1})\bm{G}_{2,k}]^\top) (\bm{I}_{p_{\alpha^k_{(2)}}p_{\alpha^k_{(3)}}\cdots p_{\alpha^k_{(M)}}}\otimes(\bm{G}_{1,k}\bm{O}_{1,k})^\top).
\end{align*}
This computation shows that the matrix $\bm{O}_{1,k}$ can be absorbed into $\bm{G}_{1,k}$ and $\bm{G}_{2,k}$ through the following change of variables:
\begin{equation}
	\label{appendix:alg1-g2}
	\widetilde{\bm{G}}_{1,k} = \bm{G}_{1,k} \bm{O}_{1,k}, \qquad 
	\bar{\bm{G}}_{2,k} = (\bm{I}_{p_{\alpha^k_{(2)}}} \otimes \bm{O}_{1,k}^{-1}) \bm{G}_{2,k},
\end{equation}
where $\bar{\bm{G}}_{2,k}$ is an intermediate form of the second component. In the subsequent steps, we apply a similar change of variables recursively:
\begin{equation}
	\label{appendix:alg1-gm}
	\widetilde{\bm{G}}_{m,k} = \bar{\bm{G}}_{m,k} \bm{O}_{m,k}, \qquad 
	\bar{\bm{G}}_{m+1,k} = (\bm{I}_{p_{\alpha^k_{(m+1)}}} \otimes \bm{O}_{m,k}^{-1}) \bm{G}_{m+1,k},
\end{equation}
for $2\leq m\leq M-1$, and finally define $\widetilde{\bm{G}}_{M,k} = \bar{\bm{G}}_{M,k} \bm{O}_{M,k}$. 
By successively applying these transformations, each $\bm{G}_{m,k}$ is mapped to $\widetilde{\bm{G}}_{m,k}$ through a sequence of invertible matrix multiplications involving Kronecker products with identity matrices. This construction guarantees that the full product on the left-hand side of \eqref{alg:orth} is equal to the transformed product on the right-hand side, up to a scaling by the inverse transpose of the final matrix $\bm{O}_{M,k}^\top$. This completes the proof.

\subsection{Algorithm for SSVD}
\label{appendix:alg-SSVD}
In this subsection, we propose an algorithm to orthogonalize all the components in $\bm{\Lambda}_x$ and $\bm{\Lambda}_y$ required in Section~\ref{subsec:algorithms} of the manuscript. We apply \eqref{alg:orth} to orthogonalize the components $\{ \{ \bm{G}_{m,k} \}_{m=1}^M \}_{k=1}^{K_x}$, and the specific update rule of each orthogonal component follows \eqref{appendix:alg1-g2} or \eqref{appendix:alg1-gm}. The matrices concerned, i.e., $\widetilde{\bm{G}}_{m,k}$'s and $\bm{O}_{m,k}$'s, are obtained via SVD.
After obtaining the orthogonalized components, we turn to the product $\bm{\Theta}_l \bm{\Lambda}_x^\top$ ($l=1,2,\ldots,L$), which can be written as
\begin{align} \label{eq:appendix-orth-decomp-theta}
	\begin{bmatrix}
		\bm{\Theta}_{l,1}  & \cdots & \bm{\Theta}_{l,K_x} 
	\end{bmatrix}
	\begin{bmatrix}
		(\bm{O}_{M,1}^\top)^{-1}\bm{\Lambda}_{x,1}^\top\\  \vdots \\ (\bm{O}_{M,K_x}^\top)^{-1}\bm{\Lambda}_{x,K_x}^\top
	\end{bmatrix} = 
	\begin{bmatrix}
		\bm{\Theta}_{l,1}(\bm{O}_{M,1}^\top)^{-1} & \cdots & \bm{\Theta}_{l,K_x} (\bm{O}_{M,K_x}^\top)^{-1}
	\end{bmatrix}
	\begin{bmatrix}
		\bm{\Lambda}_{x,1}^\top\\ \vdots \\ \bm{\Lambda}_{x,K_x}^\top
	\end{bmatrix}.
\end{align}
This shows that the effect of the scaling matrices $\{ (\bm{O}_{M,k}^\top)^{-1} \}_{k=1}^{K_x}$ can be absorbed into the coefficient matrices $\bm{\Theta}_{l,k}$ for each lag $l$ and action $k$. Therefore, we can enforce orthogonality on all components $\{ \bm{G}_{m,k} \}_{m=1}^M$ of $\bm{\Lambda}_x$ without affecting the model structure.
A similar procedure can be applied to the components $\{ \{ \bm{H}_{n,k^\prime} \}_{n=1}^N \}_{k^\prime=1}^{K_y}$ of $\bm{\Lambda}_y$ to achieve orthogonality, and by viewing the updated matrix $\bm{\Theta}$ as a whole and transposing $\bm{\Lambda}_y \bm{\Theta}$, \eqref{eq:appendix-orth-decomp-theta} can be used analogously to absorb the scaling matrices $\{ (\bm{O}_{N,k^\prime}^\top)^{-1} \}_{k^\prime=1}^{K_y}$ into $\bm{\Theta}$.
We summarize this sequential orthogonalization process into the following algorithm, denoted by $\text{SSVD}(\cdot)$.

\begin{algorithm}[H]
	\caption{SSVD procedure for orthogonalizing core components}
	\label{alg:ssvd}
	\begin{spacing}{1.667}
	\begin{algorithmic}[1]
		\STATE \textbf{Input:} $\{\{\bm{G}_{m,k}\}_{m=1}^M\}_{k=1}^{K_x}, \{\{\bm{H}_{n,k'}\}_{n=1}^N\}_{k'=1}^{K_y}$ and $\bm{\Theta}=(\bm{\Theta}_1,\dots,\bm{\Theta}_L)$.
		\FOR{$k = 1$ to $K_x$}
		\FOR{$m = 1$ to $M$}\label{line:ssvd-orth-onset}
		\STATE Set $\bar{\bm{G}}_{m,k} = (\bm{I}_{q_{\alpha^k_{(m)}}}\otimes\bm{O}_{G,m-1,k}^{-1})\bm{G}_{m,k}$ \textbf{if} $m>1$; \textbf{else} $\bar{\bm{G}}_{m,k} = \bm{G}_{m,k}$.
		\STATE Conduct SVD on $\bar{\bm{G}}_{m,k}$ as $\bar{\bm{G}}_{m,k} = \bm{U}_{G,m,k}\bm{\Sigma}_{G,m,k}\bm{V}^\top_{G,m,k}$.
		\STATE Set $\widetilde{\bm{G}}_{m,k} = \bm{U}_{G,m,k}$ and $\bm{O}_{G,m,k} = (\bm{\Sigma}_{G,m,k}\bm{V}^\top_{G,m,k})^{-1}$. \label{line:setO-G}
		\ENDFOR \label{line:ssvd-orth-end}
		\STATE Update $\bm{\Theta}^{(1)}_{l,k} = \bm{\Theta}_{l,k}(\bm{O}_{G,M,k}^\top)^{-1}$ for each $l\in \{1,2,\dots,L\}$ with the form of \eqref{eq:appendix-orth-decomp-theta}. \label{line:upd-theta-x}
		\ENDFOR
		\FOR{$k' = 1$ to $K_y$}
		\STATE Repeat lines \ref{line:ssvd-orth-onset}--\ref{line:ssvd-orth-end} for $\{ \{ \bm{H}_{n,k'} \}_{n=1}^N \}_{k'=1}^{K_y}$ to get $\{ \{ \widetilde{\bm{H}}_{n,k'}\}_{n=1}^N \}_{k'=1}^{K_y}$ and $\bm{O}_{H,N,k'}$.
		\STATE Update $[(\bm{\Theta}^{(2)})^\top]_{k'} = [(\bm{\Theta}^{(1)})^\top]_{k'}(\bm{O}_{H,N,k'}^\top)^{-1}$ similar to line \ref{line:upd-theta-x}.
		\ENDFOR
		\STATE \textbf{Return} $\{ \{ \widetilde{\bm{G}}_{m,k}\}_{m=1}^M \}_{k=1}^{K_x}$, $\{ \{ \widetilde{\bm{H}}_{n,k'}\}_{n=1}^N \}_{k'=1}^{K_y}$, and $\bm{\Theta}^{(2)}$.
	\end{algorithmic}
	\end{spacing}
\end{algorithm}
\begin{remark}\label{remark:rank-deficient-pad}
	In the unlikely event that any $\bar{\bm{G}}_{m,k}$ or $\bar{\bm{H}}_{n,k'}$ is rank-deficient (take $\bar{\bm{G}}_{m,k}$ for example and omit $k$ for convenience), we pad zero columns to the corresponding $\bm{U}_{m}$ to ensure that the output component $\widetilde{\bm{G}}_{m}$ has the same dimensions as the input components $\bm{G}_{m}$, and we add rows to $\bm{O}_{G,m}$ to make it square and invertible. Line \ref{line:setO-G} can be so modified if necessary, and the successive procedure of Section \ref{appendix:proof-orth-eq} still applies. To remove the zero columns in $\widetilde{\bm{G}}_{m}$, we notice that $(\bm{I}_{p_{\alpha_{(m+2)}}\cdots p_{\alpha_{(M)}}}\otimes\widetilde{\bm{G}}_{m+1}^\top)(\bm{I}_{p_{\alpha_{(m+1)}}p_{\alpha_{(m+2)}}\cdots p_{\alpha_{(M)}}}\otimes\widetilde{\bm{G}}_{m}^\top)=\bm{I}_{p_{\alpha_{(m+2)}}\cdots p_{\alpha_{(M)}}}\otimes[\widetilde{\bm{G}}_{m+1}^\top (\bm{I}_{p_{\alpha_{(m+1)}}}\otimes\widetilde{\bm{G}}_{m}^\top)]$, and we remove the zero rows in $\bm{I}_{p_{\alpha_{(m+1)}}}\otimes\widetilde{\bm{G}}_{m}^\top$ and the corresponding columns in $\widetilde{\bm{G}}_{m+1}^\top$. Finally, we re-run Algorithm \ref{alg:ssvd} to restore the orthogonality of all components. 
\end{remark}

\subsection{Proof of Proposition \ref{prop:additive}}
\label{appendix:proof-additive}
We aim to show that there exist $\bm{r}_{k,3} = \bm{r}_{k,1}+\bm{r}_{k,2}$ (in terms of elementwise ordering) and a matrix $\bm{O}$ such that $\widetilde{\bm{\Lambda}}_x(\alpha_k, \bm{r}_{k,3}) = \bm{O}\left[\widetilde{\bm{\Lambda}}_x(\alpha_k, \bm{r}_{k,1}), \widetilde{\bm{\Lambda}}_x(\alpha_k, \bm{r}_{k,2})\right]$, that is, absent orthogonality constraints, identical action orders can be merged with additive $\bm{r}_k$'s.

The proof is constructive. We assume without loss of generality that $\alpha_k = (1,2,\cdots, M)$ and omit the subscript $k$. Let $\bm{G}_{m,1}\in\mathbb{O}^{p_m r_{1,m-1}\times r_{1,m}}$ and $\bm{G}_{m,2}\in\mathbb{O}^{p_m r_{2,m-1}\times r_{2,m}}$ be the respective core matrices corresponding to the reversed construction of $\widetilde{\bm{\Lambda}}_x(\alpha, \bm{r}_1)$ and $\widetilde{\bm{\Lambda}}_x(\alpha, \bm{r}_2)$, i.e., the left hand side of \eqref{eq:TT-ours-proof}. We construct the new core matrices $\bm{G}_{m}^\top = \left[\begin{smallmatrix}
\bm{G}_{m,1}^\top &  \\
 & \bm{G}_{m,2}^\top \\
\end{smallmatrix}\right]\in\mathbb{R}^{(r_{1,m} + r_{2,m}) \times p_m (r_{1,m-1} + r_{2,m-1})}$ for $m=2,\ldots,M$, and $\bm{G}_{1}^\top = \left[\begin{smallmatrix}
\bm{G}_{1,1}^\top  \\
\bm{G}_{1,2}^\top \\
\end{smallmatrix}\right]\in\mathbb{R}^{(r_{1,1} + r_{2,1}) \times p_1}$, and the preliminary row block
\begin{align*}
	\widetilde{\bm{\Lambda}}_x^\top(\alpha, \bm{r}_1+\bm{r}_2)=\begin{bmatrix}
		\bm{G}_{M,1}^\top &  \\
		& \bm{G}_{M,2}^\top \\
	\end{bmatrix} &\left(\begin{bmatrix}
		\bm{G}_{M-1,1}^\top &  \\
		& \bm{G}_{M-1,2}^\top \\
	\end{bmatrix} \otimes \bm{I}_{p_M} \right)\cdots\\
	&\left(\begin{bmatrix}
		\bm{G}_{2,1}^\top &  \\
		& \bm{G}_{2,2}^\top \\
	\end{bmatrix}\otimes \bm{I}_{p_3\cdots p_M} \right)	\left(\begin{bmatrix}
		\bm{G}_{1,1}^\top  \\
		\bm{G}_{1,2}^\top \\
	\end{bmatrix}\otimes \bm{I}_{p_2\cdots p_M}\right)\bm{T}(\alpha).
\end{align*}

The equivalence can be verified by expanding the above expression by block matrix multiplication rules. Then, we convert the reversed form back to the standard form, i.e., the right hand side of \eqref{eq:TT-ours-proof}, according to Section \ref{appendix:proof-TT}.
The matrix $\bm{O}$ shall be chosen according to Section \ref{appendix:alg-SSVD} to ensure orthogonality of the new core matrices.

\subsection{Proof of Remark \ref{remark:TT}}
\label{appendix:proof-TT}
Remark \ref{remark:TT} states that the proposed model generalizes the Tensor Train (TT) decomposition when $K_y = K_x = 1$ with other appropriate specifications. Here, we provide a detailed proof of this claim by demonstrating how the TT decomposition can be represented within our model framework.
We begin by recalling the representation of TT decomposition. For a $(N+M)$-th order tensor $\cm{A}\in \mathbb{R}^{q_1\times\cdots\times q_N\times p_1\times\cdots\times p_M}$, its entry can be expressed in the TT format as
\begin{equation*}
	\label{TT-format}
	\cm{A}_{j_1\ldots j_Ni_1\ldots i_M}=\bm{H}_1(j_1)\cdots\bm{H}_N(j_N)\bm{G}_1(i_1)\cdots\bm{G}_M(i_M),
\end{equation*}
where $\bm{H}_n(j_n)$ is an $s_{n-1}\times s_{n}$ matrix for $1\leq n\leq N$, $\bm{G}_m(i_m)$ is an $r_m\times r_{m+1}$ matrix for $1\leq m\leq M$, $s_0=r_{M+1}=1$ and $s_N=r_1$.
To obtain the TT cores, we stack the matrices $\bm{H}_n(j_n)$ into a third-order tensor $\cm{H}_n\in\mathbb{R}^{s_{n-1}\times q_n\times s_n}$ for $2\leq n\leq N$, and similarly stack $\bm{G}_m(i_m)$ into $\cm{G}_m\in\mathbb{R}^{r_m\times p_m\times r_{m+1}}$ for $1\leq m\leq M-1$. Moreover, let $\bm{H}_1 = (\bm{H}_1^\top(1), \ldots, \bm{H}_1^\top(q_1))^\top\in\mathbb{R}^{q_1\times s_1}$ and $\bm{G}_M = (\bm{G}_M(1), \ldots, \bm{G}_M(p_M))^\top\in\mathbb{R}^{p_M\times r_M}$.

According to Lemma 3.1 in \cite{zhou2022optimal}, Proposition 1 in \cite{si2024efficient}, and \cite{qin2025computational}, the $N$-th sequential matricization of $\cm{A}$ with TT ranks $(s_1, \ldots, s_N, r_2, \ldots, r_M)$, admits the following decomposition:
\begin{equation*}
	\label{eq:TTdecomposition}
	[\cm{A}]_N=(\bm{I}_{q_2\cdots q_N}\otimes\bm{H}_1)\cdots(\bm{I}_{q_N}\otimes[\cm{H}_{N-1}]_2)[\cm{H}_N]_2 \bm{\Theta}_{\mathrm{TT}}[\cm{G}_1]_1 ([\cm{G}_2]_1\otimes\bm{I}_{p_1})\cdots(\bm{G}_{M}^\top\otimes\bm{I}_{p_1\cdots p_{M-1}}),
\end{equation*}
where $\bm{H}_1\in\mathbb{O}^{q_1\times s_1}$, $[\cm{H}_n]_2\in\mathbb{O}^{s_{n-1}q_n\times s_n}$ for $2\leq n \leq N$, $[\cm{G}_m]_1^\top\in\mathbb{O}^{r_{m+1}p_m\times r_m}$ for $1\leq m\leq M-1$, $\bm{G}_M\in\mathbb{O}^{p_M\times r_M}$, and $\bm{\Theta}_{\mathrm{TT}}\in\mathbb{R}^{s_N\times r_1}$ is a diagonal weight matrix.

To connect this decomposition with our model, consider the case where $K_y = K_x = 1$, and $\mathcal{P}_y(N, K_y)=\{(1,2,\ldots, N)\}$, $\mathcal{P}_x(M, K_x) = \{ (M,M-1,\ldots, 1) \}$. We claim that
\begin{equation}
	\label{eq:TT-ours}
	\begin{split}
			(\bm{I}_{q_2\cdots q_N}\otimes\bm{H}_1)\cdots(\bm{I}_{q_N}\otimes[\cm{H}_{N-1}]_2)[\cm{H}_N]_2&\in \mathcal{R}_y(\{(s_1,\ldots, s_N)\}, \mathcal{P}_y(N, K_y)),\\
			\left([\cm{G}_1]_1 ([\cm{G}_2]_1\otimes\bm{I}_{p_1})\cdots(\bm{G}_{M}^\top\otimes\bm{I}_{p_1\cdots p_{M-1}})\right)^\top &\in\mathcal{R}_x(\{(r_M,\ldots, r_1)\}, \mathcal{P}_x(M, K_x)).
		\end{split}
\end{equation}
Consequently, $\cm{A}\in \mathcal{S}_{\mathrm{LR}}(\{(r_M,\ldots, r_1)\}, \{(s_1,\ldots, s_N)\}, \mathcal{P}_x(M, K_x), \mathcal{P}_y(N, K_y))$, which corresponds to a special case of our proposed regression model, where $\bm{\Theta}_{\mathrm{TT}}$ is diagonal. 

It remains to verify \eqref{eq:TT-ours}. We only need to show $\left([\cm{G}_1]_1 ([\cm{G}_2]_1\otimes\bm{I}_{p_1})\cdots (\bm{G}_{M}^\top\otimes\bm{I}_{p_1\cdots p_{M-1}})\right)^\top \in\mathcal{R}_x(\{(r_M,\ldots, r_1)\}, \mathcal{P}_x(M, K_x))$, and by changing the notations and indexing to conform with our settings, the target of proof becomes that,
for all $\{\bm{G}_m\in\mathbb{O}^{p_m r_{m-1}\times r_m}\}_{m=1}^M$ ($r_0\defeq1$), there exist $\{\bm{H}_m\in\mathbb{O}^{p_m r_{m-1}\times r_m}\}_{m=1}^M$ and a permutation matrix $\bm{T}_M$ such that for any $\bm{x}_M\in\mathbb{R}^{p_1 p_2 \cdots p_M}$,
\begin{align}\label{eq:TT-ours-proof}
\begin{split}
	& \bm{G}_M^\top (\bm{G}_{M-1}^\top \otimes\bm{I}_{p_M}) \cdots (\bm{G}_2^\top \otimes\bm{I}_{p_3\cdots p_M})(\bm{G}_1^\top \otimes\bm{I}_{p_2\cdots p_M})\bm{x}_M \\
	=~&\bm{H}_M^\top (\bm{I}_{p_M}\otimes\bm{H}_{M-1}^\top) \cdots (\bm{I}_{p_3\cdots p_M}\otimes\bm{H}_2^\top)(\bm{I}_{p_2\cdots p_M}\otimes\bm{H}_1^\top)\bm{T}_M\bm{x}_M.
\end{split}
\end{align}
We make the following construction to restore the original algebraic structure of our model.
For $m=1, 2,\ldots,M$, let $\cm{H}_m\in\mathbb{R}^{r_{m-1} \times p_m \times r_{m}}$ be the tensor such that if $\cm{G}_m\in\mathbb{R}^{r_{m}\times p_m \times r_{m-1}}$ satisfies $[\cm{G}_m]_1 = \bm{G}_m^\top$, then $\cm{G}_m(i,j,k)=\cm{H}_m(k,j,i)$ holds for all $i\leq r_{m}, j\leq p_m, k\leq r_{m-1}$, and we construct $\bm{H}_m = [\cm{H}_m]_2$. Here we use the tuple inside the parentheses to locate the element of a matrix or a tensor.
 Equivalently,
\begin{equation} \label{eq:gh-relationship}
	\bm{H}_m^\top(i,(j-1)r_{m-1}+k)=\bm{G}_m^\top(i,j+(k-1)p_m).
\end{equation}

Specifically, $\bm{H}_1=\bm{G}_1$. Note that this construction preserves orthogonality, i.e., $\bm{H}_m\in\mathbb{O}^{p_m r_{m-1}\times r_m}$ for all $m$, because only the ordering of the rows of $\bm{G}_m$ is changed to form $\bm{H}_m$. 

Let $\cm{X}_m\in\mathbb{R}^{p_m\times p_{m-1}\times\cdots\times p_1}$ ($m \leq M$) satisfy $\vectorize(\cm{X}_M) = \bm{x}_M$, and $\widetilde {\cm{X}}_m\in\mathbb{R}^{p_1\times p_2 \times\cdots\times p_m}$ be the tensor satisfying $\widetilde{\cm{X}}_m(i_1,i_2,\ldots,i_m) = \cm{X}_m(i_m,i_{m-1},\ldots,i_1)$ for all $i_1,\dots,i_m$, and $\bm{T}_m$ be the permutation matrix such that $\vectorize(\widetilde{\cm{X}}_m) = \bm{T}_m\bm{x}_m$. Some properties of $\bm{T}_m$ follow this construction: If we take the $j$-th row vector of $[\cm{X}_{m+1}]_1\in \mathbb{R}^{p_{m+1}\times p_{m}\cdots p_1}$ as $[\cm{X}_{m+1}]_1^\top(:,j)$ and the $j$-th column of $[\widetilde{\cm{X}}_{m+1}]_{m} \in \mathbb{R}^{p_1 p_2\cdots p_{m}\times p_{m+1}}$ as $[\widetilde{\cm{X}}_{m+1}]_{m}(:,j)$ for $j\leq p_{m+1}$, then we have
\begin{equation} \label{eq:permuting-reduced-dim}
	[\cm{X}_{m+1}]_1^\top(:,j) = \bm{T}_{m}^{-1} [\widetilde{\cm{X}}_{m+1}]_{m}(:,j)
\end{equation}
as both sides equal $\vectorize(\cm{X}_{m+1}(j,:,:,\ldots,:))$. 

We now corroborate \eqref{eq:TT-ours-proof} by induction. When $M=1$, we have $\widetilde{\cm{X}}_1 = \cm{X}_1$, and hence $\bm{T}_1=\bm{I}_{p_1}$, implying $\bm{G}_1^\top = \bm{H}_1^\top\bm{T}_1$.
Assume \eqref{eq:TT-ours-proof} holds for $M=m$ with the above construction. We will show it also holds for $M=m+1$. Note that
\begin{align*}
	& \bm{G}_{m+1}^\top (\bm{G}_m^\top \otimes\bm{I}_{p_{m+1}}) \cdots (\bm{G}_2^\top \otimes\bm{I}_{p_3\cdots p_{m+1}})(\bm{G}_1^\top \otimes\bm{I}_{p_2\cdots p_{m+1}})\bm{x}_{m+1} \\
	=~& \bm{G}_{m+1}^\top [\bm{G}_m^\top(\bm{G}_{m-1}^\top\otimes \bm{I}_{p_m})\cdots (\bm{G}_1^\top\otimes\bm{I}_{p_2\cdots p_m})\otimes \bm{I}_{p_{m+1}}] \vectorize(\cm{X}_{m+1})\\
	=~& \bm{G}_{m+1}^\top \vectorize(\bm{I}_{p_{m+1}}[\cm{X}_{m+1}]_1(\bm{G}_m^\top(\bm{G}_{m-1}^\top\otimes \bm{I}_{p_m})\cdots (\bm{G}_1^\top\otimes\bm{I}_{p_2\cdots p_m}))^\top) \\
	=~& \bm{G}_{m+1}^\top \vectorize((\bm{G}_m^\top(\bm{G}_{m-1}^\top\otimes \bm{I}_{p_m})\cdots (\bm{G}_1^\top\otimes\bm{I}_{p_2\cdots p_m})[\cm{X}_{m+1}]_1^\top)^\top),
\end{align*}
and similarly,
\begin{align*}
	& \bm{H}_{m+1}^\top (\bm{I}_{p_{m+1}}\otimes\bm{H}_m^\top) \cdots (\bm{I}_{p_3\cdots p_{m+1}}\otimes\bm{H}_2^\top)(\bm{I}_{p_2\cdots p_{m+1}}\otimes\bm{H}_1^\top)\bm{T}_{m+1}\bm{x}_{m+1} \\
	=~& \bm{H}_{m+1}^\top \vectorize(\bm{H}_m^\top(\bm{I}_{p_m}\otimes\bm{H}_{m-1}^\top)\cdots (\bm{I}_{p_m\cdots p_2}\otimes\bm{H}_1^\top)[\widetilde{\cm{X}}_{m+1}]_m \bm{I}_{p_{m+1}})\\
	=~& \bm{H}_{m+1}^\top \vectorize(\bm{H}_m^\top(\bm{I}_{p_m}\otimes\bm{H}_{m-1}^\top)\cdots (\bm{I}_{p_m\cdots p_2}\otimes\bm{H}_1^\top) \bm{T}_m [\cm{X}_{m+1}]_1^\top),
\end{align*}
where the last equality follows from \eqref{eq:permuting-reduced-dim}.
But by the induction hypothesis, we have
\begin{equation*}
	\bm{G}_m^\top(\bm{G}_{m-1}^\top\otimes \bm{I}_{p_m})\cdots (\bm{G}_1^\top\otimes\bm{I}_{p_2\cdots p_m}) = \bm{H}_m^\top(\bm{I}_{p_m}\otimes\bm{H}_{m-1}^\top)\cdots (\bm{I}_{p_m\cdots p_2}\otimes\bm{H}_1^\top) \bm{T}_m
\end{equation*}
and thereby we proceed to establish for any $\bm{M}\in \mathbb{R}^{p_{m+1}\times r_m}$,
\begin{equation*}
	\bm{G}_{m+1}^\top \vectorize(\bm{M}) = \bm{H}_{m+1}^\top \vectorize(\bm{M}^\top).
\end{equation*}

Inspecting the $i$-th element of both sides for $i\leq r_{m+1}$, we have
\begin{align*}
	 \bm{G}_{m+1}^\top(i,:) \vectorize(\bm{M}) &= \sum_{j=1}^{p_{m+1}}\sum_{k=1}^{r_m} \bm{G}_{m+1}^\top(i,j+(k-1)p_{m+1})\bm{M}(j,k), \\
	 \bm{H}_{m+1}^\top(i,:) \vectorize(\bm{M}^\top) &= \sum_{j=1}^{p_{m+1}}\sum_{k=1}^{r_m} \bm{H}_{m+1}^\top(i,(j-1)r_m+k)\bm{M}^\top(k,j).
\end{align*}
By \eqref{eq:gh-relationship}, these two quantities are equal, and hence the induction is complete.

\section{Proofs of Two Theorems} \label{appendix:sec-theory}
This section provides the detailed proofs of propositions and theorems in Section \ref{sec:method} and \ref{sec:theory} of the manuscript.

\subsection{Proof of Theorem \ref{thm:reg}}
\begin{proof}
	Denote sets
	\begin{align*}
		&\mathcal{W}_\mathrm{LR}(\{\bm{r}_{k}\}_{k=1}^{K_x}, \{\bm{s}_{k^\prime}\}_{k^\prime=1}^{K_y}, \mathcal{P}_x(M, K_x), \mathcal{P}_y(N, K_y)) = \{\bm{\Delta}  \in \mathbb{R}^{\prod_{n=1}^{N} q_n \times \prod_{m=1}^{M} p_m}:\\
		&\hspace{10mm}\bm{\Delta} = \bm{\Lambda}_{y,1}\bm{\Theta}_1\bm{\Lambda}_{x,1}^\top - \bm{\Lambda}_{y,2}\bm{\Theta}_2\bm{\Lambda}_{x,2}^\top; \bm{\Lambda}_{y,1}, \bm{\Lambda}_{y,2}\in\mathcal{R}_y(\{\bm{s}_{k^\prime}\}_{k^\prime=1}^{K_y}, \mathcal{P}_y(N, K_y)), \\
		&\hspace{10mm} \bm{\Lambda}_{x,1}, \bm{\Lambda}_{x,2}\in\mathcal{R}_x(\{\bm{r}_{k}\}_{k=1}^{K_x},  \mathcal{P}_x(M, K_x)), 
		\bm{\Theta}_1, \bm{\Theta}_2\in\mathbb{R}^{s_\mathrm{tot}\times r_\mathrm{tot}}, \norm{\bm{\Theta}_1}_2\leq g_1, \norm{\bm{\Theta}_2}_2\leq g_1 \},
	\end{align*}
	abbreviated to $\mathcal{W}_\mathrm{LR}$. Further denote
	\begin{align*}
		&\mathcal{W}_\mathrm{LR}(\gamma) = \{\bm{\Delta} \in \mathcal{W}_\mathrm{LR}: \norm{\bm{\Delta}}_{\mathrm{F}} \geq \gamma \} \quad \text{and}\quad
		\mathcal{W}_\mathrm{LR}^\prime(\gamma) = \{\bm{\Delta}^\prime = \frac{\bm{\Delta}}{\norm{\bm{\Delta}}_\Fr}: \bm{\Delta}\in \mathcal{W}_\mathrm{LR}(\gamma) \},
	\end{align*}
	for some $\gamma\in(0,1)$. We also let $\bm{y}_t \defeq \vectorize(\cm{Y}_t)$, $\bm{x}_t \defeq \vectorize(\cm{X}_t)$, and $\bm{e}_t \defeq \vectorize(\cm{E}_t)$.
	
	Recall the quadratic loss function
	\[
	\mathcal{L}_T(\cm{A}) 
	= \frac{1}{T}\sum_{t=1}^T \norm{\vectorize(\cm{Y}_t) - [\cm{A}]_N \vectorize(\cm{X}_t)}_2^2
	= \frac{1}{T}\sum_{t=1}^T \norm{\bm{y}_t - [\cm{A}]_N \bm{x}_t}_2^2.
	\]
	By the optimality of $\cm{\widehat{A}}_{\mathrm{LR}}$, we have
	\begin{align*}
		&\frac{1}{T}\sum_{t=1}^T\norm{\bm{y}_t-[\cm{\widehat{A}}_{\mathrm{LR}}]_N\bm{x}_t}_2^2\leq
		\frac{1}{T}\sum_{t=1}^T\norm{\bm{y}_t-[\cm{A}_\mathrm{LR}^\ast]_N\bm{x}_t}_2^2\\
		\Rightarrow~&\frac{1}{T}\sum_{t=1}^T\norm{\bm{\Delta}\bm{x}_t}_2^2
		\leq
		\frac{2}{T}\sum_{t=1}^T\langle\bm{e}_t,\bm{\Delta}\bm{x}_t\rangle
	\end{align*}
	where $\bm{\Delta}=[\cm{\widehat{A}}_{\mathrm{LR}}]_N-[\cm{A}_\mathrm{LR}^\ast]_N\in\mathcal{W}_\mathrm{LR}$.
	Multiplying both sides by the non-negative indicator function $\mathbbm{1}_{\{\norm{\bm{\Delta}}_\Fr \geq \gamma\}}$ for some $\gamma\in(0,1)$ to be specified later, we have
	\begin{equation} \label{eq:reg-basic-ineq-indicator}
		\frac{1}{T}\sum_{t=1}^T\norm{\bm{\Delta}\bm{x}_t}_2^2 \mathbbm{1}_{\{\norm{\bm{\Delta}}_\Fr \geq \gamma\}}
		\leq
		\frac{2}{T}\sum_{t=1}^T\langle\bm{e}_t,\bm{\Delta}\bm{x}_t\rangle \mathbbm{1}_{\{\norm{\bm{\Delta}}_\Fr \geq \gamma\}}.
	\end{equation}
	Denoting $\bm{\Delta}_{\geq\gamma} = \bm{\Delta} \mathbbm{1}_{\{\norm{\bm{\Delta}}_\Fr \geq \gamma\}}\in \mathcal{W}_\mathrm{LR}(\gamma)\cup\{\bm{0}\}$, and also noticing that $\mathbbm{1}_{\{\norm{\bm{\Delta}}_\Fr \geq \gamma\}}\in \{0,1\}$, we can rewrite \eqref{eq:reg-basic-ineq-indicator} as
	\begin{equation} \label{eq:reg-basic-ineq-reduced}
		\frac{1}{T}\sum_{t=1}^T\norm{\bm{\Delta}_{\geq\gamma}\bm{x}_t}_2^2
		\leq
		\frac{2}{T}\sum_{t=1}^T\langle\bm{e}_t,\bm{\Delta}_{\geq\gamma}\bm{x}_t\rangle,
	\end{equation}
	which yields the deterministic bound
	\begin{equation} \label{eq:reg-deterministic-bound}
		\norm{\bm{\Delta}_{\geq\gamma}}_\Fr \inf_{\bm{\Delta}^\prime \in \mathcal{W}_\mathrm{LR}^\prime(\gamma)}\sum_{t=1}^T \norm{\bm{\Delta}^\prime \bm{x}_t}_2^2
		\leq
		2\sup_{\bm{\Delta}^\prime \in \mathcal{W}_\mathrm{LR}^\prime(\gamma)}\sum_{t=1}^T \langle \bm{e}_t, \bm{\Delta}^\prime \bm{x}_t \rangle.
	\end{equation}
	Combining Lemmas \ref{lemma:reg-RSC} and \ref{lemma:reg-DB}, we can show when $T \gtrsim \max(1, {\sigma^4}/{c_x^2})(d_\mathrm{LR} d_\mathrm{LR}^\prime(\gamma))$, with probability at least $1 - 2\exp\left\{ -c_1d_\mathrm{LR} d_\mathrm{LR}^\prime(\gamma) \right\}$, the following event $\mathcal{E}_{\mathrm{LR}}(\gamma)$ holds:
	\begin{align*}
		\frac{c_x}{2} \norm{\bm{\Delta}_{\geq\gamma}}_{\mathrm{F}}
		\leq 2C\kappa \sigma \sqrt{\frac{d_\mathrm{LR} d_\mathrm{LR}^\prime(\gamma)}{T}} 
	\end{align*}
	where
	\begin{equation}\label{eq:def-d-lr}
		\begin{aligned}
			&d_\mathrm{LR} = 2K_x r_{\max}^2  \sum_{m=1}^{M}p_m + 2K_y s_{\max}^2 \sum_{n=1}^{N}q_n + 2K_xK_yr_{\max}s_{\max},\\
			&d_\mathrm{LR}^\prime(\gamma) 
			= \log \left(\frac{(M\vee N)\sqrt{[(K_ys_{\max})\wedge(K_xr_{\max})]K_xK_y (r_{\max}\vee s_{\max})}}{\gamma}\right),\\
			&d_\mathrm{LR}^\prime
			= d_\mathrm{LR}^\prime(1)
			= \log \left((M\vee N)\sqrt{[(K_ys_{\max})\wedge(K_xr_{\max})]K_xK_y (r_{\max}\vee s_{\max})}\right),
		\end{aligned}
	\end{equation}
	and $C$, $c_1$, $c_2$, $c_3$ are some positive constants. Note that $\sigma^2 \geq \norm{\bm{x}_t}_{\psi_2}^2 = \sup_{\norm{\bm{u}}=1}\norm{\langle \bm{u}, \bm{x}_t \rangle}_{\psi_2}^2 \gtrsim \sup_{\norm{\bm{u}}=1}\norm{\langle \bm{u}, \bm{x}_t \rangle}_{L_2}^2 \geq c_x$, and therefore $\max(1, {\sigma^4}/{c_x^2}) \asymp {\sigma^4}/{c_x^2}$.

	Under the event $\mathcal{E}_{\mathrm{LR}}(\gamma)$, it holds that
	\begin{align*}
		\norm{\bm{\Delta}}_\Fr = \norm{\bm{\Delta}_{\geq\gamma}+\bm{\Delta}{\mathbbm{1}_{\{\norm{\bm{\Delta}}_\Fr < \gamma\}}}}_\Fr \leq \gamma + \norm{\bm{\Delta}_{\geq\gamma}}_\Fr
		\lesssim \gamma + \frac{\kappa \sigma}{c_x} \sqrt{\frac{d_\mathrm{LR} d_\mathrm{LR}^\prime(\gamma)}{T}}.
	\end{align*}

	Take $\gamma = \exp(d'_{\mathrm{LR}})/T \ll \sqrt{\frac{d_\mathrm{LR}}{T}}$. With the convexity of $\log$ function, the sample size requirement can be relaxed to $T \gtrsim \frac{\sigma^4}{c_x^2}d_\mathrm{LR} \log(\frac{\sigma^4}{c_x^2}d_\mathrm{LR})$, and with probability at least $1-2\exp\{-c_1 d_\mathrm{LR} \log T\}$,
	\[
	\norm{\bm{\Delta}}_{\mathrm{F}}\lesssim \frac{\kappa \sigma}{c_x}\sqrt{\frac{d_\mathrm{LR} \log T}{T}}.
	\]
\end{proof}

\subsection{Proof of Theorem \ref{thm:autoreg}}
\begin{proof}
	Denote sets
	\begin{align*}
		&\mathcal{W}_\mathrm{AR}(\{\bm{r}_{k}\}_{k=1}^{K_x}, \{\bm{s}_{k^\prime}\}_{k^\prime=1}^{K_y}, \mathcal{P}_x(N, K_x), \mathcal{P}_y(N, K_y),L) = \{\bm{\Delta} \in \mathbb{R}^{\prod_{n=1}^{N} q_n \times \prod_{n=1}^{N} q_n}:\\
		&\hspace{10mm}\bm{\Delta} = \bm{\Lambda}_{y,1}\bm{\Theta}_1(\bm{I}_L\otimes\bm{\Lambda}_{x,1}^\top) - \bm{\Lambda}_{y,2}\bm{\Theta}_2(\bm{I}_L\otimes\bm{\Lambda}_{x,2}^\top), \bm{\Lambda}_{y,1}, \bm{\Lambda}_{y,2}\in\mathcal{R}_y(\{\bm{s}_{k^\prime}\}_{k^\prime=1}^{K_y}, \mathcal{P}_y(N, K_y)), \\
		&\hspace{10mm} \bm{\Lambda}_{x,1},\bm{\Lambda}_{x,2}\in\mathcal{R}_x(\{\bm{r}_{k}\}_{k=1}^{K_x}, \mathcal{P}_x(N, K_x)),
		\bm{\Theta}_1, \bm{\Theta}_2\in\mathbb{R}^{s_\mathrm{tot}\times r_\mathrm{tot}L}, \norm{\bm{\Theta}_1}_2\leq g_2, \norm{\bm{\Theta}_2}_2\leq g_2 \},
	\end{align*}
	abbreviated to $\mathcal{W}_\mathrm{AR}$. Further denote
	\begin{align*}
		&\mathcal{W}_\mathrm{AR}(\gamma) = \{\bm{\Delta} \in \mathcal{W}_\mathrm{AR}: \norm{\bm{\Delta}}_{\mathrm{F}} \geq \gamma \} \quad \text{and}\quad
		\mathcal{W}_\mathrm{AR}^\prime(\gamma) = \{\bm{\Delta}^\prime = \frac{\bm{\Delta}}{\norm{\bm{\Delta}}_\Fr}: \bm{\Delta}\in \mathcal{W}_\mathrm{AR}(\gamma) \},
	\end{align*}
	for some $\gamma\in(0,1)$, and let $\bm{y}_t \defeq \vectorize(\cm{Y}_t)$, $\bm{x}_t \defeq \vectorize\left(\vectorize(\cm{Y}_{t-1}), \ldots, \vectorize(\cm{Y}_{t-L})\right)$, $\bm{e}_t \defeq \vectorize(\cm{E}_t)$.	
	
	By the optimality of estimator $\cm{\widehat{A}}_{\mathrm{AR}}$, similar to the proof of Theorem \ref{thm:reg}, we could obtain
	\begin{align*}
		\frac{1}{T}\sum_{t=1}^T\norm{\bm{\Delta}\bm{x}_t}_2^2
		\leq
		\frac{2}{T}\sum_{t=1}^T\langle\bm{e}_t,\bm{\Delta}\bm{x}_t\rangle,
	\end{align*}
	which implies that deterministically,
	\begin{equation} \label{eq:ar-deterministic-bound}
		\norm{\bm{\Delta}_{\geq\gamma}}_\Fr \inf_{\bm{\Delta}^\prime \in \mathcal{W}_\mathrm{AR}^\prime(\gamma)}\sum_{t=1}^T \norm{\bm{\Delta}^\prime \bm{x}_t}_2^2
		\leq
		2\sup_{\bm{\Delta}^\prime \in \mathcal{W}_\mathrm{AR}^\prime(\gamma)}\sum_{t=1}^T \langle \bm{e}_t, \bm{\Delta}^\prime \bm{x}_t \rangle.
	\end{equation}
	where $\bm{\Delta}=[\cm{\widehat{A}}_{\mathrm{AR}}]_N-[\cm{A}_\mathrm{AR}^\ast]_N \in\mathcal{W}_\mathrm{AR}$, and $\bm{\Delta}_{\geq\gamma} = \bm{\Delta} \mathbbm{1}_{\{\norm{\bm{\Delta}}_\Fr \geq \gamma\}}\in \mathcal{W}_\mathrm{AR}(\gamma)\cup\{\bm{0}\}$.
	Applying Lemmas \ref{lemma:AR-RSC} and \ref{lemma:AR-DB} to \eqref{eq:ar-deterministic-bound}, we can show when $T \gtrsim \max(\kappa^4\kappa_{U,B}^2/\kappa_{L,A}^2,1) d_\mathrm{AR}d_\mathrm{AR}^\prime(\gamma)$, with probability at least $1-c_1\exp\left\{-c_2d_\mathrm{AR}d_\mathrm{AR}^\prime(\gamma)\right\}$, the following event $\mathcal{E}_{\mathrm{AR}}(\gamma)$ holds: 
	\begin{align*}
		\frac{\kappa_{L,A}}{16} \norm{\bm{\Delta}_{\geq\gamma}}_\Fr
		\leq 2C\kappa^2\kappa_{U,B}'\sqrt{\frac{d_\mathrm{AR}d_\mathrm{AR}^\prime(\gamma)}{T}}
		\label{eq:thm2-prob}
	\end{align*}
	where
	\begin{equation}\label{eq:def-d-ar}
		\begin{aligned}
			&d_\mathrm{AR} = 2(K_x r_{\max}^2 + K_y s_{\max}^2) \sum_{n=1}^{N}q_n + 2LK_xK_yr_{\max}s_{\max},\\
			&d_\mathrm{AR}^\prime(\gamma) = \log\left(\frac{N\sqrt{[(K_ys_{\max})\wedge(LK_xr_{\max})]LK_xK_y (r_{\max}\vee s_{\max})}}{\gamma}\right),\\
			&d_\mathrm{AR}^\prime = d_\mathrm{AR}^\prime(1) = \log\left(N\sqrt{[(K_ys_{\max})\wedge(LK_xr_{\max})]LK_xK_y (r_{\max}\vee s_{\max})}\right),\\
			&\kappa_{U, B}=C_{e}/\mu_{\mathrm{min}}(\bm{B}^{\ast}), ~\kappa_{L, A}=c_{e}/\mu_{\mathrm{max}}(\cm{A}^\ast), ~\kappa^\prime_{U, B}=C_{e}/\mu^{1/2}_{\mathrm{min}}(\bm{B}^{\ast}),
		\end{aligned}
	\end{equation}
	and $C$, $c_1$, $c_2$ are some positive constants. By Lemma \ref{lemma:AR-spectral}, $\max(\kappa^4\kappa_{U,B}^2/\kappa_{L,A}^2,1) \asymp  \kappa^4\kappa_{U,B}^2/\kappa_{L,A}^2$.

	Under the event $\mathcal{E}_{\mathrm{AR}}(\gamma)$, it holds that
	\begin{align*}
		\norm{\bm{\Delta}}_\Fr = \norm{\bm{\Delta}_{\geq\gamma}+\bm{\Delta}{\mathbbm{1}_{\{\norm{\bm{\Delta}}_\Fr < \gamma\}}}}_\Fr \leq \gamma + \norm{\bm{\Delta}_{\geq\gamma}}_\Fr
		\lesssim \gamma + \frac{\kappa^2\kappa_{U,B}'}{\kappa_{L,A}} \sqrt{\frac{d_\mathrm{AR}d_\mathrm{AR}^\prime(\gamma)}{T}}.
	\end{align*}

	Take $\gamma = \exp (d^\prime_\mathrm{AR})/T$. Whenever $T \gtrsim \frac{\kappa^4\kappa_{U,B}^2}{\kappa_{L,A}^2} d_\mathrm{AR}\log(\frac{\kappa^4\kappa_{U,B}^2}{\kappa_{L,A}^2} d_\mathrm{AR})$, with probability at least $1-c_1\exp\left\{-c_2d_\mathrm{AR}\log T \right\}$,
	\[
	\norm{\bm{\Delta}}_{\mathrm{F}}\lesssim \frac{\kappa^2\kappa_{U,B}'}{\kappa_{L,A}} \sqrt{\frac{d_\mathrm{AR}\log T}{T}}.
	\]
\end{proof}

\section{Auxiliary Lemmas}  \label{appendix:sec-lemma}
This section provides the auxiliary lemmas that are used in establishing the theorems in Section \ref{appendix:sec-theory} and their proofs. We first introduce the concepts of Gaussian width and Talagrand's $\gamma_{\alpha}$ functional, which will be frequently utilized in the upcoming lemmas.
\begin{definition}[Gaussian width, Definition 7.5.1 of \cite{vershynin2019high}]
	The Gaussian width of a subset $T \subset \mathbb{R}^n$ is defined as 
	\[
	\omega(T) \defeq \mathbb{E} \sup_{\bm{x} \in T} \langle \bm{g}, \bm{x} \rangle 
	\quad \text{where} \quad
	\bm{g} \sim N(\bm{0}, \bm{I}_n).
	\]
\end{definition}

\begin{definition}[Talagrand's $\gamma_{\alpha}$ functional, \cite{dirksenTailBoundsGeneric2015a}]
	Let $(T, d)$ be a metric space and $\alpha > 0$. A sequence of subsets $\left(T_k\right)_{k=0}^\infty$ of $T$ is called an admissible sequence if the cardinalities of $T_k$ satisfy $|T_0| = 1$ and $|T_k| \leq 2^{2^k}$ for all $k \geq 1$. The $\gamma_\alpha$ functional of $T$ is defined as 
	\[
	\gamma_\alpha(T,d) = \inf_{(T_k)}\sup_{t \in T} \sum_{k=0}^{\infty} 2^{k/\alpha}d(t, T_k)
	\]
	where the infimum is with respect to all admissible sequences.
\end{definition}
\begin{remark}
This definition is consistent with Definition 8.5.1 of \cite{vershynin2019high} when $\alpha=2$, motivated by the covering number argument. Some of the references in this paper define the $\gamma_\alpha$ functional as in \cite{talagrand2005generic}, based on the notion of ultrametricity. These two definitions are equivalent up to a constant factor depending only on $\alpha$ according to \cite{talagrand2001majorizing}.
\end{remark}
\begin{lemma}[\textbf{Covering number}] \label{lemma:covering}
	This lemma derives the covering number for a series of parameter space that are involved in the subsequent proof.
	\begin{itemize}
		\item[(a)] The cardinality of the $\epsilon$-covering net of $\mathcal{R}_x(\{\bm{r}_{k}\}_{k=1}^{K_x}, \mathcal{P}_x(M, K_x))$, denoted as $\mathcal{N}(\mathcal{R}_x, \norm{\cdot}_{\mathrm{F}}, \epsilon)$, satisfies
		\begin{align*}
			\mathcal{N}(\mathcal{R}_x, \norm{\cdot}_{\mathrm{F}}, \epsilon) \leq 
			\left(1 + \frac{2M \sqrt{K_x r_{\max}}}{\epsilon}\right)^{K_x r_{\max}^2 \sum_{m=1}^{M}p_m}.
		\end{align*}
		
		\item[(b)] The cardinality of the $\epsilon$-covering net of $\mathcal{R}_y(\{\bm{s}_{k^\prime}\}_{k^\prime=1}^{K_y}, \mathcal{P}_y(N, K_y))$, denoted as $\mathcal{N}(\mathcal{R}_y, \norm{\cdot}_{\mathrm{F}}, \epsilon)$, satisfies
		\begin{align*}
			\mathcal{N}(\mathcal{R}_y, \norm{\cdot}_{\mathrm{F}}, \epsilon) \leq 
			\left(1 + \frac{2N \sqrt{K_y s_{\max}}}{\epsilon}\right)^{K_y s_{\max}^2 \sum_{n=1}^{N}q_n}.
		\end{align*}
		
		\item[(c)] The cardinality of the $\epsilon$-covering net of $\mathcal{W}_\mathrm{AR}^\prime(\gamma)$, denoted as $\mathcal{N}(\mathcal{W}_\mathrm{AR}^\prime(\gamma), \norm{\cdot}_{\mathrm{F}}, \epsilon)$, with $\gamma \in (0,1)$, satisfies
		\begin{align*}
			\mathcal{N}(\mathcal{W}_\mathrm{AR}^\prime(\gamma), \norm{\cdot}_{\mathrm{F}}, \epsilon) \leq \left(1 + \frac{24g_2^\prime N\sqrt{LK_xK_y (r_{\max}\vee s_{\max})(s_\mathrm{tot}\wedge (Lr_\mathrm{tot}))}}{\gamma \epsilon}\right)^{d_\mathrm{AR}},
		\end{align*}
		where $g_2^\prime = \sqrt{s_\mathrm{tot}\wedge (Lr_\mathrm{tot})} g_2$, and $d_\mathrm{AR} = 2(K_x r_{\max}^2 + K_y s_{\max}^2) \sum_{n=1}^{N}q_n + 2LK_xK_yr_{\max}s_{\max}$.
		
		\item[(d)] The cardinality of the $\epsilon$-covering net of $\mathcal{W}_\mathrm{LR}^\prime$, denoted as $\mathcal{N}(\mathcal{W}_\mathrm{LR}^\prime, \norm{\cdot}_{\mathrm{F}}, \epsilon)$, satisfies
		\begin{align*}
			\mathcal{N}(\mathcal{W}_\mathrm{LR}^\prime(\gamma), \norm{\cdot}_{\mathrm{F}}, \epsilon) \leq \left(1 + \frac{24g_1^\prime (M\vee N)\sqrt{K_xK_y (r_{\max}\vee s_{\max})(s_\mathrm{tot}\wedge r_\mathrm{tot})}}{\gamma\epsilon}\right)^{d_\mathrm{LR}},
		\end{align*}
		where $g_1^\prime = \sqrt{s_\mathrm{tot}\wedge r_\mathrm{tot}}g_1$, and $d_\mathrm{LR} = 2K_x r_{\max}^2  \sum_{m=1}^{M}p_m + 2K_y s_{\max}^2 \sum_{n=1}^{N}q_n + 2K_xK_yr_{\max}s_{\max}$.
		
	\end{itemize}
\end{lemma}

\begin{proof}
	(a) Denote the $\epsilon/(M\sqrt{r_{\mathrm{tot}}})$-covering net of $\mathbb{O}^{p\times r}$ by $\bar{\mathbb{O}}^{p\times r}$. 
	By Lemma 7 in \citet{zhang2018tensor}, the cardinality of $\bar{\mathbb{O}}^{p\times r}$ satisfies $|\bar{\mathbb{O}}^{p\times r}| \leq ((1+2M\sqrt{r_{\mathrm{tot}}})/\epsilon)^{pr}$.
	Then for any $\bm{G}\in\mathbb{O}^{p\times r}$, there exists a $\bar{\bm{G}}\in\bar{\mathbb{O}}^{p\times r}$ such that $\norm{ \bm{G} - \bar{\bm{G}} }_{\mathrm{F}} \leq \epsilon/(M\sqrt{r_{\mathrm{tot}}})$.
	
	Set the $\epsilon$-covering net of $\mathcal{R}_x(\{\bm{r}_{k}\}_{k=1}^{K_x}, \mathcal{P}_x(M, K_x))$, with abbreviation $\mathcal{R}_x$, to be 
	\begin{align}
		\bar{\mathcal{R}}_x(\epsilon) = \{
		&\bar{\bm{\Lambda}}_x \in \mathbb{R}^{\prod_{m=1}^{M}p_m \times r_{\mathrm{tot}}}:\notag\\
		&\bar{\bm{\Lambda}}_x^\top = 
		\begin{bmatrix}
			\bar{\bm{G}}_{M,1}^\top (\bm{I}_{p_{\alpha^1_{(M)}}}\otimes\bar{\bm{G}}_{M-1,1}^\top) \cdots (\bm{I}_{p_{\alpha^1_{(2)}}\cdots p_{\alpha^1_{(M)}}}\otimes\bar{\bm{G}}_{1,1}^\top) \bm{T}_x(\alpha_1)\notag\\
			\bar{\bm{G}}_{M,2}^\top (\bm{I}_{p_{\alpha^2_{(M)}}}\otimes\bar{\bm{G}}_{M-1,2}^\top) \cdots (\bm{I}_{p_{\alpha^2_{(2)}}\cdots p_{\alpha^2_{(M)}}}\otimes\bar{\bm{G}}_{1,2}^\top) \bm{T}_x(\alpha_2)\\
			\cdots\\
			\bar{\bm{G}}_{M,K_x}^\top (\bm{I}_{p_{\alpha^{K_x}_{(M)}}}\otimes\bar{\bm{G}}_{M-1,K_x}^\top) \cdots (\bm{I}_{p_{\alpha^{K_x}_{(2)}}\cdots p_{\alpha^{K_x}_{(M)}}}\otimes\bar{\bm{G}}_{1,K_x}^\top) \bm{T}_x(\alpha_{K_x})
		\end{bmatrix},\notag\\
		&\bar{\bm{G}}_{m,k}\in\bar{\mathbb{O}}^{r_{m-1,k}p_{\alpha^k_{(m)}}\times r_{m,k}}, \text{ for } 1 \leq m \leq M, 1 \leq k \leq K_x\}.\notag
	\end{align}
	It remains to show that for any $\bm{\Lambda}_x \in \mathcal{R}_x$, there exists a $\bar{\bm{\Lambda}}_x \in \bar{\mathcal{R}}_x(\epsilon)$ such that $\norm{ \bm{\Lambda}_x - \bar{\bm{\Lambda}}_x }_{\mathrm{F}} \leq \epsilon$. Let $\phi(\bm{G}_{1,k}, \ldots, \bm{G}_{M,k}) = \bm{G}_{M,k}^\top (\bm{I}_{p_{\alpha^k_{(M)}}}\otimes\bm{G}_{M-1,k}^\top) \cdots (\bm{I}_{p_{\alpha^k_{(2)}}\cdots p_{\alpha^k_{(M)}}}\otimes\bm{G}_{1,k}^\top) \bm{T}_x(\alpha_k)$. Note that
	\begin{align*}
		\norm{ \bm{\Lambda}_x - \bar{\bm{\Lambda}}_x }_{\mathrm{F}}^2
		= &  \sum_{k=1}^{K_x} \norm{\phi(\bm{G}_{1,k}, \ldots, \bm{G}_{M,k}) - \phi(\bar{\bm{G}}_{1,k}, \ldots, \bar{\bm{G}}_{M,k})}_{\mathrm{F}}^2 \\
		\leq & \sum_{k=1}^{K_x}( \sum_{m=1}^{M} \| \phi(\bar{\bm{G}}_{1,k}, \ldots, \bar{\bm{G}}_{m-1,k}, \bm{G}_{m,k}, \bm{G}_{m+1,k}, \ldots, \bm{G}_{M,k}) \\
		&\hspace{15mm}- \phi(\bar{\bm{G}}_{1,k}, \ldots, \bar{\bm{G}}_{m-1,k}, \bar{\bm{G}}_{m,k}, \bm{G}_{m+1,k}, \ldots, \bm{G}_{M,k})\|_{\mathrm{F}})^2\\
		= & \sum_{k=1}^{K_x}( \sum_{m=1}^{M} \|\phi(\bar{\bm{G}}_{1,k}, \ldots, \bar{\bm{G}}_{m-1,k}, \bm{G}_{m,k} - \bar{\bm{G}}_{m,k}, \bm{G}_{m+1,k}, \ldots, \bm{G}_{M,k})\|_{\mathrm{F}})^2.
	\end{align*}
	In the following, we will bound each term of the last equality in the right hand side. For $m = 1,2,\ldots, M-1$,
	\begin{align*}
		&\|\phi(\bar{\bm{G}}_{1,k}, \ldots, \bar{\bm{G}}_{m-1,k}, \bm{G}_{m,k} - \bar{\bm{G}}_{m,k}, \bm{G}_{m+1,k}, \ldots, \bm{G}_{M,k})\|_{\mathrm{F}}\\
		=~ & \|\bm{G}_{M,k}^\top \cdots (\bm{I}_{p_{\alpha^k_{(m+1)}}\cdots p_{\alpha^k_{(M)}}}\otimes(\bm{G}_{m,k}^\top - \bar{\bm{G}}_{m,k}^\top))\cdots (\bm{I}_{p_{\alpha^k_{(2)}}\cdots p_{\alpha^k_{(M)}}}\otimes\bar{\bm{G}}_{1,k}^\top) \bm{T}_x(\alpha_k) \|_{\mathrm{F}}\\
		\leq~ & \|\bm{G}_{M,k}^\top\|_{\mathrm{F}} \|\bm{I}_{p_{\alpha^k_{(M)}}}\otimes \bm{G}_{M-1,k}^\top\|_2 \cdots \|\bm{I}_{p_{\alpha^k_{(m+1)}}\cdots p_{\alpha^k_{(M)}}}\otimes(\bm{G}_{m,k}^\top - \bar{\bm{G}}_{m,k}^\top)\|_2 \\
		&\cdots \|\bm{I}_{p_{\alpha^k_{(2)}}\cdots p_{\alpha^k_{(M)}}}\otimes\bar{\bm{G}}_{1,k}^\top\|_2 \|\bm{T}_x(\alpha_k)\|_2\\
		\leq~ & \sqrt{r_{M,k}} \|\bm{G}_{m,k} - \bar{\bm{G}}_{m,k}\|_2 \leq \sqrt{r_{M,k}} \|\bm{G}_{m,k} - \bar{\bm{G}}_{m,k}\|_{\mathrm{F}},
	\end{align*}
	where the first inequality follows from $\|\bm{A}\bm{B}\|_{\mathrm{F}}\leq\|\bm{A}\|_{\mathrm{F}}\|\bm{B}\|_2$ for any matrices $\bm{A}\in\mathbb{R}^{p\times r}$ and $\bm{B}\in\mathbb{R}^{r\times q}$. The second inequality uses $\|\bm{I}_q\otimes\bm{A}\|_2 = \|\bm{A}\|_2\leq 1$ for any orthogonal matrix $\bm{A}$. When $m=M$, we can bound the term in a similar way,
	\begin{align*}
		&\|\phi(\bar{\bm{G}}_{1,k}, \ldots, \bar{\bm{G}}_{M-1,k}, \bm{G}_{M,k} - \bar{\bm{G}}_{M,k})\|_{\mathrm{F}}\\
		=~ & \|(\bm{G}_{M,k}^\top - \bar{\bm{G}}_{M,k}^\top) (\bm{I}_{p_{\alpha^k_{(M)}}}\otimes\bar{\bm{G}}_{M-1,k}^\top) \cdots (\bm{I}_{p_{\alpha^k_{(2)}}\cdots p_{\alpha^k_{(M)}}}\otimes\bar{\bm{G}}_{1,k}^\top) \bm{T}_x(\alpha_k) \|_{\mathrm{F}}\\
		\leq~ & \|\bm{G}_{M,k}^\top - \bar{\bm{G}}_{M,k}^\top\|_{\mathrm{F}} \|\bm{I}_{p_{\alpha^k_{(M)}}}\otimes \bar{\bm{G}}_{M-1,k}^\top\|_2 \cdots \|\bm{I}_{p_{\alpha^k_{(2)}}\cdots p_{\alpha^k_{(M)}}}\otimes\bar{\bm{G}}_{1,k}^\top\|_2 \|\bm{T}_x(\alpha_k)\|_2\\
		\leq~ & \|\bm{G}_{M,k}^\top - \bar{\bm{G}}_{M,k}^\top\|_{\mathrm{F}} \leq \sqrt{r_{M,k}} \|\bm{G}_{M,k}^\top - \bar{\bm{G}}_{M,k}^\top\|_{\mathrm{F}}.
	\end{align*}
	Thus,
	\begin{align*}
		\norm{ \bm{\Lambda}_x - \bar{\bm{\Lambda}}_x }_{\mathrm{F}}^2
		\leq & \sum_{k=1}^{K_x}( \sum_{m=1}^{M} \sqrt{r_{M,k}} \|\bm{G}_{m,k} - \bar{\bm{G}}_{m,k}\|_{\mathrm{F}})^2
		\leq \sum_{k=1}^{K_x}( \sum_{m=1}^{M} \epsilon\sqrt{r_{M,k}}/(M\sqrt{r_{\mathrm{tot}}}))^2\\
		\leq & \sum_{k=1}^{K_x}(\epsilon\sqrt{r_{M,k}}/\sqrt{r_{\mathrm{tot}}})^2 \leq \epsilon^2.
	\end{align*}
	So the cardinality of $\bar{\mathcal{R}}_x(\epsilon)$ is
	\begin{align*}
		\mathcal{N}(\mathcal{R}_x, \norm{\cdot}_{\mathrm{F}}, \epsilon) \leq & \prod_{k=1}^{K_x}\prod_{m=1}^{M}
		\left(\frac{1+2M\sqrt{r_{\mathrm{tot}}}}{\epsilon}\right)^{r_{m-1,k}p_{\alpha_k(m)}r_{m,k}}\\
		\leq & \left(1 + \frac{2M \sqrt{K_x r_{\max}}}{\epsilon}\right)^{K_x r_{\max}^2 \sum_{m=1}^{M}p_m}.
	\end{align*}
	(b) The proof of this lemma is identical to the proof of part (a).
	
	(c) Denote the radius-$\nu$ ball by $\mathcal{B}_2^d(\nu) = \{\bm{b} \in \mathbb{R}^d: \lVert \bm{b} \rVert_2 \leq \nu\}$. By Assumption \ref{assump:auto-core}, $\norm{\bm{\Theta}_\mathrm{AR}}_\mathrm{F}\leq\sqrt{s_\mathrm{tot}\wedge (Lr_\mathrm{tot})} g_2\defeq g_2^\prime$. Set the $\epsilon/(3\sqrt{K_xK_y})$-covering net of $\mathcal{B}_2^{s_{\mathrm{tot}}r_{\mathrm{tot}}L}(g_2^\prime)$ by $\bar{\mathcal{B}}^{s_{\mathrm{tot}}r_{\mathrm{tot}}L}$ with cardinality $|\bar{\mathcal{B}}^{s_{\mathrm{tot}}r_{\mathrm{tot}}L}|\leq (1+6g_2^\prime\sqrt{K_xK_y}/\epsilon)^{s_{\mathrm{tot}}r_{\mathrm{tot}}L}$.
	We first construct the $\epsilon$-covering net of $\mathcal{S}_{\mathrm{AR}}(\{\bm{r}_{k}\}_{k=1}^{K_x}, \{\bm{s}_{k^\prime}\}_{k^\prime=1}^{K_y}, \mathcal{P}_x(M, K_x), \mathcal{P}_y(N, K_y),L)$, with abbreviation $\mathcal{S}_{\mathrm{AR}}$, to be
	\begin{align*}
		\bar{\mathcal{S}}_{\mathrm{AR}}(\epsilon) = \{
		&\bar{\cm{A}} \in \mathbb{R}^{q_1 \times \cdots \times q_N \times p_1 \times \cdots \times p_M \times L}:
		[\bar{\cm{A}}]_N = \bar{\bm{\Lambda}}_y \bar{\bm{\Theta}}_{\mathrm{AR}} (\bm{I}_L\otimes\bar{\bm{\Lambda}}_x^\top),\notag\\
		&\bar{\bm{\Lambda}}_y \in \bar{\mathcal{R}}_y(\epsilon/(3g_2\sqrt{K_x})), \bar{\bm{\Lambda}}_x \in \bar{\mathcal{R}}_x(\epsilon/(3g_2\sqrt{LK_y})),\\
		&\bar{\bm{\Theta}}_{\mathrm{AR}} \in\mathbb{R}^{s_{\mathrm{tot}}\times r_{\mathrm{tot}}L}\text{ and }\vectorize(\bar{\bm{\Theta}}_{\mathrm{AR}})\in\bar{\mathcal{B}}^{s_{\mathrm{tot}}r_{\mathrm{tot}}L}\}.\notag
	\end{align*}
	It remains to show that for any $\cm{A} \in \mathcal{S}_{\mathrm{AR}}$, there exists an $\bar{\cm{A}} \in \bar{\mathcal{S}}_{\mathrm{AR}}(\epsilon)$ such that $\norm{ \cm{A} - \bar{\cm{A}} }_{\mathrm{F}} \leq \epsilon$. Note that
	\begin{align*}
		\norm{ \cm{A} - \bar{\cm{A}} }_{\mathrm{F}}
		& = \norm{\bm{\Lambda}_y \bm{\Theta}_{\mathrm{AR}} (\bm{I}_L\otimes\bm{\Lambda}_x^\top) - \bar{\bm{\Lambda}}_y \bar{\bm{\Theta}}_{\mathrm{AR}} (\bm{I}_L\otimes\bar{\bm{\Lambda}}_x^\top)}_{\mathrm{F}}\\
		& \leq \norm{\bm{\Lambda}_y - \bar{\bm{\Lambda}}_y}_{\mathrm{F}} \norm{\bm{\Theta}_{\mathrm{AR}}}_2 \norm{\bm{I}_L\otimes\bm{\Lambda}_x^\top}_2
		+ \norm{\bar{\bm{\Lambda}}_y}_2 \norm{\bm{\Theta}_{\mathrm{AR}} - \bar{\bm{\Theta}}_{\mathrm{AR}}}_{\mathrm{F}} \norm{\bm{I}_L\otimes\bm{\Lambda}_x^\top}_2\\
		& + \norm{\bar{\bm{\Lambda}}_y}_2 \norm{\bar{\bm{\Theta}}_{\mathrm{AR}}}_2 \norm{\bm{I}_L\otimes(\bm{\Lambda}_x^\top- \bar{\bm{\Lambda}}_x^\top)}_{\mathrm{F}}\\
		& \leq g_2\sqrt{K_x} \norm{\bm{\Lambda}_y - \bar{\bm{\Lambda}}_y}_{\mathrm{F}} + \sqrt{K_xK_y}\norm{\bm{\Theta}_{\mathrm{AR}} - \bar{\bm{\Theta}}_{\mathrm{AR}}}_{\mathrm{F}} + g_2\sqrt{LK_y} \norm{\bm{\Lambda}_x - \bar{\bm{\Lambda}}_x}_{\mathrm{F}}\\
		& \leq \epsilon,
	\end{align*}
	where the second inequality uses $\norm{\bm{\Theta}_{\mathrm{AR}}}_2\leq g_2$ and $\norm{\bar{\bm{\Theta}}_{\mathrm{AR}}}_2\leq g_2$ by assumption \ref{assump:auto-core}. And we have $\norm{\bar{\bm{\Lambda}}_y}_2\leq\sqrt{K_y}$ and $\norm{\bm{I}_L\otimes\bm{\Lambda}_x^\top}_2=\norm{\bm{\Lambda}_x^\top}_2\leq\sqrt{K_x}$ given that $\bar{\bm{\Lambda}}_y$ and $\bm{\Lambda}_x$ are composed of some orthogonal matrices.
	By Lemma \ref{lemma:covering}(a) and (b), the cardinality of $\bar{\mathcal{S}}_{\mathrm{AR}}(\epsilon)$ is
	\begin{align*}
		\mathcal{N}(\mathcal{S}_{\mathrm{AR}}, \norm{\cdot}_{\mathrm{F}}, \epsilon) \leq & \left(1 + \frac{6g_2N \sqrt{LK_xK_y r_{\max}}}{\epsilon}\right)^{K_x r_{\max}^2 \sum_{n=1}^{N}q_n} \left(1 + \frac{6g_2^\prime\sqrt{K_xK_y}}{\epsilon}\right)^{LK_xK_yr_{\max}s_{\max}} \\
		& \left(1 + \frac{6g_2N \sqrt{K_xK_y s_{\max}}}{\epsilon}\right)^{K_y s_{\max}^2 \sum_{n=1}^{N}q_n}\\
		\leq & \left(1 + \frac{6g_2^\prime N\sqrt{LK_xK_y (r_{\max}\vee s_{\max})}}{\epsilon}\right)^{(K_x r_{\max}^2 + K_y s_{\max}^2) \sum_{n=1}^{N}q_n + LK_xK_yr_{\max}s_{\max}}.
	\end{align*}
	Now we are ready to construct the $\epsilon$-covering net of $\mathcal{W}_\mathrm{AR}$. Note that for any $\bm{\Delta}\in\mathcal{W}_\mathrm{AR}$, we have $\bm{\Delta} = [\cm{A}_1]_N - [\cm{A}_2]_N$ with $\cm{A}_1, \cm{A}_2 \in \mathcal{S}_{\mathrm{AR}}$. Then the covering net can be constructed as
	\begin{align}
		\bar{\mathcal{W}}_{\mathrm{AR}} = \{ 
		&\bar{\bm{\Delta}} \in \mathbb{R}^{\prod_{n=1}^{N} q_n \times L\prod_{m=1}^{M} p_m}:
		\bar{\bm{\Delta}} = [\bar{\cm{A}}_1]_N - [\bar{\cm{A}}_2]_N,\bar{\cm{A}}_1, \bar{\cm{A}}_2\in \bar{\mathcal{S}}_{\mathrm{AR}}(\epsilon/2)\}.\notag
	\end{align}
	It can be easily verified that $\norm{\bm{\Delta} - \bar{\bm{\Delta}}}_\mathrm{F}\leq \norm{[\cm{A}_1]_N - [\bar{\cm{A}}_1]_N}_\mathrm{F} + \norm{[\cm{A}_2]_N - [\bar{\cm{A}}_2]_N}_\mathrm{F}\leq \epsilon$. And the cardinality of $\bar{\mathcal{W}}_{\mathrm{AR}}$ is given by
	\begin{align}
		\nonumber
		&~\mathcal{N}(\mathcal{W}_\mathrm{AR}, \norm{\cdot}_{\mathrm{F}}, \epsilon) \\ \nonumber
		\leq & ~\mathcal{N}\left(\mathcal{S}_{\mathrm{AR}}, \norm{\cdot}_{\mathrm{F}}, \frac{\epsilon}{2}\right)^2\\
		\leq & \left(1 + \frac{12g_2^\prime(M\vee N)\sqrt{LK_xK_y (r_{\max}\vee s_{\max})}}{\epsilon}\right)^{2(K_x r_{\max}^2 + K_y s_{\max}^2) \sum_{n=1}^{N}q_n + 2LK_xK_yr_{\max}s_{\max}}.
		\label{eq:covering-ar-full}
	\end{align}

	Note that $\mathcal{W}_\mathrm{AR}^\prime(\gamma) = \{\bm{\Delta}/\norm{\bm{\Delta}}_\Fr : \bm{\Delta}\in\mathcal{W}_\mathrm{AR}, \norm{\bm{\Delta}}_\mathrm{F} > \gamma\} \subset \mathcal{W}_\mathrm{AR}/\gamma \defeq \{\bm{\Delta}/\gamma : \bm{\Delta}\in\mathcal{W}_\mathrm{AR}\}$.
	Each element in $\mathcal{W}_\mathrm{AR}/\gamma$ is a scaled version of an element in $\mathcal{W}_\mathrm{AR}$ by a factor $1/\gamma$, which can be multiplied to $\bm{\Theta}_\mathrm{AR}$ and $\bar{\bm{\Theta}}_{\mathrm{AR}}$ with their upper bound $g_2$ replaced by $g_2/\gamma$ and other components unchanged. 
	Thus, the covering number of $\mathcal{W}_\mathrm{AR}/\gamma$ can be obtained by replacing $g_2^\prime$ with $g_2^\prime/\gamma$ in \eqref{eq:covering-ar-full}.

	Denote $\mathcal{D}(\cdot, \norm{\cdot}, \epsilon)$ as the $\epsilon$-packing number as in Definition 5.4 of \cite{wainwright2019high}. By the relationship between covering number and packing number (\cite{wainwright2019high}, Lemma 5.5), we have $\mathcal{N}(\mathcal{W}_\mathrm{AR}^\prime(\gamma), \norm{\cdot}_{\mathrm{F}}, \epsilon) \leq \mathcal{D}(\mathcal{W}_\mathrm{AR}^\prime(\gamma), \norm{\cdot}_{\mathrm{F}}, \epsilon) \leq \mathcal{D}(\mathcal{W}_\mathrm{AR}/\gamma, \norm{\cdot}_{\mathrm{F}}, \epsilon) \leq \mathcal{N}(\mathcal{W}_\mathrm{AR}/\gamma, \norm{\cdot}_{\mathrm{F}}, \epsilon/2)$, yielding the desired result.

	(d) This is similar to part (c) with $L=1$ and $g_1^\prime\defeq\sqrt{s_\mathrm{tot}\wedge r_\mathrm{tot}}g_1$.
\end{proof}

\begin{lemma}[\textbf{Truncated Regression RSC}]
	\label{lemma:reg-RSC}
	Suppose Assumptions \ref{assump:reg-input}, \ref{assump:reg-error}, \ref{assump:reg-core} hold with $T \gtrsim \frac{\sigma^4}{c_x^2} (d_\mathrm{LR} d_\mathrm{LR}^\prime(\gamma))$ for some $\gamma \in (0,1)$. Then
	\[
	\frac{1}{2}c_x\norm{\bm{\Delta}}_{\mathrm{F}}^2\leq\frac{1}{T}\sum_{t=1}^T\norm{\bm{\Delta}\bm{x}_t}_2^2\leq 2C_x\norm{\bm{\Delta}}_{\mathrm{F}}^2 \quad\text{for all } \bm{\Delta}\in \mathcal{W}_\mathrm{LR}(\gamma)
	\]
	with probability at least 
	$1-\exp\left\{-c_1d_\mathrm{LR} d_\mathrm{LR}^\prime(\gamma)\right\}$, where $c_1>0$ is a constant, and $d_\mathrm{LR}$, $d_\mathrm{LR}^\prime(\gamma)$ are defined in \eqref{eq:def-d-lr}.
\end{lemma}
\begin{proof}
	This lemma establishes the probabilistic uniform convexity condition for the empirical loss function confined on the set $\mathcal{W}_\mathrm{LR}(\gamma)$. By the definition of $\mathcal{W}_\mathrm{LR}(\gamma)$ and $\mathcal{W}_\mathrm{LR}^\prime(\gamma)$, it suffices to show that $\sup_{\bm{\Delta} \in \mathcal{W}_\mathrm{LR}^\prime(\gamma)} \sum_{t=1}^T\norm{\bm{\Delta}\bm{x}_t}_2^2/T \leq 2C_x$ and $\inf_{\bm{\Delta} \in \mathcal{W}_\mathrm{LR}^\prime(\gamma)} \sum_{t=1}^T\norm{\bm{\Delta}\bm{x}_t}_2^2/T \geq c_x/2$ hold with high probability. Truncating $\mathcal{B}_\Fr^{Q\times \prod_{m=1} ^{M}p_m}(\gamma) \defeq \{\bm{\Delta} \in \mathbb{R}^{Q\times \prod_{m=1} ^{M}p_m}: \lVert \bm{\Delta} \rVert_\mathrm{F} \leq \gamma\}$ out from the restricted strong convexity (RSC) condition avoids the anomalous behavior of the normalized process when $\bm{\Delta}$ approaches the origin.

	We first bound the $\psi_1$-norm of $\norm{\bm{\Delta}_0\bm{x}_{t}}_2^2$ for a fixed $\bm{\Delta}_0\in\mathbb{R}^{Q\times \prod_{m=1}^{M}p_m}$ with unit Frobenius norm. Note that $\norm{\langle \bm{\Delta}_0^j, \bm{x}_t\rangle}_{\psi_2}^2 \leq \sigma^2 \norm{\bm{\Delta}_0^j}_2^2$, where $\bm{\Delta}_0^j$ is the $j$-th row vector of $\bm{\Delta}_0$. Hence,
	\begin{align*}
		\norm{\norm{\bm{\Delta}_0\bm{x}_t}_2^2}_{\psi_1} 
		& = \norm{\sum_{j=1}^Q \langle \bm{\Delta}_0^j, \bm{x}_t\rangle^2}_{\psi_1} 
		\leq \sum_{j=1}^Q \norm{\langle \bm{\Delta}_0^j, \bm{x}_t\rangle^2}_{\psi_1} 
		= \sum_{j=1}^Q \norm{\langle \bm{\Delta}_0^j, \bm{x}_t\rangle}_{\psi_2}^2 \\
		& \leq \sigma^2 \sum_{j=1}^Q \norm{\bm{\Delta}_0^j}_\Fr^2
		=\sigma^2 \norm{\bm{\Delta}_0}_\Fr^2
		=\sigma^2,
	\end{align*}
	which implies that $\norm{\norm{\bm{\Delta}_0\bm{x}_t}_2}_{\psi_2} \leq \sigma$.

	To proceed from the pointwise sub-Gaussian property to a uniform concentration over the set $\mathcal{W}_\mathrm{LR}^\prime(\gamma)$, we utilize the generic chaining technique in Theorem 10 from \cite{banerjee2015estimation}.
	Specifically, we consider the following class of functions indexed by $\bm{\Delta} \in \mathcal{W}_\mathrm{LR}^\prime(\gamma)$:
	\begin{equation*}
		F = \left\{
		f_{\bm{\Delta}}\left\vert \bm{\Delta} \in \mathcal{W}_\mathrm{LR}^\prime(\gamma),
		f_{\bm{\Delta}}: \mathbb{R}^{\prod_{m=1}^{M}p_m}\ni \bm{x}\mapsto \frac{1}{\sqrt{\trace(\bm{\Delta}\bm{\Sigma}_x\bm{\Delta}^\top)}} 
		\norm{\bm{\Delta}\bm{x}}_2 \right.
		\right\}.
	\end{equation*}
	Let $(\Omega, \mu)$ be the probability space where $\bm{x}_t$'s are defined. For any $f_{\bm{\Delta}} \in F$,
	\[
	\norm{f_{\bm{\Delta}}}_{L_2(\mu)}^2 = \frac{1}{\trace(\bm{\Delta}\bm{\Sigma}_x\bm{\Delta}^\top)} \mathbb{E}[\trace(\bm{x}_t^\top\bm{\Delta}^\top\bm{\Delta}\bm{x}_t)] = 1,
	\]
	so that $F$ is a subset of the unit sphere of $L_2(\mu)$, i.e., $F \subseteq S_{L_2(\mu)}$.

	Meanwhile, $\sup_{f_{\bm{\Delta}} \in F} \norm{f_{\bm{\Delta}}}_{\psi_2} = \sup_{\bm{\Delta} \in \mathcal{W}_\mathrm{LR}^\prime(\gamma)} \norm{\frac{\norm{\bm{\Delta}\bm{x}_{t}}_2} {\sqrt{\trace(\bm{\Delta}\bm{\Sigma}_x\bm{\Delta}^\top)}}}_{\psi_2}\leq\sigma /\sqrt{c_x}$.
	As a result,
	\[
	\gamma_2(F, \norm{\cdot}_{\psi_2}) \leq 
	\sigma \gamma_2(F, \norm{\cdot}_{L_2(\mu)}) / \sqrt{c_x} \leq
	c_1\sigma \omega(\mathcal{W}_\mathrm{LR}^\prime(\gamma)) / \sqrt{c_x}.
	\]
	In the context of Theorem 10 from \cite{banerjee2015estimation}, let $\sqrt{T} \geq 2c_1c_2\sigma^2 \omega(\mathcal{W}_\mathrm{LR}^\prime(\gamma)) /c_x$ and $\theta = 1/2$, then the condition 
	$\theta \sqrt{T} \geq c_2\sigma \gamma_2(F, \norm{\cdot}_{\psi_2}) /\sqrt{c_x}$
	is satisfied. Therefore, by Theorem 10 from \cite{banerjee2015estimation}, we have
	\[
	\sup_{\bm{\Delta} \in \mathcal{W}_\mathrm{LR}^\prime(\gamma)}
	\left| 
	\frac{1}{T} \frac{1}{\trace(\bm{\Delta}\bm{\Sigma}_x\bm{\Delta}^\top)}
	\sum_{t=1}^{T} \norm{\bm{\Delta}\bm{x}_{t}}_2^2 - 1 
	\right| \leq \frac{1}{2}
	\]
	with probability at least $1 - \exp\left\{-\eta c_x^2 T / \sigma^4\right\}$, where $\eta$ is some positive constant.

	But by Assumption \ref{assump:reg-input} and Rayleigh's quotient, $\trace(\bm{\Delta}\bm{\Sigma}_x\bm{\Delta}^\top)=\sum_{j=1}^Q \bm{\Delta}^j{}^\top \bm{\Sigma}_x \bm{\Delta}^j\in [c_x, C_x]$ for any $\bm{\Delta} \in \mathcal{W}_\mathrm{LR}^\prime(\gamma)$. Putting together the pieces, we have when $T \gtrsim \sigma^4 \omega^2(\mathcal{W}_\mathrm{LR}^\prime(\gamma)) / c_x^2$,
	\[
	\frac{1}{2} c_x \leq \inf_{\bm{\Delta} \in \mathcal{W}_\mathrm{LR}^\prime(\gamma)} \frac{1}{T} \sum_{t=1}^T \norm{\bm{\Delta}\bm{x}_{t}}_2^2
	\leq \sup_{\bm{\Delta} \in \mathcal{W}_\mathrm{LR}^\prime(\gamma)} \frac{1}{T} \sum_{t=1}^T \norm{\bm{\Delta}\bm{x}_{t}}_2^2
	\leq \frac{3}{2}C_x
	\]
	with probability at least $1 - \exp\left\{-\eta c_x^2 T / \sigma^4\right\}$.

	Finally, we study the Gaussian width $\omega(\mathcal{W}_\mathrm{LR}^\prime(\gamma))$. 
	By Dudley's inequality and Lemma \ref{lemma:covering}(d), we have
	\begin{align}
		\nonumber
		\omega(\mathcal{W}_\mathrm{LR}^\prime(\gamma)) 
		& \leq C_1\int_{0}^{2} \sqrt{\log \mathcal{N} \left(\mathcal{W}_\mathrm{LR}^\prime(\gamma), \norm{\cdot}_2, \epsilon \right)} \mathrm{d}\epsilon \\ \nonumber
		& \leq C_2 \int_{0}^{2}\left[d_\mathrm{LR}
		\log \left(\frac{24(M\vee N)\sqrt{[(K_ys_{\max})\wedge(K_xr_{\max})]K_xK_y (r_{\max}\vee s_{\max})}}{\gamma\epsilon}\right)
		\right]^{1/2} \mathrm{d} \epsilon \\ \nonumber
		& \leq C_3 d_\mathrm{LR}^{1/2} 
		\left[\log\left(\frac{(M\vee N)\sqrt{[(K_ys_{\max})\wedge(K_xr_{\max})]K_xK_y (r_{\max}\vee s_{\max})}}{\gamma}\right)
		\right]^{1/2} \\
		& = C_3 \sqrt{d_\mathrm{LR} d_\mathrm{LR}^\prime(\gamma)},
		\label{eq:reg-GW}
	\end{align}
	where $d_\mathrm{LR}, d_\mathrm{LR}^\prime(\gamma)$ are defined in \eqref{eq:def-d-lr}. The third inequality follows from Mill's ratio after integration by parts. 
	Then, when $T \gtrsim (\sigma^4/c_x^2) (d_\mathrm{LR} d_\mathrm{LR}^\prime(\gamma))$, with probability at least $1 - \exp\left\{-c_1d_\mathrm{LR} d_\mathrm{LR}^\prime(\gamma)\right\}$, 
	\begin{align*}
		\frac{c_x}{2} \norm{\bm{\Delta}}_\Fr^2
		\leq \frac{1}{T} \sum_{t=1}^T \norm{\bm{\Delta}\bm{x}_{t}}_2^2
		\leq 2 C_x\norm{\bm{\Delta}}_\Fr^2
	\end{align*}
	holds simultaneously for all $\bm{\Delta}\in\mathcal{W}_\mathrm{LR}(\gamma)$.
\end{proof}

\begin{lemma}[\textbf{Truncated regression deviation bound}]
	\label{lemma:reg-DB}
	Suppose that Assumptions \ref{assump:reg-input}, \ref{assump:reg-error}, \ref{assump:reg-core} hold. Then, when $T \gtrsim d_\mathrm{LR} d_\mathrm{LR}^\prime(\gamma)$,
	\[
	\sup_{\bm{\Delta}\in\mathcal{W}_\mathrm{LR}^\prime(\gamma)}\frac{1}{T}\sum_{t=1}^T\langle\bm{e}_t,\bm{\Delta}\bm{x}_{t}\rangle \leq C\kappa \sigma \sqrt{d_\mathrm{LR} d_\mathrm{LR}^\prime(\gamma) /T}
	\]
	with probability at least $1 - \exp\left\{-cd_\mathrm{LR} d_\mathrm{LR}^\prime(\gamma) \right\}$,
	where $d_\mathrm{LR}, d_\mathrm{LR}^\prime(\gamma)$ are defined in \eqref{eq:def-d-lr}, and $C, c$ are some positive constants.
\end{lemma}
\begin{proof}
	The key approach to proving this lemma is to utilize Lemma \ref{lemma:chaining} for the random process $\bm{W}(\bm{\Delta}):=\frac{1}{T}\sum_{t=1}^T\langle\bm{e}_t,\bm{\Delta}\bm{x}_{t}\rangle$. To apply Lemma \ref{lemma:chaining},
	we first need to verify the mixed tail condition.
	For any $\bm{\Delta}_a, \bm{\Delta}_b \in\mathcal{W}_\mathrm{LR}^\prime(\gamma) \cup \{\bm{0}\}$ and $0 < \lambda \leq c / \left(\kappa \sigma \norm{(\bm{\Delta}_a - \bm{\Delta}_b) / T}_\op\right)$, we have
	\begin{align*}
		\mathbb{P} \left\{\bm{W}(\bm{\Delta}_a) - \bm{W}(\bm{\Delta}_b) \geq u \right\}
		& \leq \exp(-\lambda u) \mathbb{E} \left[ \exp \left(\lambda \sum_{t=1}^{T}\langle \bm{e}_t,(\bm{\Delta}_a - \bm{\Delta}_b)\bm{x}_{t} / T\rangle \right) \right]\\
		& = \exp(-\lambda u) \prod_{t=1}^{T} \mathbb{E}_x \left\{ \mathbb{E}_e \left[ \exp \left(\lambda \langle \bm{e}_t,(\bm{\Delta}_a - \bm{\Delta}_b)\bm{x}_{t}/T\rangle \right) | \bm{x}_t \right]\right\}\\
		& \leq \exp(-\lambda u) \prod_{t=1}^{T} \mathbb{E} \left[ \exp \left( C_1 \lambda^2 \kappa^2 \norm{(\bm{\Delta}_a - \bm{\Delta}_b)\bm{x}_t/T}_2^2 \right) \right] \\
		& \leq \exp \left( C_2 \lambda^2 T \kappa^2 \sigma^2 \norm{ (\bm{\Delta}_a - \bm{\Delta}_b) / T}_\Fr^2 - \lambda u \right),
	\end{align*}
	where the first inequality is a direct conclusion from Markov's inequality and the definition of $\bm{W}(\bm{\Delta})$, the first equality is due to the independence condition in Assumptions \ref{assump:reg-input} and \ref{assump:reg-error}, and the last inequality is from \cite{vershynin2019high}, Exercise 6.2.6. We have also used the fact that, conditional on $\bm{x}_{t}$, $\langle\bm{e}_t, (\bm{\Delta}_a - \bm{\Delta}_b)\bm{x}_t / T \rangle$ is a sub-Gaussian random variable with parameter $\kappa^2\norm{(\bm{\Delta}_a - \bm{\Delta}_b)\bm{x}_t / T}_2^2 $.
	By choosing $\lambda$ according to the sub-exponential tail bound, we get
	\begin{align*}
		\mathbb{P} \left\{\bm{W}(\bm{\Delta}_a) - \bm{W}(\bm{\Delta}_b) \geq u \right\}
		\leq \exp \left\{ - \min \left( \frac{u^2}{C_3^2\kappa^2 \sigma^2\norm{\bm{\Delta}_a - \bm{\Delta}_b}_\Fr^2 / T }, \frac{u}{C_3 \kappa \sigma \norm{\bm{\Delta}_a - \bm{\Delta}_b}_\Fr / T} \right) \right\}.
	\end{align*}
	The other direction of the tail bound can be easily verified with the same approach.
	
	By the resulting fixed-point mixed tail for $\bm{W}(\bm{\Delta})$, we let $d_1(\bm{\Delta}_a, \bm{\Delta}_b) = C_3 \kappa \sigma \norm{\bm{\Delta}_a - \bm{\Delta}_b}_\Fr / T$, and $d_2(\bm{\Delta}_a, \bm{\Delta}_b)=C_3 \kappa \sigma \norm{\bm{\Delta}_a - \bm{\Delta}_b}_\Fr / \sqrt{T}$.
	By Theorem 8.6.1 of \cite{vershynin2019high} and Eq. (46) and Lemma 2.7 of \cite{melnyk2016estimating}, we have
	\begin{align*}
		&\gamma_2(\mathcal{W}_\mathrm{LR}^\prime(\gamma), d_2) \leq \left(C_3  \kappa \sigma / \sqrt{T}\right) \gamma_2(\mathcal{W}_\mathrm{LR}^\prime(\gamma), \norm{\cdot}_\Fr) \leq \left(C_5 \kappa \sigma / \sqrt{T}\right) \omega(\mathcal{W}_\mathrm{LR}^\prime(\gamma)), \\
		&\gamma_1(\mathcal{W}_\mathrm{LR}^\prime(\gamma), d_1) \leq  \left(C_3  \kappa \sigma / T\right) \gamma_2^2(\mathcal{W}_\mathrm{LR}^\prime(\gamma), \norm{\cdot}_\Fr) \leq \left(C_6  \kappa \sigma / T \right) \omega^2(\mathcal{W}_\mathrm{LR}^\prime(\gamma)), \\
		&\mathrm{Diam}_2(\mathcal{W}_\mathrm{LR}^\prime(\gamma)) =  2 C_3  \kappa \sigma / \sqrt{T}, \quad
		\mathrm{Diam}_1(\mathcal{W}_\mathrm{LR}^\prime(\gamma)) = 2C_3  \kappa \sigma / T.
	\end{align*}
	Note that adding the origin $\bm{0}$ to the set $\mathcal{W}_\mathrm{LR}^\prime(\gamma)$ does not change its diameter or its Gaussian width, as $\bm{0}$ is already in the convex hull of $\mathcal{W}_\mathrm{LR}^\prime(\gamma)$.
	Let $u = \omega(\mathcal{W}_\mathrm{LR}^\prime(\gamma))$ given by \eqref{eq:reg-GW}. Substituting all above terms into Lemma \ref{lemma:chaining}, we establish that when $T \gtrsim d_\mathrm{LR} d_\mathrm{LR}^\prime(\gamma)$,
	\begin{align*}
		\label{eq:reg-dev-prob1}
		\sup_{\bm{\Delta}\in\mathcal{W}_\mathrm{LR}^\prime(\gamma)}\bm{W}(\bm{\Delta}) \leq C\kappa \sigma \sqrt{d_\mathrm{LR} d_\mathrm{LR}^\prime(\gamma) /T}
	\end{align*}
	with probability at least $1 - \exp\{ -cd_\mathrm{LR} d_\mathrm{LR}^\prime(\gamma) \}$.
\end{proof}

\begin{lemma}[\textbf{Truncated Autoregression RSC}]
	\label{lemma:AR-RSC} 
	Suppose Assumptions \ref{assump:errorauto}, \ref{assump:stationarity}, \ref{assump:auto-core} hold. When $T\gtrsim (\kappa^4\kappa_{U,B}^2 /\kappa_{L,A}^2)d_\mathrm{AR}d_\mathrm{AR}^\prime(\gamma)$, we have
	\[
	\frac{\kappa_{L, A}}{16}
	\|\bm{\Delta}\|_{\mathrm{F}}^2
	\leq \frac{1}{T}\sum_{t=1}^T\norm{\bm{\Delta}\bm{x}_{t}}_2^2
	\leq4\kappa_{U, A}\|\bm{\Delta}\|_{\mathrm{F}}^2 \quad \text{for all } \bm{\Delta}\in\mathcal{W}_\mathrm{AR}(\gamma)
	\]
	with probability at least $1-2\exp\left\{-c d_\mathrm{AR}d_\mathrm{AR}^\prime(\gamma)\right\}$, where $c>0$ is a constant, $d_\mathrm{AR}$, $d_\mathrm{LR}^\prime(\gamma)$ are defined in \eqref{eq:def-d-ar}, and $\kappa_{U,B}=C_{e}/\mu_{\mathrm{min}}(\bm{B}^{\ast})$, $\kappa_{L, A}=c_{e}/\mu_{\mathrm{max}}(\cm{A}^\ast)$, $\kappa_{U, A}=C_{e}/\mu_{\mathrm{min}}(\cm{A}^{\ast})$.
\end{lemma}
\begin{proof}
	We define the semi-norm $\norm{\bm{\Delta}}_T=\sqrt{\frac{1}{T}\sum_{t=1}^T\norm{\bm{\Delta}\bm{x}_{t}}_2^2}$. It is sufficient to show that $\sup_{\bm{\Delta}\in\mathcal{W}_\mathrm{AR}^\prime(\gamma)}\norm{\bm{\Delta}}_{T}^2 \leq4\kappa_{U, A}$ and $\inf_{\bm{\Delta}\in\mathcal{W}_\mathrm{AR}^\prime(\gamma)}\norm{\bm{\Delta}}_{T}^2 \geq\kappa_{L, A} / 16$ hold with high probability. 
	
	For any coefficient tensor $\cm{A}\in\mathcal{S}_{\mathrm{AR}}$, denote $\bm{A} = [\cm{A}]_N$. Then we have $\bm{A} = (\bm{A}_1, \dots, \bm{A}_L)$ where $\bm{A}_l = [\cm{A}_l]_N \in \mathbb{R}^{Q\times Q}$ and $Q=\prod_{n=1}^N q_n$.
	We first rewrite the VAR($L$) representation $\bm{y}_t = \sum_{l=1}^{L}\bm{A}_l \bm{y}_{t-l} + \bm{e}_t$ into a VAR($1$) form, $\bm{x}_t = \bm{B} \bm{x}_{t-1}+\bm{\zeta}_t$. From \cite{luetkepohl2005intro} we know $\{\bm{x}_t\}$ is also stationary. Also as a centered process, by Wold representation theorem, $\{\bm{x}_t\}$ admits of a VMA($\infty$) representation, given by $\bm{x}_t = \sum_{j=0}^{\infty}\bm{B}^{j}\bm{\zeta}_{t-j}$ or $\bm{z}=\bm{P}\bm{e}$, where
	\begin{equation} \label{eq:def-B-P}
	\bm{B}=
	\begin{bmatrix}
		\bm{A}_1 & \bm{A}_2 & \cdots &  \bm{A}_{L-1} &\bm{A}_L \\
		\bm{I}_Q &\bm{O} & \cdots & \bm{O} & \bm{O}\\
		\bm{O} & \bm{I}_Q & \cdots & \bm{O} & \bm{O} \\
		\vdots & \vdots & \ddots & \vdots & \vdots \\
		\bm{O} & \bm{O} & \cdots & \bm{I}_Q & \bm{O} 
	\end{bmatrix},
	\quad
	\bm{P}=
	\begin{bmatrix}
		\bm{I}_{QL} & \bm{B} & \bm{B}^2 & \cdots & \bm{B}^{T-1} & \cdots\\
		\bm{O} & \bm{I}_{QL} & \bm{B} & \cdots & \bm{B}^{T-2} &\cdots  \\
		\bm{O} & \bm{O} & \bm{I}_{QL} & \cdots & \bm{B}^{T-3} & \cdots\\
		\vdots & \vdots & \vdots & \ddots & \vdots &\cdots\\
		\bm{O} & \bm{O} & \bm{O} & \cdots & \bm{I}_{QL} & \cdots
	\end{bmatrix},
	\end{equation}
	$\bm{\zeta}_t = (\bm{e}_{t-1}^\top, \bm{0}^\top, \dots, \bm{0}^\top)^\top \in \mathbb{R}^{QL}$, $\bm{e} = (\bm{\zeta}_{{T}}^\top, \bm{\zeta}_{{T}-1}^\top, \dots, \bm{\zeta}_1^\top, \dots)^\top $ and $\bm{z} = (\bm{x}_{{T}}^\top, \bm{x}_{{T}-1}^\top, \dots, $ $\bm{x}_1^\top)^\top \in \mathbb{R}^{TQL}$. Assumption \ref{assump:stationarity} ensures that the coefficients $B^j$'s are absolutely summable and thus $\bm{P}$ and the process are well-defined \citep[Section~2.1.1]{luetkepohl2005intro}.
	By Assumption \ref{assump:errorauto}, we have $\bm{e}_t = \bm{\Sigma}_e^{1/2}\bm{\xi}_t$ and thus $\bm{e}=(\bm{I}_{\infty}\otimes\bm{\Sigma}_e^{1/2})\bar{\bm{\xi}}$, where $\bar{\bm{\xi}} = (\bar{\bm{\xi}}_{T}^\top,  \bar{\bm{\xi}}_{T-1}^\top, \dots, \bar{\bm{\xi}}_1^\top,\dots)^\top$ and $\bar{\bm{\xi}}_t = (\bm{\xi}_{t-1}^\top, \bm{0}^\top, \dots, \bm{0}^\top)^\top\in\mathbb{R}^{QL}$.
	
	Similar to the proof of Lemma \ref{lemma:reg-DB}, we consider a random process $\bm{W}(\bm{\Delta}) = \norm{\bm{\Delta}}_T^2 -\mathbb{E}\norm{\bm{\Delta}}_T^2$ and apply the conclusion in Lemma \ref{lemma:chaining}. We first verify the mixed tail condition for $\mathcal{W}_\mathrm{AR}^\prime(\gamma)$. For any $\bm{\Delta}_a, \bm{\Delta}_b\in\mathcal{W}_\mathrm{AR}^\prime(\gamma)\subset\mathcal{W}_\mathrm{AR}$, we have 
	\[\left|\bm{W}(\bm{\Delta}_a) - \bm{W}(\bm{\Delta}_b) \right| = \left|\left(\norm{\bm{\Delta}_a}_T^2 - \norm{\bm{\Delta}_b}_T^2\right) -\mathbb{E}\left(\norm{\bm{\Delta}_a}_T^2 - \norm{\bm{\Delta}_b}_T^2\right)\right|.\]
	Note that
	\begin{align*}
		\norm{\bm{\Delta}_a}_T^2 &= \frac{1}{T}\sum_{t=1}^T\norm{\bm{\Delta}_a\bm{x}_{t}}_2^2 =\frac{1}{T}\bm{z}^\top\left(\bm{I}_{T}\otimes(\bm{\Delta}_a^\top \bm{\Delta}_a)\right)\bm{z} \\
		&=\frac{1}{T}\bar{\bm{\xi}}^\top(\bm{I}\otimes\bm{\Sigma}_e^{1/2})^\top\bm{P}^\top\left(\bm{I}_{T}\otimes(\bm{\Delta}_a^\top \bm{\Delta}_a)\right)\bm{P}(\bm{I}\otimes\bm{\Sigma}_e^{1/2})\bar{\bm{\xi}}\\
		&=\bar{\bm{\xi}}^\top\left[\frac{1}{T}(\bm{I}\otimes\bm{\Sigma}_e^{1/2})^\top\bm{P}^\top\left(\bm{I}_{T}\otimes(\bm{\Delta}_a^\top \bm{\Delta}_a)\right)\bm{P}(\bm{I}\otimes\bm{\Sigma}_e^{1/2})\right]\bar{\bm{\xi}}.
	\end{align*}
	Similar transformation can be applied to $\norm{\bm{\Delta}_b}_T^2$. So we have
	\begin{align*}
		\norm{\bm{\Delta}_a}_T^2 - \norm{\bm{\Delta}_b}_T^2 = \bar{\bm{\xi}}^\top\left[\frac{1}{T}(\bm{I}\otimes\bm{\Sigma}_e^{1/2})^\top\bm{P}^\top\left(\bm{I}_{T}\otimes(\bm{\Delta}_a^\top \bm{\Delta}_a - \bm{\Delta}_b^\top \bm{\Delta}_b)\right)\bm{P}(\bm{I}\otimes\bm{\Sigma}_e^{1/2})\right]\bar{\bm{\xi}}.
	\end{align*}
	Denote $\bm{M} = \bm{\Delta}_a^\top \bm{\Delta}_a - \bm{\Delta}_b^\top \bm{\Delta}_b$. Given that $\bm{\Delta}_a, \bm{\Delta}_b \in\mathcal{W}_\mathrm{AR}^\prime(\gamma)$ and hence that $\norm{\bm{\Delta}_a}_\op\leq \norm{\bm{\Delta}_a}_\mathrm{F}=1$, $\norm{\bm{\Delta}_b}_\op\leq \norm{\bm{\Delta}_b}_\mathrm{F}=1$, we have $\norm{\bm{M}}_\mathrm{F} = \norm{(\bm{\Delta}_a  - \bm{\Delta}_b)^\top \bm{\Delta}_a + \bm{\Delta}_b^\top (\bm{\Delta}_a - \bm{\Delta}_b)}_{\mathrm{F}} \leq 2\norm{\bm{\Delta}_a  - \bm{\Delta}_b}_{\mathrm{F}}$.
	Further denote $\bm{Q} = (\bm{I}\otimes\bm{\Sigma}_e^{1/2})^\top\bm{P}^\top(\bm{I}_{T} \otimes \bm{M}) \bm{P}(\bm{I}\otimes\bm{\Sigma}_e^{1/2}) / T$. Then, we can upper bound the operator norm and Frobenius norm of $\bm{Q}$ by
	\begin{align*}
		\norm{\bm{Q}}_{\op}
		&=\norm{\frac{1}{T}(\bm{I}\otimes\bm{\Sigma}_e^{1/2})^\top\bm{P}^\top\left(\bm{I}_{T}\otimes\bm{M}\right)\bm{P}(\bm{I}\otimes\bm{\Sigma}_e^{1/2})}_{\op}\\
		&\leq\frac{1}{{T}}\norm{\bm{I}\otimes\bm{\Sigma}_e^{1/2}}_{\op}^2\norm{\bm{P}}_{\op}^2\norm{\bm{I}_{T}\otimes\bm{M}}_{\op}\\ 
		&\leq\frac{1}{{T}}C_{e}\lambda_{\mathrm{max}}(\bm{P}\bm{P}^\top)\norm{\bm{M}}_{\mathrm{F}} \leq\frac{2}{{T}}C_{e}\lambda_{\mathrm{max}}(\bm{P}\bm{P}^\top)\norm{\bm{\Delta}_a - \bm{\Delta}_b}_{\mathrm{F}}.\\
		\norm{\bm{Q}}_{\mathrm{F}}^2& = \norm{\frac{1}{T}(\bm{I}\otimes\bm{\Sigma}_e^{1/2})^\top\bm{P}^\top\left(\bm{I}_{T}\otimes\bm{M}\right)\bm{P}(\bm{I}\otimes\bm{\Sigma}_e^{1/2})}_{\mathrm{F}}^2\\
		& \leq \frac{1}{T^2} \norm{\bm{I}\otimes\bm{\Sigma}_e^{1/2}}_{\op}^4\norm{\bm{P}}_{\op}^4 \norm{\bm{I}_T\otimes\bm{M}}_{\mathrm{F}}^2\\
		&= \frac{1}{T^2} C_{e}^2\lambda_{\mathrm{max}}(\bm{P}\bm{P}^\top)^2
		\norm{\bm{I}_{T}}_{\mathrm{F}}^2  \norm{\bm{M}}_{\mathrm{F}}^2 \leq \frac{4}{{T}}C_{e}^2\lambda_{\mathrm{max}}(\bm{P}\bm{P}^\top)^2 \norm{\bm{\Delta}_a  - \bm{\Delta}_b}_{\mathrm{F}}^2.
	\end{align*}
	
	With the upper bound of $\norm{\bm{Q}}_{\op}$ and $\norm{\bm{Q}}_{\mathrm{F}}^2$, the tail probability for the difference process holds for any $u>0$ as follows:
	\begin{align*}
		&\mathbb{P}\left\{\left| \bm{W}(\bm{\Delta}_a) - \bm{W}(\bm{\Delta}_b)\right|\geq u\right\}\\
		=~ & \mathbb{P}\left\{\left|\bar{\bm{\xi}}^\top\bm{Q}\bar{\bm{\xi}} -\mathbb{E}\left(\bar{\bm{\xi}}^\top\bm{Q}\bar{\bm{\xi}}\right)\right|\geq u\right\} \\
		=~ & \mathbb{P}\left\{\left|\tilde{\bm{\xi}}^\top\tilde{\bm{Q}}\tilde{\bm{\xi}} -\mathbb{E}\left(\tilde{\bm{\xi}}^\top\tilde{\bm{Q}}\tilde{\bm{\xi}}\right)\right|\geq u\right\} \\
		\leq~ & 2\exp\left\{-c\min\left(\frac{u}{\kappa^2\|\tilde{\bm{Q}}\|_{\mathrm{op}}}, \frac{u^2}{\kappa^4\|\tilde{\bm{Q}}\|_{\mathrm{F}}^2}\right)\right\} \\
		\leq~ & 2\exp\left\{-c\min\left(\frac{u}{2\kappa^2\kappa_{U, B}\norm{\bm{\Delta}_a-\bm{\Delta}_b}_{\mathrm{F}}/{T}},\frac{u^2}{(2\kappa^2\kappa_{U, B}\norm{\bm{\Delta}_a-\bm{\Delta}_b}_{\mathrm{F}}/\sqrt{{T}})^2}\right)\right\}, \label{eq:ar-rsc-hansonwright}
	\end{align*}
	where $\kappa_{U,B}=C_{e}\lambda_{\mathrm{max}}(\bm{P}\bm{P}^\top)$, $\tilde{\bm{\xi}} = (\bm{\xi}_T^\top,\dots,\bm{\xi}_1^\top)^\top$ contains all non-zero entries of $\bm{\xi}$, and $\tilde{\bm{Q}}$ is the corresponding sub-matrix of $\bm{Q}$. The first inequality is the Hanson-Wright inequality \citep[Theorem~6.2.1]{vershynin2019high} as the entries of $\tilde{\bm{\xi}}$ are independent $\kappa^2$-sub-Gaussian random variables, and the second inequality is due to $\|\tilde{\bm{Q}}\|_{\mathrm{F}} \leq \norm{\bm{Q}}_{\mathrm{F}}$ and $\|\tilde{\bm{Q}}\|_{\mathrm{op}} \leq \norm{\bm{Q}}_{\mathrm{op}}$.

	Also, for the mixed tail for $\bm{W}(\bm{\Delta})$, we define $d_1(\bm{\Delta}_a, \bm{\Delta}_b) = 2c^\prime\kappa^2\kappa_{U, B}\norm{\bm{\Delta}_a-\bm{\Delta}_b}_{\mathrm{F}}/T$, and $d_2(\bm{\Delta}_a, \bm{\Delta}_b)=2c^{\prime\prime}\kappa^2\kappa_{U, B}\norm{\bm{\Delta}_i-\bm{\Delta}_j}_{\mathrm{F}} / \sqrt{T}$.
	By Theorem 8.6.1 of \cite{vershynin2019high} and Eq.~(46) and Lemma 2.7 of \cite{melnyk2016estimating}, we have
	\begin{equation*}\label{RSC-gamma-fcn}
		\begin{split}
			&\gamma_2(\mathcal{W}_\mathrm{AR}^\prime(\gamma),d_2)=\left(2c^{\prime\prime}\kappa^2\kappa_{U,B}/\sqrt{T}\right)\gamma_2(\mathcal{W}_\mathrm{AR}^\prime(\gamma),\norm{\cdot}_{\mathrm{F}})
			\leq\left(c_2\kappa^2\kappa_{U,B}/\sqrt{T}\right)\cdot\omega(\mathcal{W}_\mathrm{AR}^\prime(\gamma)),\\
			&\gamma_1(\mathcal{W}_\mathrm{AR}^\prime(\gamma),d_1)=\left(2c^\prime\kappa^2\kappa_{U,B}/{T}\right)\cdot\gamma_1(\mathcal{W}_\mathrm{AR}^\prime(\gamma),\norm{\cdot}_{\mathrm{F}}) \leq\left(c_1\kappa^2\kappa_{U,B}/T\right)\cdot\omega^2(\mathcal{W}_\mathrm{AR}^\prime(\gamma)),\\
			&\mathrm{Diam}_2(\mathcal{W}_\mathrm{AR}^\prime(\gamma))= 4c^{\prime\prime}\kappa^2\kappa_{U, B}/\sqrt{T},\quad
			\mathrm{Diam}_1(\mathcal{W}_\mathrm{AR}^\prime(\gamma)) = 4c^\prime\kappa^2\kappa_{U, B}/T.
		\end{split}
	\end{equation*}
	
	On the other hand, by Dudley's inequality and Lemma \ref{lemma:covering}(c), we have
	\begin{align}\label{eq:omega-ar}
		\omega(\mathcal{W}_\mathrm{AR}^\prime(\gamma)) & \leq C_1\int_{0}^{2}\sqrt{\log\mathcal{N}\left(\mathcal{W}_\mathrm{AR}^\prime(\gamma), \norm{\cdot}_{\mathrm{F}},\epsilon\right)}\mathrm{d}\epsilon \nonumber\\
		& \leq C_2\int_{0}^{2}
		\left[d_\mathrm{AR}\log\left(\frac{24N\sqrt{[(K_ys_{\max})\wedge(LK_xr_{\max})]LK_xK_y (r_{\max}\vee s_{\max})}}{\gamma\epsilon}\right)\right]^{1/2}\mathrm{d}\epsilon\nonumber\\
		& \leq C_3d_\mathrm{AR}^{1/2} 
		\left[\log\left(\frac{N\sqrt{[(K_ys_{\max})\wedge(LK_xr_{\max})]LK_xK_y (r_{\max}\vee s_{\max})}}{\gamma}\right)\right]^{1/2}\nonumber\\
		& = C_3\sqrt{d_\mathrm{AR}d_\mathrm{AR}^\prime(\gamma)}
	\end{align}
	where $d_\mathrm{AR}, d_\mathrm{AR}^\prime(\gamma)$ are defined in \eqref{eq:def-d-ar}.
	
	Let $u = \omega(\mathcal{W}_\mathrm{AR}^\prime(\gamma))$. Substituting all above terms into Lemma \ref{lemma:chaining}, we have 
	\begin{align*}
		&\mathbb{P}\left\{ \sup_{\bm{\Delta}\in\mathcal{W}_\mathrm{AR}^\prime(\gamma)}\left|\norm{\bm{\Delta}}_T^2 -\mathbb{E}\norm{\bm{\Delta}}_T^2\right| =
		\sup_{\bm{\Delta}\in\mathcal{W}_\mathrm{AR}^\prime(\gamma)}\left| \bm{W}(\bm{\Delta}) \right| > c_3\kappa^2\kappa_{U,B}\left(\sqrt{\frac{d_\mathrm{AR}d_\mathrm{AR}^\prime(\gamma)}{T}}+\frac{d_\mathrm{AR}d_\mathrm{AR}^\prime(\gamma)}{T}\right)\right\} \\
		&\hspace{330pt}\leq 2\exp\left\{-c d_\mathrm{AR}d_\mathrm{AR}^\prime(\gamma)\right\}.
	\end{align*}
	
	We now turn to the upper and lower bound for $\mathbb{E}\norm{\bm{\Delta}}_T^2$. According to \cite{basu2015regularized} (Proposition 2.3 and Equation (2.6)), we have
	\begin{align*}
		\lambda_{\mathrm{min}}\{\mathbb{E}(\bm{x}_t \bm{x}_t^\top)\} \geq \lambda_{\mathrm{min}}(\bm{\Sigma}_{e}) / \mu_{\mathrm{max}}(\cm{A}^\ast) \quad \text{and} \quad
		\lambda_{\mathrm{max}}\{\mathbb{E}(\bm{x}_t \bm{x}_t^\top)\} \leq \lambda_{\mathrm{max}}(\bm{\Sigma}_{e}) / \mu_{\mathrm{min}}(\cm{A}^\ast).
	\end{align*}
	 We denote $\kappa_{L, A}=\lambda_{\mathrm{min}}(\bm{\Sigma}_{e}) / \mu_{\mathrm{max}}(\cm{A}^\ast)$ and $\kappa_{U, A}=\lambda_{\mathrm{max}}(\bm{\Sigma}_{e}) / \mu_{\mathrm{min}}(\cm{A}^\ast)$. Note that
	\begin{align*}\label{AR-RSC-expectation}
		 \mathbb{E}\norm{\bm{\Delta}}_T^2 =& \mathbb{E}\left[\frac{1}{T} \sum_{t=1}^{T} \mathrm{vec}(\bm{\Delta}^\top)^\top (\bm{I}_Q \otimes \bm{x}_t \bm{x}_t^\top)\mathrm{vec}(\bm{\Delta}^\top)\right] = \mathrm{vec}(\bm{\Delta}^\top)^\top (\bm{I}_Q \otimes \mathbb{E} [\bm{x}_t \bm{x}_t^\top])\mathrm{vec}(\bm{\Delta}^\top),
	\end{align*}
	and by Rayleigh's quotient, $\kappa_{L, A} \leq \mathbb{E} \norm{\bm{\Delta}}_T^2 \leq \kappa_{U, A} $ for all $\bm{\Delta}\in\mathcal{W}_\mathrm{AR}^\prime(\gamma)$.
	
	Note that
	\begin{equation*}
		\inf_{\bm{\Delta}\in\mathcal{W}_\mathrm{AR}^\prime(\gamma)} \norm{\bm{\Delta}}_T^2 \geq \inf_{\bm{\Delta}\in\mathcal{W}_\mathrm{AR}^\prime(\gamma)}\mathbb{E}\norm{\bm{\Delta}}_T^2 - \sup_{\bm{\Delta}\in\mathcal{W}_\mathrm{AR}^\prime(\gamma)}\left|\norm{\bm{\Delta}}_T^2 - \mathbb{E}\norm{\bm{\Delta}}_T^2\right| \geq \kappa_{L,A} - \sup_{\bm{\Delta}\in\mathcal{W}_\mathrm{AR}^\prime}\left|\norm{\bm{\Delta}}_T^2 - \mathbb{E}\norm{\bm{\Delta}}_T^2\right|
	\end{equation*}
	and vice versa for the supremum. We get the uniform concentration inequality
	\begin{align*}
		\mathbb{P}\Biggl\{ & \kappa_{L,A} - c_3\kappa^2\kappa_{U,B}\left(\sqrt{\frac{d_\mathrm{AR}d_\mathrm{AR}^\prime(\gamma)}{T}} + \frac{d_\mathrm{AR}d_\mathrm{AR}^\prime(\gamma)}{T}\right)\leq\inf_{\bm{\Delta}\in\mathcal{W}_\mathrm{AR}^\prime(\gamma)}\norm{\bm{\Delta}}_T^2\leq\sup_{\bm{\Delta}\in\mathcal{W}_\mathrm{AR}^\prime(\gamma)}\norm{\bm{\Delta}}_T^2 \\
		& \leq \kappa_{U, A} + c_3\kappa^2\kappa_{U,B}\left(\sqrt{\frac{d_\mathrm{AR}d_\mathrm{AR}^\prime(\gamma)}{T}} + \frac{d_\mathrm{AR}d_\mathrm{AR}^\prime(\gamma)}{T}\right)\Biggl\} \geq 1-2\exp\left\{-c d_\mathrm{AR}d_\mathrm{AR}^\prime(\gamma)\right\}.
	\end{align*}
	By letting $T\gtrsim \max\{\kappa^4\kappa_{U,B}^2 /\kappa_{L,A}^2,\kappa^2\kappa_{U,B} /\kappa_{L,A}\} d_\mathrm{AR}d_\mathrm{AR}^\prime(\gamma)$, we have
	\begin{equation*}
		\frac{\kappa_{L,A}}{16}\leq\inf_{\bm{\Delta}\in\mathcal{W}_\mathrm{AR}^\prime(\gamma)}\norm{\bm{\Delta}}_T^2\leq\sup_{\bm{\Delta}\in\mathcal{W}_\mathrm{AR}^\prime(\gamma)}\norm{\bm{\Delta}}_T^2\leq 4\kappa_{U, A}
	\end{equation*}
	with probability at least $1-2\exp\left\{-c d_\mathrm{AR}d_\mathrm{AR}^\prime(\gamma)\right\}$.

	By lemma \ref{lemma:AR-spectral}, we could relax $\lambda_{\mathrm{max}}(\bm{P}\bm{P}^\top)$ to $1/\mu_{\mathrm{min}}(\cm{B}^\ast)$ where $\cm{B}^\ast(z) \vcentcolon= \bm{I}_{QL} - \bm{B}^\ast z$ and $\mu_{\mathrm{\min}}(\cm{B}^\ast) = \min_{|z|=1}\lambda_{\mathrm{min}}(\bar{\cm{B}^\ast}(z)$ $\cm{B}^\ast(z))$.
	And since $\kappa \gtrsim \norm{\bm{\xi}_{ti}}_{L_2}=1$, and $\kappa_{U,B} > \kappa_{L,A}$ by lemma \ref{lemma:AR-spectral}, the sample size requirement simplifies to $T\gtrsim (\kappa^4\kappa_{U,B}^2 /\kappa_{L,A}^2)d_\mathrm{AR}d_\mathrm{AR}^\prime(\gamma)$.
\end{proof}

\begin{lemma}[\textbf{Autoregression deviation bound}]
	\label{lemma:AR-DB}
	Suppose that Assumptions \ref{assump:errorauto}, \ref{assump:stationarity}, \ref{assump:auto-core} hold. Then, when $T\gtrsim d_\mathrm{AR}d_\mathrm{AR}^\prime(\gamma)$, 
	\[
	\sup_{\bm{\Delta}\in \mathcal{W}_\mathrm{AR}^\prime(\gamma)} \frac{1}{T}\sum_{t=1}^T\langle \bm{e}_{t},\bm{\Delta}\bm{x}_t\rangle \leq C\kappa^2\kappa_{U,B}'\sqrt{d_\mathrm{AR}d_\mathrm{AR}^\prime(\gamma)/T}
	\]
	with probability at least $1-\exp\{-c d_\mathrm{AR}d_\mathrm{AR}^\prime(\gamma)\}$, where $d_\mathrm{AR}$, $d_\mathrm{AR}^\prime(\gamma)$, $\kappa_{U, B}^\prime$ are defined in \eqref{eq:def-d-ar}, and $C, c$ are some positive constants.
\end{lemma}
\begin{proof}
	Similar to the proof of Lemma \ref{lemma:reg-DB}, we consider the random process $\bm{W}(\bm{\Delta}) \defeq \frac{1}{T}\sum_{t=1}^T\langle \bm{e}_{t},\bm{\Delta}\bm{x}_t\rangle$ and apply the conclusion in Lemma \ref{lemma:chaining}. We first need to verify the mixed tail condition. For any $\bm{\Delta}_a, \bm{\Delta}_b\in\mathcal{W}_\mathrm{AR}^\prime(\gamma)$ and $\lambda>0$, we have
	\begin{align*}
		\mathbb{P} \left\{\left|\bm{W}(\bm{\Delta}_a) - \bm{W}(\bm{\Delta}_b)\right| \geq u \right\}
		& \leq \exp(-\lambda u) \mathbb{E} \left[ \exp \left(\lambda \sum_{t=1}^{T}\langle \bm{e}_t,(\bm{\Delta}_a - \bm{\Delta}_b)\bm{x}_{t} / T\rangle \right) \right]\\
		& = \exp(-\lambda u) \mathbb{E} \left[ \exp \left(\lambda \sum_{t=1}^{T}\langle  \bm{\xi}_t, \bm{\Sigma}_e^{1/2}(\bm{\Delta}_a - \bm{\Delta}_b)\bm{x}_{t} / T\rangle \right) \right]\\
		& \leq \exp(-\lambda u) \mathbb{E} \left[ \exp\left(C^\prime\lambda^2\kappa^2 \sum_{t=1}^{T}\|\bm{\Sigma}_e^{1/2}(\bm{\Delta}_a - \bm{\Delta}_b)\bm{x}_t / T\|_2^{2}\right)\right] \\
		& \leq \exp(-\lambda u) \mathbb{E} \left[ \exp\left(C^\prime\lambda^2\kappa^2C_{e}\sum_{t=1}^{T}\|(\bm{\Delta}_a - \bm{\Delta}_b)\bm{x}_t / T\|_2^{2}\right)\right],
	\end{align*}
	where the first inequality is a direct conclusion from Markov's inequality, the equality follows from Assumption \ref{assump:errorauto}, and the second inequality uses Lemma \ref{lemma:martingale}.

	Denote $\bm{M}=(\bm{\Delta}_a - \bm{\Delta}_b)/T$. By a similar argument to Lemma \ref{lemma:AR-RSC}, we can rewrite 
	\begin{align*}
		\sum_{t=1}^{T}\|\bm{M}\bm{x}_t\|_2^{2} =\bar{\bm{\xi}}^\top\left((\bm{I}\otimes\bm{\Sigma}_e^{1/2})^\top\bm{P}^\top(\bm{I}_{T}\otimes\bm{M})^\top(\bm{I}_{T}\otimes\bm{M})\bm{P}(\bm{I}\otimes\bm{\Sigma}_e^{1/2})\right)\bar{\bm{\xi}}.
	\end{align*}
	Let $\bm{Q}_1 = (\bm{I}_{T}\otimes\bm{M})\bm{P}(\bm{I}\otimes\bm{\Sigma}_e^{1/2})$, we can upper bound $\norm{\bm{Q}_1}_{\op}$ and $\norm{\bm{Q}_1}_{\mathrm{F}}$ by
	\begin{align*}\label{lemma4:op}
		\norm{\bm{Q}_1}_{\op}&=\norm{(\bm{I}_{T}\otimes\bm{M})\bm{P}(\bm{I}\otimes\bm{\Sigma}_e^{1/2})}_{\op}
		\leq\norm{\bm{I}_{T}\otimes\bm{M}}_{\op}\norm{\bm{P}}_{\op}\norm{\bm{I}\otimes\bm{\Sigma}_e^{1/2}}_{\op}\\ 
		&\leq\frac{1}{{T}}C_{e}^{1/2}\lambda_{\mathrm{max}}^{1/2}(\bm{P}\bm{P}^\top)\norm{\bm{\Delta}_a  - \bm{\Delta}_b}_{\mathrm{F}}.\\
		\norm{\bm{Q}_1}_{\mathrm{F}}^2&=\norm{(\bm{I}_{T}\otimes\bm{M})\bm{P}(\bm{I}\otimes\bm{\Sigma}_e^{1/2})}_{\mathrm{F}}^2 \leq \norm{\bm{I}_{T}\otimes\bm{M}}_{\mathrm{F}}^2\norm{\bm{P}}_{\op}^2 \norm{\bm{I}\otimes\bm{\Sigma}_e^{1/2}}_{\op}^2\\
		&= C_{e}\lambda_{\mathrm{max}}(\bm{P}\bm{P}^\top)
		\norm{\bm{I}_{T}}_{\mathrm{F}}^2  \norm{\bm{M}}_{\mathrm{F}}^2 \leq \frac{1}{{T}}C_{e}\lambda_{\mathrm{max}}(\bm{P}\bm{P}^\top) \norm{\bm{\Delta}_a - \bm{\Delta}_b}_{\mathrm{F}}^2.
	\end{align*}
	Note that $\sum_{t=1}^{T}\|\bm{M}\bm{x}_t\|_2^{2} = \norm{\bm{Q}_1\bar{\bm{\xi}}}_2^2$, and therefore for any $0 < \lambda \leq c_1 / (\kappa^2C_{e}^{1/2}\norm{\bm{Q}_1}_{\op})$,
	\begin{align*}
		\mathbb{P} \left\{\left|\bm{W}(\bm{\Delta}_a) - \bm{W}(\bm{\Delta}_b)\right| \geq u \right\} & \leq \exp(-\lambda u) \mathbb{E} \left[ \exp\left(C^\prime\lambda^2\kappa^2C_{e}\sum_{t=1}^{T}\|\bm{M}\bm{x}_t\|_2^{2}\right)\right]\\
		& = \exp(-\lambda u) \mathbb{E} \left[ \exp\left(C^\prime\lambda^2\kappa^2C_{e} \norm{\bm{Q}_1\bar{\bm{\xi}}}_2^2 \right)\right]\\
		& \leq \exp\left(C\lambda^2\kappa^4C_{e} \norm{\bm{Q}_1}_\mathrm{F}^2 -\lambda u \right).
	\end{align*}
	The last inequality is due to \cite{vershynin2019high}, Exercise 6.2.6 and ignoring all zero entries. By choosing $\lambda$ similar to a sub-exponential tail, we have
	\begin{align*}
		&\mathbb{P} \left\{\left|\bm{W}(\bm{\Delta}_a) - \bm{W}(\bm{\Delta}_b)\right| \geq u \right\} \\ \leq&\exp\left\{-c'\min\left(\frac{u}{\kappa^2C_{e}^{1/2}\norm{\bm{Q}_1}_{\op}}, \frac{u^2}{\kappa^4C_{e}\norm{\bm{Q}_1}_{\mathrm{F}}^2}\right)\right\}\\
		\leq&\exp\left\{-\min\left(\frac{u}{c^{\prime\prime}\kappa^2\kappa_{U,B}'\norm{\bm{\Delta}_a  - \bm{\Delta}_b}_{\mathrm{F}}/T}, \frac{u^2}{\left(c^{\prime\prime}\kappa^2\kappa_{U,B}' \norm{\bm{\Delta}_a  - \bm{\Delta}_b}_{\mathrm{F}}/\sqrt{T}\right)^2}\right)\right\},
	\end{align*}
	where $\kappa_{U,B}'=C_{e}\lambda_{\mathrm{max}}^{1/2}(\bm{P}\bm{P}^\top)$, and the second inequality follows from the upper bounds of $\norm{\bm{Q}_1}_{\op}$ and $\norm{\bm{Q}_1}_{\mathrm{F}}$. 
	By the resulting mixed tail bound for $\bm{W}(\bm{\Delta})$, we let $d_1(\bm{\Delta}_a, \bm{\Delta}_b) = c^{\prime\prime}\kappa^2\kappa_{U,B}'\norm{\bm{\Delta}_a  - \bm{\Delta}_b}_{\mathrm{F}}/T$ and $d_2(\bm{\Delta}_a, \bm{\Delta}_b) = c^{\prime\prime}\kappa^2\kappa_{U,B}'\norm{\bm{\Delta}_a  - \bm{\Delta}_b}_{\mathrm{F}}/\sqrt{T}$. By Theorem 8.6.1 of \cite{vershynin2019high} and Eq. (46) and Lemma 2.7 of \cite{melnyk2016estimating}, we have
	\begin{align*}
		&\gamma_2(\mathcal{W}_\mathrm{AR}^\prime(\gamma),d_2)=\left(c^{\prime\prime}\kappa^2\kappa_{U,B}'/\sqrt{T}\right)\cdot\gamma_2(\mathcal{W}_\mathrm{AR}^\prime(\gamma),\norm{\cdot}_{\mathrm{F}})
		\leq\left(c_2\kappa^2\kappa_{U,B}'/\sqrt{T}\right)\cdot\omega(\mathcal{W}_\mathrm{AR}^\prime(\gamma)),\\
		&\gamma_1(\mathcal{W}_\mathrm{AR}^\prime(\gamma),d_1)=\left(c^{\prime\prime}\kappa^2\kappa_{U,B}'/T\right)\cdot\gamma_1(\mathcal{W}_\mathrm{AR}^\prime(\gamma),\norm{\cdot}_{\mathrm{F}})  \leq\left(c_1\kappa^2\kappa_{U,B}'/T\right)\cdot\omega^2(\mathcal{W}_\mathrm{AR}^\prime(\gamma)),\\
		&\mathrm{Diam}'_2({\mathcal{W}_\mathrm{AR}^\prime(\gamma)})= 2c^{\prime\prime}\kappa^2\kappa_{U,B}'/\sqrt{T},\quad
		\mathrm{Diam}_1'(\mathcal{W}_\mathrm{AR}^\prime(\gamma)) = 2c^{\prime\prime}\kappa^2\kappa_{U,B}'/T.
	\end{align*}
	By \eqref{eq:omega-ar}, we have $\omega(\mathcal{W}_\mathrm{AR}^\prime(\gamma))\lesssim d_\mathrm{AR}^{1/2}d_\mathrm{AR}^{\prime1/2}$, with $d_\mathrm{AR} = 2(K_x r_{\max}^2 + K_y s_{\max}^2) \sum_{n=1}^{N}q_n + 2LK_xK_yr_{\max}s_{\max}$ and $d_\mathrm{AR}^\prime = \log\left(N\sqrt{[(K_ys_{\max})\wedge(LK_xr_{\max})]LK_xK_y (r_{\max}\vee s_{\max})}\right)$.
	
	Let $u=\omega(\mathcal{W}_\mathrm{AR}^\prime(\gamma))$. Substituting all above terms into Lemma \ref{lemma:chaining}, we have when $T\gtrsim d_\mathrm{AR}d_\mathrm{AR}^\prime(\gamma)$,
	\begin{equation*}
		\sup_{\bm{\Delta}\in\mathcal{W}_\mathrm{AR}^\prime(\gamma)}\bm{W}(\bm{\Delta}) \leq C\kappa^2\kappa_{U,B}'\sqrt{d_\mathrm{AR}d_\mathrm{AR}^\prime(\gamma)/T}
	\end{equation*}
	with probability at least $1-\exp\{-c d_\mathrm{AR}d_\mathrm{AR}^\prime(\gamma)\}$.
	By lemma \ref{lemma:AR-spectral}, we could relax $\lambda_{\mathrm{max}}(\bm{P}\bm{P}^\top)$ to $1/\mu_{\mathrm{min}}(\cm{B}^\ast)$ where $\cm{B}^\ast(z) \vcentcolon= \bm{I}_{QL} - \bm{B}^\ast z$ and $\mu_{\mathrm{\min}}(\cm{B}^\ast) = \min_{|z|=1}\lambda_{\mathrm{min}}(\bar{\cm{B}^\ast}(z)$ $\cm{B}^\ast(z))$.
\end{proof}

\begin{lemma}[Spectral properties of VAR($L$) process]
	\label{lemma:AR-spectral}
	Suppose that Assumptions \ref{assump:errorauto} and \ref{assump:stationarity} hold. With the notations of Lemma \ref{lemma:AR-RSC}, we have
	\begin{itemize}
		\item[(a)] \begin{equation}
	\lambda_{\max}(\bm{P}\bm{P}^\top) \leq 1/\mu_{\mathrm{min}}(\cm{B}^\ast)
	\end{equation}
	\item[(b)] \begin{equation} \label{eq:spectral-norm-inequality}
		\mu_{\mathrm{min}}(\cm{B}^\ast) \leq \mu_{\mathrm{max}}(\cm{A}^\ast)
	\end{equation}
	\end{itemize}
\end{lemma}
\begin{proof}
	(a) Given an arbitrary vector $\bm{e} = (\bm{e}_T^\top, \bm{e}_{T-1}^\top, \dots, \bm{e}_1^\top, \dots)^\top$ where $\bm{e}_t\in\mathbb{R}^{QL}$, the $n$-th block of $\bm{P}^\top\bm{e}$ is $\sum_{j=-\infty}^{n} (\bm{B}^\top)^{n-j}\bm{e}_{j}$ for $n=0,1,\dots, T$. Then, we observe that $\bm{P}^\top\bm{e}$ is part of $\{(\bm{B}^\top)^k\}\ast \{\bm{e}_{k}\} = \{\sum_{j=-\infty}^{k} (\bm{B}^\top)^{k-j}\bm{e}_{j}\}_{k=0}^\infty$, where the asterisk denotes the convolution operator.
	The Fourier transform of $\{(\bm{B}^\top)^k\}\ast \{\bm{e}_{k}\}$ is given by $\mathcal{F}(\{(\bm{B}^\top)^k\}\ast\{\bm{e}_{k}\})(\theta) = (\sum_{k=0}^{\infty} (\bm{B}^\top)^k e^{-i k \theta})\mathcal{F}(\{\bm{e}_{k}\})(\theta) = (\bm{I}_Q-\bm{B}^\top e^{-i\theta})^{-1}\mathcal{F}(\{\bm{e}_{k}\})(\theta)$.
	Hence, by Parseval's theorem, we have
	\begin{align*}
		\norm{\bm{P}^\top\bm{e}}_2^2  &\leq \norm{\{(\bm{B}^\top)^k\}\ast \{\bm{e}_{k}\}}_2^2 \\
		&\leq \frac{1}{2\pi}\int_{0}^{2\pi} \norm{\mathcal{F}(\{(\bm{B}^\top)^k\})(\theta)}_{\op}^2 \norm{\mathcal{F}(\{\bm{e}_k\})(\theta)}_2^2 \mathrm{d}\theta \\
		& = \frac{1}{2\pi}\int_{0}^{2\pi} \lambda_{\max}\left((\bm{I}_{QL}-\bm{B}^\top e^{-i\theta})^{-1}(\bm{I}_{QL}-\bm{B}e^{i\theta})^{-1}\right) \norm{\mathcal{F}(\{\bm{e}_k\})(\theta)}_2^2 \mathrm{d}\theta \\
		& = \frac{1}{2\pi}\int_{0}^{2\pi} (1/\lambda_{\min}\left((\bm{I}_{QL}-\bm{B}^\top e^{-i\theta})(\bm{I}_{QL}-\bm{B}e^{i\theta})\right)) \norm{\mathcal{F}(\{\bm{e}_k\})(\theta)}_2^2 \mathrm{d}\theta \\
		& = \frac{1}{2\pi}\int_{0}^{2\pi} (1/\lambda_{\min}(\bar{\cm{B}}(e^{i\theta})\cm{B}(e^{i\theta}))) \norm{\mathcal{F}(\{\bm{e}_k\})(\theta)}_2^2 \mathrm{d}\theta\\
		& \leq (1/\mu_{\mathrm{min}}(\cm{B}^\ast)) \cdot \norm{\bm{e}}_2^2,
	\end{align*}
	and therefore, $\lambda_{\max}(\bm{P}\bm{P}^\top) \leq 1/\mu_{\mathrm{min}}(\cm{B}^\ast)$.

	(b) By relevant definitions,
	\begin{align*}
		\mu_{\mathrm{max}}(\cm{A}^\ast) = \max_{|z|=1} \lambda_{\mathrm{max}}(\bar{\cm{A}}(z)\cm{A}(z))&\geq \norm{\bm{I}_Q-\bm{A}_1-\cdots-\bm{A}_L}_\op^2 \\
		&= \sup_{\norm{\bm{u}}=1}\norm{\bm{u}-\bm{A}_1\bm{u}-\cdots-\bm{A}_L\bm{u}}_2^2 \\
		&= \sup_{\norm{\bm{u}}=1}\norm{(\bm{I}_{QL}-\bm{B})(\bm{u}^\top, \bm{u}^\top, \dots, \bm{u}^\top)^\top}_2^2\\
		& \geq L \lambda_{\min}((\bm{I}_{QL}-\bm{B})^\top (\bm{I}_{QL}-\bm{B}))\\
		& > \mu_{\mathrm{min}}(\cm{B}^\ast),
	\end{align*}
	concluding the proof.
\end{proof}

\begin{lemma}[\textbf{Chaining}, Theorem 3.5 in \citet{dirksenTailBoundsGeneric2015a}]
	\label{lemma:chaining} 
	Suppose that the random process $\bm{W}(\bm{\Delta})_{\bm{\Delta}\in\mathcal{W}^\prime}$ has a mixed tail in the sense that for all $\bm{\Delta}_a, \bm{\Delta}_b \in \mathcal{W}^\prime$ and $u > 0$,
	\begin{equation}
	\label{eq:mix-tail}
	\mathbb{P}\left\{\left|\bm{W}(\bm{\Delta}_a) - \bm{W}(\bm{\Delta}_b)\right|\geq u\right\}\notag \leq 2\exp\left\{-\min\left(\frac{u^2}{d_2(\bm{\Delta}_a, \bm{\Delta}_ b)^2}, \frac{u}{d_1(\bm{\Delta}_a, \bm{\Delta}_b)}\right)\right\}
	\end{equation}
	where $d_1$ and $d_2$ are two semi-metrics on $\mathcal{W}^\prime$. Assume that $\mathcal{W}^\prime$ is bounded with respect to both $d_1$ and $d_2$, and that there exists a zero point $\bm{\Delta}_0\in\mathcal{W}^\prime$ such that $\bm{W}(\bm{\Delta}_0)=0$ with probability 1.
	Then, there exists a constant $C>0$ such that for any $u\geq 1$,
	\begin{align*}
		\mathbb{P}\left\{\sup_{\bm{\Delta}\in\mathcal{W}^\prime}\left|\bm{W}(\bm{\Delta})\right| > C\left(\gamma_2(\mathcal{W}^\prime,d_2) + \gamma_1(\mathcal{W}^\prime,d_1) + u\mathrm{Diam}_2({\mathcal{W}^\prime}) + u^2\mathrm{Diam}_1(\mathcal{W}^\prime)\right) \right\}\\
		 \leq 2\exp(-u^2),
	\end{align*}
	where $\mathrm{Diam}_2({\mathcal{W}^\prime})$ and $\mathrm{Diam}_1(\mathcal{W}^\prime)$ are the diameters of $\mathcal{W}^\prime$ with respect to the semi-metric $d_2$ and $d_1$, respectively. 
\end{lemma}
\begin{proof}
	Substituting $\sqrt{u} d_2(\bm{\Delta}_a, \bm{\Delta}_b) + ud_1(\bm{\Delta}_a, \bm{\Delta}_b)$ for $u$ in the condition of this lemma, the mixed tail condition set out in Equation (3.8) of \citet{dirksenTailBoundsGeneric2015a} is satisfied. Then the conclusion follows directly from Theorem 3.5 of \citet{dirksenTailBoundsGeneric2015a} by setting the initial point $\bm{\Delta}_0$ to be the one such that $\bm{W}(\bm{\Delta}_0)=0$ with probability 1. 
\end{proof}

\begin{lemma}[\textbf{Martingale inequality}]\label{lemma:martingale} 
	Let $\{\mathcal{F}_t, t\in\mathbb{Z}\}$ be a filtration. Suppose that $\{\bm{w}_t\}$ and $\{\bm{e}_t\}$ are processes taking values in $\mathbb{R}^n$, and for each integer $t$, $\bm{w}_t$ is $\mathcal{F}_{t-1}$-measurable, $\bm{e}_t$ is $\mathcal{F}_{t}$-measurable, and $\bm{e}_t\mid\mathcal{F}_{t-1}$ is mean-zero and $\kappa^2$-sub-Gaussian distributed. Then for any $\lambda>0$, we have 
	\begin{equation*}
		\mathbb{E}\left[\exp\left(\lambda\sum_{t=1}^{T}\langle\bm{w}_t,\bm{e}_t \rangle\right)\right] \leq \mathbb{E}\left[\exp\left(C\lambda^2\kappa^2\sum_{t=1}^{T}\|\bm{w}_{t}\|_2^{2}\right)\right],
	\end{equation*}
	where $C$ is some positive constants.
\end{lemma}
\begin{proof}
	By the tower rule, we have
	\begin{align*}
		\mathbb{E}\left[\exp\left(\lambda\sum_{t=1}^{T}\langle\bm{w}_t,\bm{e}_t \rangle\right)\right] 
		&= \mathbb{E}\left[\mathbb{E}\left[ \exp\left(\lambda\sum_{t=1}^{T} \langle\bm{w}_t, \bm{e}_t\rangle \right)\mid\mathcal{F}_{T-1}\right]\right]\\
		&= \mathbb{E}\left[\exp\left(\lambda\sum_{t=1}^{T-1}\langle\bm{w}_t, \bm{e}_t \rangle\right) \mathbb{E}\left[\exp\left(\lambda \langle\bm{w}_{T}, \bm{e}_T \rangle\right) \mid\mathcal{F}_{T-1}\right]\right]\\
		&\leq \mathbb{E}\left[\exp\left(c^\prime\lambda^2\kappa^2\|\bm{w}_{T}\|_2^{2}\right) \mathbb{E}\left[\exp\left(\lambda\sum_{t=1}^{T-1}\langle\bm{w}_t, \bm{e}_t \rangle\right)\right] \right],
	\end{align*}
	where the above inequality follows from the the fact that $\langle \bm{w}_{T},\bm{e}_T\rangle \mid \mathcal{F}_{T-1}$ is mean-zero and $\kappa^2\|\bm{w}_{T}\|^2_2$-sub-Gaussian, since $\bm{e}_T\mid \mathcal{F}_{t-1}$ is mean-zero and $\kappa^2$-sub-Gaussian. Thus we have $\mathbb{E}\left[ \exp\left(\lambda\langle\bm{w}_{T}, \bm{e}_T \rangle \right) \mid\mathcal{F}_{T-1}\right] \leq \exp\left(c^\prime\lambda^2\kappa^2\|\bm{w}_{T}\|_2^{2}\right)$.
	Then induction on $t$ gives the conclusion.
\end{proof}

\section{Setups and Results of the Third Simulation Studies in Section \ref{sec:simulation}} \label{appendix:simu}
The third simulation compares our feature extracting methods with two traditional ones, hierarchical factor models (HFM) and higher-order factor models (HOFM), under the new model-designing framework. 
We consider two scenarios for the coefficient tensor $\cm{A}_\mathrm{AR}$: (i) it follows HOFM with both $\bm{\Lambda}_x$ and $\bm{\Lambda}_y$ being the matrix product of three different block diagonal matrices; and (ii) it follows our method with two action orders $\{ (1,2,3), (2,1,3) \}$ for both $\bm{\Lambda}_x$ and $\bm{\Lambda}_y$.
For both scenarios, we set the ranks as $r_{n,k} = 4$ and $s_{n,k'} = 3$ for all $n$, $k$ and $k'$, and all component and coefficient matrices are generated as in the second experiment; specifically, the construction of the loading matrices $\bm{\Lambda}_y$ and $\bm{\Lambda}_x$ for both our method and HOFM is given as follows:
\begin{itemize}
	\item For our model, we set $(K_y, K_x) = (2, 2)$, with action sets $\mathcal{P}_y(3, 2) = \mathcal{P}_x(3, 2) = \{ (1,2,3), (2,1,3) \}$, and generate loading components $\{ \{ \bm{G}_{n,k} \}_{n=1}^3 \}_{k=1}^2$ and $\{ \{ \bm{H}_{n,k^\prime} \}_{n=1}^3 \}_{k^\prime=1}^2$ by applying QR decomposition to random matrices with i.i.d. standard normal entries. Then $\bm{\Lambda}_y$ and $\bm{\Lambda}_x$ for our model are computed according to \eqref{eq:lambda-x}.
	\item For HOFM, since it performs feature extraction in a sequential manner and only considers one action order $(1,2,3)$, we simply generate the loading components $\{ \bm{G}_{n} \}_{n=1}^3$ such that $\bm{G}_{1} = \text{diag}(\bm{G}_1^{(1)}, \bm{G}_1^{(2)}, \ldots, \bm{G}_1^{(p_2p_3)})$, and $\bm{G}_{2} = \text{diag}(\bm{G}_2^{(1)}, \bm{G}_2^{(2)}, \ldots, \bm{G}_2^{(p_3)})$. Here, $\bm{G}_1^{(i)}\in\mathbb{R}^{r\times p_1}$, $\bm{G}_2^{(j)}\in\mathbb{R}^{r\times p_2r}$, and $\bm{G}_3\in\mathbb{R}^{r\times p_3r}$ are orthogonal matrices for $1\leq i\leq q_2q_3$ and $1\leq j\leq q_3$, generated by QR decomposition of random matrices with i.i.d. standard normal entries. Similarly, we generate the loading components $\{ \bm{H}_{n} \}_{n=1}^3$ and compute $\bm{\Lambda}_y=\bm{H}_1\bm{H}_2\bm{H}_3$ and $\bm{\Lambda}_x=\bm{G}_1\bm{G}_2\bm{G}_3$ accordingly.
\end{itemize}
The coefficient matrix $\bm{\Theta}_\mathrm{AR}$ is generated in the same way as in the first simulation experiment. 
The data generating process is model \eqref{model:AR(P)} with order $L=1$ and tensor mode $N=3$, i.e. $\cm{Y}_t\in\mathbb{R}^{q_1\times q_2\times q_3}$.
We set $(q_1, q_2, q_3) = (10, 11, 12)$, and error terms $\{\cm{E}_t\}$ have $i.i.d.$ standard normal entries. The coefficient tensor $\cm{A}_\mathrm{AR} = \bm{\Lambda}_y \bm{\Theta}_\mathrm{AR} \bm{\Lambda}_x^\top$ is rescaled such that $\| \cm{A}_\mathrm{AR} \|_{\mathrm{F}} < 1$,

\begin{figure}[t!]
	\centering
	\includegraphics[width=0.5\textwidth]{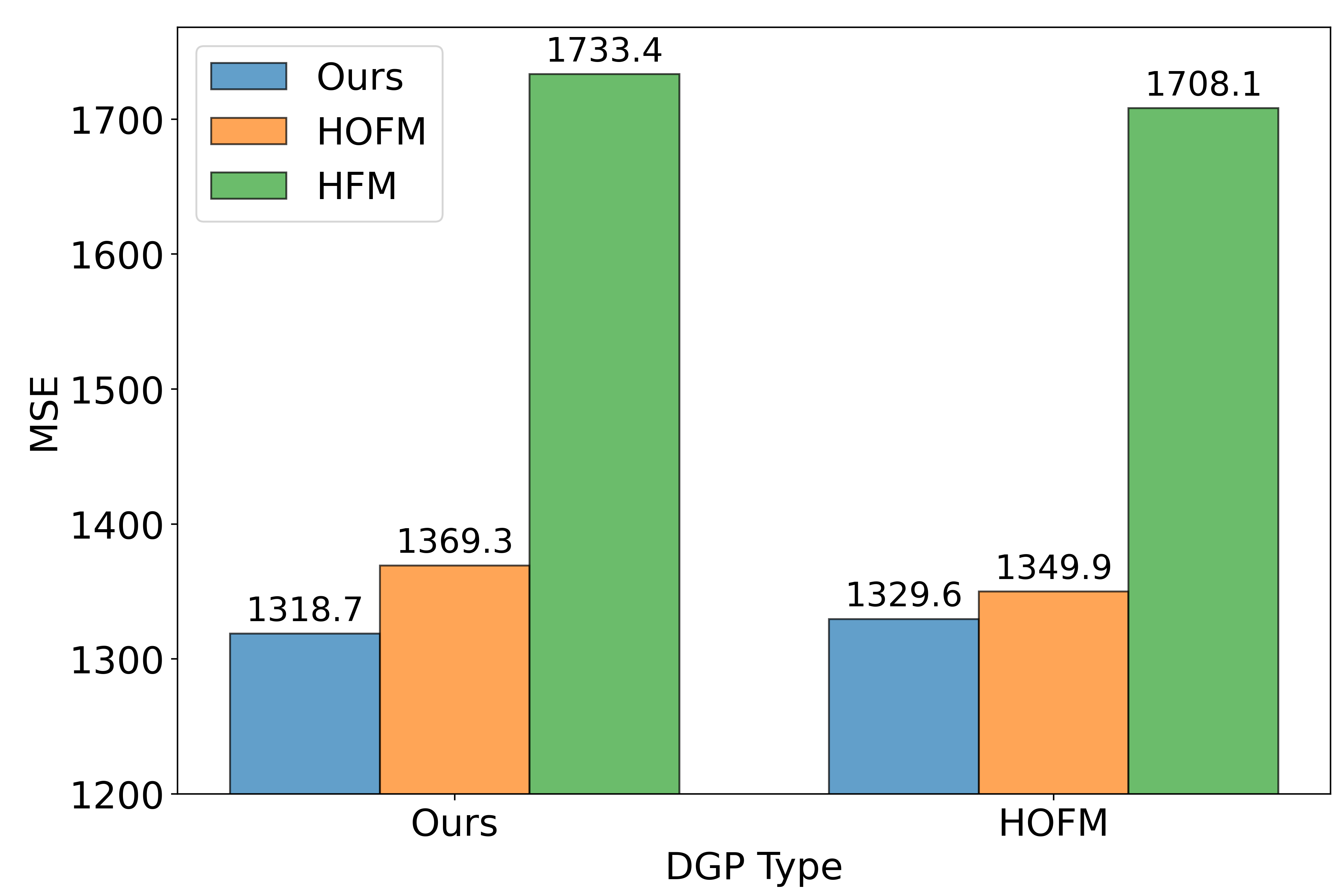}
	\caption{Average rolling forecast MSEs for HOFM, HFM, and our proposed method under our new supervised modeling framework.}
	\label{fig:sim3}
\end{figure}

Algorithm \ref{alg:als-auto-seq} is employed to estimate our models, while an alternating least squares algorithm is used for models with HFM and HOFM being feature extracting tools.
One-step ahead forecasting is used for the evaluation, and Figure \ref{fig:sim3} gives the MSEs, averaged over 500 replications. 
Our model consistently outperforms both HFM and HOFM, regardless of whether the data are generated by HOFM or our method. This may be because the fact that both HFM and HOFM involve a large number of parameters. 
As a result, the efficiency of our model in handling hierarchical structures can be well demonstrated by capturing shared components within each mode under high-dimensional settings.


\section{Additional Analysis for the Personality‑120 Data} \label{appendix:real-data}
We first offer an interpretation of Item 58 for the first action order $\alpha_1=(3,2,1)$ as introduced in Section \ref{sec:real-data}, following the way Item 46 is analyzed in the main text.
For elderly females, the important questions include Question 3 and 9 of Neuroticism trait, as well as Question 4, 7, and 13 of Openness trait. These questions are summarized as follows: ``Am afraid of many things" and ``Often feel blue" exhibit positive loadings, signifying a tendency of anxiety and depression, while ``Like to get lost in thought," ``Prefer variety to routine," and ``Do not like poetry" exhibit negative loadings, signifying a loss of imagination and adventurousness. These questions, despite being scattered across different traits, can be collectively interpreted as a concern of control and security among elderly females.

\begin{figure}[t!]
	\centering
	\includegraphics[width=1\textwidth]{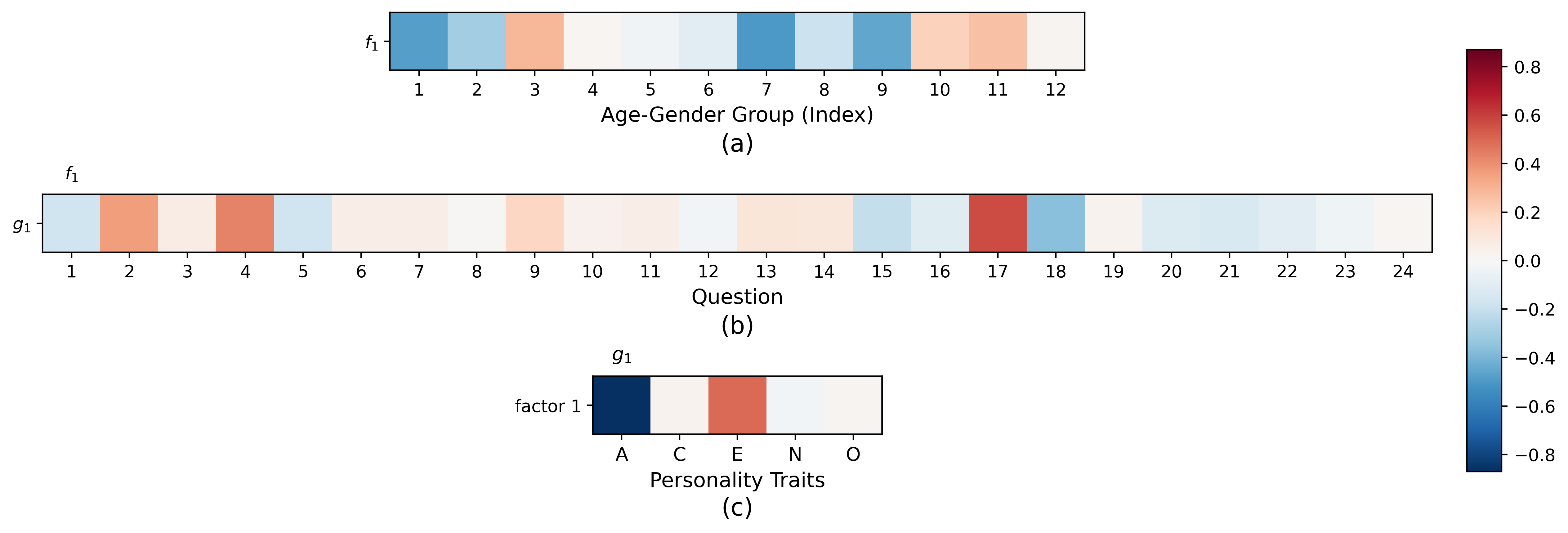}
	\caption{Heatmaps of obliquely rotated estimated component matrices and the coefficient matrix: (a) $\widehat{\bm{G}}_{1,2}^\top$, (b) $\widehat{\bm{G}}_{2,2}^\top$, (c) $\widehat{\bm{G}}_{3,2}^\top$. In (c), A, C, E, N, and O stand for Agreeableness, Conscientiousness, Extraversion, Neuroticism, and Openness, respectively.}
	\label{fig:G2s}
\end{figure}

For the second action order $\alpha_2 = (1,3,2)$, we present the obliquely rotated component matrices for both predictors and responses to illustrate the interpretability of our proposed model. Figure \ref{fig:G2s} shows the heatmaps of the estimated component matrices $\widehat{\bm{G}}_{1,2}^\top$, $\widehat{\bm{G}}_{2,2}^\top$, and $\widehat{\bm{G}}_{3,2}^\top$, whose interpretation follows the same logic as Section \ref{sec:real-data}. In Figure \ref{fig:G2s}(a), Group 1--6 represent males and Group 7--12 females. Within each group, a larger index indicates a higher age, and therefore the elderly tend to have a different attitude than the younger, especially for females.

\end{document}